\documentclass[longauth]{aa} 

\usepackage{graphicx}
\usepackage{txfonts}
\usepackage{tabularx}
\usepackage{booktabs}

\usepackage{hyperref}
\usepackage{multirow}
\usepackage{subfig}

\newcommand{\Mcoreb}{\mbox{$0.17^{+0.12}_{-0.14}$}}
\newcommand{\Mwaterb}{\mbox{$0.12^{+0.17}_{-0.10}$}}
\newcommand{\Mgasb}{\mbox{$-9.38^{+2.47}_{-2.36}$}}
\newcommand{\Fecoreb}{\mbox{$0.90^{+0.09}_{-0.08}$}}
\newcommand{\Simantleb}{\mbox{$0.39^{+0.08}_{-0.05}$}}
\newcommand{\Mgmantleb}{\mbox{$0.46^{+0.11}_{-0.10}$}}

\newcommand{\Mcorec}{\mbox{$0.14^{+0.13}_{-0.12}$}}
\newcommand{\Mwaterc}{\mbox{$0.25^{+0.22}_{-0.22}$}}
\newcommand{\Mgasc}{\mbox{$-0.81^{+0.27}_{-0.39}$}}
\newcommand{\Fecorec}{\mbox{$0.90^{+0.09}_{-0.08}$}}
\newcommand{\Simantlec}{\mbox{$0.39^{+0.08}_{-0.05}$}}
\newcommand{\Mgmantlec}{\mbox{$0.46^{+0.11}_{-0.10}$}}

\newcommand{\rev}[1]{\textbf{#1}}

\def\approxsup{%
  \def\p{%
    \setbox0=\vbox{\hbox{$>$}}%
    \ht0=0.6ex \box0 }%
  \def\s{%
    \vbox{\hbox{$\sim$}}%
  }%
  \mathrel{\raisebox{0.7ex}{%
      \mbox{$\underset{\s}{\p}$}%
    }}%
}

\begin{document}

   \title{A CHEOPS-enhanced view of the HD3167 system\thanks{This article uses data from CHEOPS program CH\_PR100008. Radial velocity and photometry data of HD\,3167 are available at the CDS via anonymous ftp to cdsarc.u-strasbg.fr (130.79.128.5) or via http://cdsarc.u-strasbg.fr/viz-bin/qcat?J/A+A/}}

   \author{
   V. Bourrier\inst{\ref{inst:1}} $^{\href{https://orcid.org/0000-0002-9148-034X}{\includegraphics[scale=0.5]{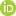}}}$,
        A. Deline\inst{\ref{inst:1}},
        A. Krenn\inst{\ref{inst:2}} $^{\href{https://orcid.org/0000-0003-3615-4725}{\includegraphics[scale=0.5]{orcid.jpg}}}$,
J. A. Egger\inst{3}, 
$^{\href{https://orcid.org/0000-0003-1628-4231}{\includegraphics[scale=0.5]{orcid.jpg}}}$,
           A.~C. Petit\inst{\ref{inst:AP}}
           $^{\href{https://orcid.org/0000-0003-1970-1790}{\includegraphics[scale=0.5]{orcid.jpg}}}$,
           L. Malavolta\inst{\ref{inst:38},\ref{inst:14}},
           M. Cretignier\inst{\ref{inst:1}},
N. Billot\inst{\ref{inst:1}} $^{\href{https://orcid.org/0000-0003-3429-3836}{\includegraphics[scale=0.5]{orcid.jpg}}}$, 
C. Broeg\inst{\ref{inst:3},\ref{inst:13}} $^{\href{https://orcid.org/0000-0001-5132-2614}{\includegraphics[scale=0.5]{orcid.jpg}}}$,
H.-G. Flor\'en\inst{\ref{inst:16}},
D. Queloz\inst{\ref{inst:39},\ref{inst:40}} $^{\href{https://orcid.org/0000-0002-3012-0316}{\includegraphics[scale=0.5]{orcid.jpg}}}$,
Y. Alibert\inst{\ref{inst:3}} $^{\href{https://orcid.org/0000-0002-4644-8818}{\includegraphics[scale=0.5]{orcid.jpg}}}$,
A. Bonfanti\inst{\ref{inst:2}} $^{\href{https://orcid.org/0000-0002-1916-5935}{\includegraphics[scale=0.5]{orcid.jpg}}}$,
A.S. Bonomo\inst{\ref{inst:51}},
J.-B. Delisle\inst{\ref{inst:1}} $^{\href{https://orcid.org/0000-0001-5844-9888}{\includegraphics[scale=0.5]{orcid.jpg}}}$,
O. D. S. Demangeon\inst{\ref{inst:5},\ref{inst:12}} $^{\href{https://orcid.org/0000-0001-7918-0355}{\includegraphics[scale=0.5]{orcid.jpg}}}$,
B.-O. Demory\inst{\ref{inst:13}} $^{\href{https://orcid.org/0000-0002-9355-5165}{\includegraphics[scale=0.5]{orcid.jpg}}}$,
X. Dumusque\inst{\ref{inst:1}},
D. Ehrenreich\inst{\ref{inst:1}} $^{\href{https://orcid.org/0000-0001-9704-5405}{\includegraphics[scale=0.5]{orcid.jpg}}}$,
R. D. Haywood\inst{\ref{inst:50},\ref{inst:50b}}$^{\href{https://orcid.org/0000-0001-9140-3574}{\includegraphics[scale=0.5]{orcid.jpg}}}$,	
S.~B Howell\inst{\ref{inst:53}} $^{\href{https://orcid.org/0000-0002-2532-2853}{\includegraphics[scale=0.5]{orcid.jpg}}}$,
M. Lendl\inst{\ref{inst:1}}, $^{\href{https://orcid.org/0000-0001-9699-1459}{\includegraphics[scale=0.5]{orcid.jpg}}}$,
A. Mortier\inst{\ref{inst:40},\ref{inst:49}}
$^{\href{https://orcid.org/0000-0001-7254-4363}{\includegraphics[scale=0.5]{orcid.jpg}}}$,
G. Nigro,
S. Salmon\inst{\ref{inst:1}} $^{\href{https://orcid.org/0000-0002-1714-3513}{\includegraphics[scale=0.5]{orcid.jpg}}}$,
S. G. Sousa\inst{\ref{inst:5}} $^{\href{https://orcid.org/0000-0001-9047-2965}{\includegraphics[scale=0.5]{orcid.jpg}}}$,
T. G. Wilson\inst{\ref{inst:4}}, $^{\href{https://orcid.org/0000-0001-8749-1962}{\includegraphics[scale=0.5]{orcid.jpg}}}$,
V. Adibekyan\inst{\ref{inst:4},\ref{inst:12}}, 
R. Alonso\inst{\ref{inst:6},\ref{inst:7}} $^{\href{https://orcid.org/0000-0001-8462-8126}{\includegraphics[scale=0.5]{orcid.jpg}}}$,
G. Anglada\inst{\ref{inst:8},\ref{inst:9}} $^{\href{https://orcid.org/0000-0002-3645-5977}{\includegraphics[scale=0.5]{orcid.jpg}}}$,
T. Bárczy\inst{\ref{inst:10}} $^{\href{https://orcid.org/0000-0002-7822-4413}{\includegraphics[scale=0.5]{orcid.jpg}}}$,
D. Barrado y Navascues\inst{\ref{inst:11}} $^{\href{https://orcid.org/0000-0002-5971-9242}{\includegraphics[scale=0.5]{orcid.jpg}}}$,
S. C. C. Barros\inst{\ref{inst:5},\ref{inst:12}} $^{\href{https://orcid.org/0000-0003-2434-3625}{\includegraphics[scale=0.5]{orcid.jpg}}}$,
W. Baumjohann\inst{\ref{inst:2}} $^{\href{https://orcid.org/0000-0001-6271-0110}{\includegraphics[scale=0.5]{orcid.jpg}}}$,
M. Beck\inst{\ref{inst:1}} $^{\href{https://orcid.org/0000-0003-3926-0275}{\includegraphics[scale=0.5]{orcid.jpg}}}$,
W. Benz\inst{\ref{inst:3},\ref{inst:13}} $^{\href{https://orcid.org/0000-0001-7896-6479}{\includegraphics[scale=0.5]{orcid.jpg}}}$,
F. Biondi\inst{\ref{inst:14}} $^{\href{https://orcid.org/0000-0002-1337-3653}{\includegraphics[scale=0.5]{orcid.jpg}}}$,
X. Bonfils\inst{\ref{inst:15}} $^{\href{https://orcid.org/0000-0001-9003-8894}{\includegraphics[scale=0.5]{orcid.jpg}}}$,			
A. Brandeker\inst{\ref{inst:16}} $^{\href{https://orcid.org/0000-0002-7201-7536}{\includegraphics[scale=0.5]{orcid.jpg}}}$,
J. Cabrera\inst{\ref{inst:18}}, 
S. Charnoz\inst{\ref{inst:19}} $^{\href{https://orcid.org/0000-0002-7442-491X}{\includegraphics[scale=0.5]{orcid.jpg}}}$,		
Sz. Csizmadia\inst{\ref{inst:18}} $^{\href{https://orcid.org/0000-0001-6803-9698}{\includegraphics[scale=0.5]{orcid.jpg}}}$,
A. Collier Cameron\inst{\ref{inst:4}} $^{\href{https://orcid.org/0000-0002-8863-7828}{\includegraphics[scale=0.5]{orcid.jpg}}}$,
M. Damasso\inst{\ref{inst:50}},		
M. B. Davies\inst{\ref{inst:20}} $^{\href{https://orcid.org/0000-0001-6080-1190}{\includegraphics[scale=0.5]{orcid.jpg}}}$,
M. Deleuil\inst{\ref{inst:21}} $^{\href{https://orcid.org/0000-0001-6036-0225}{\includegraphics[scale=0.5]{orcid.jpg}}}$,
L. Delrez\inst{\ref{inst:22},\ref{inst:23}} $^{\href{https://orcid.org/0000-0001-6108-4808}{\includegraphics[scale=0.5]{orcid.jpg}}}$,			
L. Di Fabrizio\inst{\ref{inst:52}},			
A. Erikson\inst{\ref{inst:18}}, 
A. Fortier\inst{\ref{inst:3},\ref{inst:13}} $^{\href{https://orcid.org/0000-0001-8450-3374}{\includegraphics[scale=0.5]{orcid.jpg}}}$,
L. Fossati\inst{\ref{inst:2}} $^{\href{https://orcid.org/0000-0003-4426-9530}{\includegraphics[scale=0.5]{orcid.jpg}}}$,
M. Fridlund\inst{\ref{inst:24},\ref{inst:25}} $^{\href{https://orcid.org/0000-0002-0855-8426}{\includegraphics[scale=0.5]{orcid.jpg}}}$,
D. Gandolfi\inst{\ref{inst:26}} $^{\href{https://orcid.org/0000-0001-8627-9628}{\includegraphics[scale=0.5]{orcid.jpg}}}$,
M. Gillon\inst{\ref{inst:22}} $^{\href{https://orcid.org/0000-0003-1462-7739}{\includegraphics[scale=0.5]{orcid.jpg}}}$,
M. Güdel\inst{\ref{inst:27}}, 	
K. Heng\inst{\ref{inst:13},\ref{inst:28}} $^{\href{https://orcid.org/0000-0003-1907-5910}{\includegraphics[scale=0.5]{orcid.jpg}}}$,
S. Hoyer\inst{\ref{inst:21}} $^{\href{https://orcid.org/0000-0003-3477-2466}{\includegraphics[scale=0.5]{orcid.jpg}}}$,
K. G. Isaak\inst{\ref{inst:29}} $^{\href{https://orcid.org/0000-0001-8585-1717}{\includegraphics[scale=0.5]{orcid.jpg}}}$,
L. L. Kiss\inst{\ref{inst:30},\ref{inst:32}}, 
J. Laskar\inst{\ref{inst:33}} $^{\href{https://orcid.org/0000-0003-2634-789X}{\includegraphics[scale=0.5]{orcid.jpg}}}$,
A. Lecavelier des Etangs\inst{\ref{inst:34}} $^{\href{https://orcid.org/0000-0002-5637-5253}{\includegraphics[scale=0.5]{orcid.jpg}}}$,
V. Lorenzi\inst{\ref{inst:52},\ref{inst:6}},		
C. Lovis\inst{\ref{inst:1}} $^{\href{https://orcid.org/0000-0001-7120-5837}{\includegraphics[scale=0.5]{orcid.jpg}}}$,
D. Magrin\inst{\ref{inst:14}} $^{\href{https://orcid.org/0000-0003-0312-313X}{\includegraphics[scale=0.5]{orcid.jpg}}}$,
A. Massa\inst{\ref{inst:26}},
P. F. L. Maxted\inst{\ref{inst:35}} $^{\href{https://orcid.org/0000-0003-3794-1317}{\includegraphics[scale=0.5]{orcid.jpg}}}$,
V. Nascimbeni\inst{\ref{inst:14}} $^{\href{https://orcid.org/0000-0001-9770-1214}{\includegraphics[scale=0.5]{orcid.jpg}}}$,		
G. Olofsson\inst{\ref{inst:16}} $^{\href{https://orcid.org/0000-0003-3747-7120}{\includegraphics[scale=0.5]{orcid.jpg}}}$,
R. Ottensamer\inst{\ref{inst:36}}, 
I. Pagano\inst{\ref{inst:37}} $^{\href{https://orcid.org/0000-0001-9573-4928}{\includegraphics[scale=0.5]{orcid.jpg}}}$,
E. Pallé\inst{\ref{inst:6},\ref{inst:7}} $^{\href{https://orcid.org/0000-0003-0987-1593}{\includegraphics[scale=0.5]{orcid.jpg}}}$,
G. Peter\inst{\ref{inst:17}} $^{\href{https://orcid.org/0000-0001-6101-2513}{\includegraphics[scale=0.5]{orcid.jpg}}}$,
G. Piotto\inst{\ref{inst:14},\ref{inst:38}} $^{\href{https://orcid.org/0000-0002-9937-6387}{\includegraphics[scale=0.5]{orcid.jpg}}}$,
D. Pollacco\inst{\ref{inst:28}}, 
R. Ragazzoni\inst{\ref{inst:14},\ref{inst:38}} $^{\href{https://orcid.org/0000-0002-7697-5555}{\includegraphics[scale=0.5]{orcid.jpg}}}$,
N. Rando\inst{\ref{inst:41}}, 
H. Rauer\inst{\ref{inst:18},\ref{inst:42},\ref{inst:43}} $^{\href{https://orcid.org/0000-0002-6510-1828}{\includegraphics[scale=0.5]{orcid.jpg}}}$,
I. Ribas\inst{\ref{inst:8},\ref{inst:9}} $^{\href{https://orcid.org/0000-0002-6689-0312}{\includegraphics[scale=0.5]{orcid.jpg}}}$,
N. C. Santos\inst{\ref{inst:5},\ref{inst:12}} $^{\href{https://orcid.org/0000-0003-4422-2919}{\includegraphics[scale=0.5]{orcid.jpg}}}$,
G. Scandariato\inst{\ref{inst:37}} $^{\href{https://orcid.org/0000-0003-2029-0626}{\includegraphics[scale=0.5]{orcid.jpg}}}$,
D. Ségransan\inst{\ref{inst:1}} $^{\href{https://orcid.org/0000-0003-2355-8034}{\includegraphics[scale=0.5]{orcid.jpg}}}$,
A. E. Simon\inst{\ref{inst:3}} $^{\href{https://orcid.org/0000-0001-9773-2600}{\includegraphics[scale=0.5]{orcid.jpg}}}$,
A. M. S. Smith\inst{\ref{inst:18}} $^{\href{https://orcid.org/0000-0002-2386-4341}{\includegraphics[scale=0.5]{orcid.jpg}}}$,
M. Steller\inst{\ref{inst:2}} $^{\href{https://orcid.org/0000-0003-2459-6155}{\includegraphics[scale=0.5]{orcid.jpg}}}$,
Gy. M. Szabó\inst{\ref{inst:44},\ref{inst:45}}, 
N. Thomas\inst{\ref{inst:3}}, 
S. Udry\inst{\ref{inst:1}} $^{\href{https://orcid.org/0000-0001-7576-6236}{\includegraphics[scale=0.5]{orcid.jpg}}}$,
V. Van Grootel\inst{\ref{inst:23}} $^{\href{https://orcid.org/0000-0003-2144-4316}{\includegraphics[scale=0.5]{orcid.jpg}}}$,
F. Verrecchia\inst{\ref{inst:46},\ref{inst:47}}, 
N. Walton\inst{\ref{inst:48}},
T. Beck\inst{\ref{inst:3}},
M. Buder\inst{\ref{inst:17}},
F. Ratti\inst{\ref{inst:41}}, 
B. Ulmer\inst{\ref{inst:17}},
V. Viotto\inst{\ref{inst:14}},	
          }

   \institute{Observatoire Astronomique de l'Universit\'e de Gen\`eve, Chemin Pegasi 51b, CH-1290 Versoix, Switzerland\label{inst:1}\\
              \email{vincent.bourrier@unige.ch}
         \and
            Space Research Institute, Austrian Academy of Sciences, Schmiedlstrasse 6, A-8042 Graz, Austria\label{inst:2}  \and
            Physikalisches Institut, University of Bern, Gesellschaftsstrasse 6, 3012 Bern, Switzerland\label{inst:3}  \and
\label{inst:4} Centre for Exoplanet Science, SUPA School of Physics and Astronomy, University of St Andrews, North Haugh, St Andrews KY16 9SS, UK \and
Instituto de Astrofisica e Ciencias do Espaco, Universidade do Porto, CAUP, Rua das Estrelas, 4150-762 Porto, Portugal\label{inst:5} \and
\label{inst:6} Instituto de Astrofisica de Canarias, 38200 La Laguna, Tenerife, Spain \and
\label{inst:7} Departamento de Astrofisica, Universidad de La Laguna, 38206 La Laguna, Tenerife, Spain \and
\label{inst:8} Institut de Ciencies de l'Espai (ICE, CSIC), Campus UAB, Can Magrans s/n, 08193 Bellaterra, Spain \and
\label{inst:9} Institut d'Estudis Espacials de Catalunya (IEEC), 08034 Barcelona, Spain \and
\label{inst:10} Admatis, 5. Kandó Kálmán Street, 3534 Miskolc, Hungary \and
\label{inst:11} Depto. de Astrofisica, Centro de Astrobiologia (CSIC-INTA), ESAC campus, 28692 Villanueva de la Cañada (Madrid), Spain \and
\label{inst:12} Departamento de Fisica e Astronomia, Faculdade de Ciencias, Universidade do Porto, Rua do Campo Alegre, 4169-007 Porto, Portugal \and
\label{inst:13} Center for Space and Habitability, Gesellschaftsstrasse 6, 3012 Bern, Switzerland \and
\label{inst:14} INAF, Osservatorio Astronomico di Padova, Vicolo dell'Osservatorio 5, 35122 Padova, Italy \and
\label{inst:15} Université Grenoble Alpes, CNRS, IPAG, 38000 Grenoble, France \and
\label{inst:16} Department of Astronomy, Stockholm University, AlbaNova University Center, 10691 Stockholm, Sweden \and
\label{inst:17} Institute of Optical Sensor Systems, German Aerospace Center (DLR), Rutherfordstrasse 2, 12489 Berlin, Germany \and
\label{inst:18} Institute of Planetary Research, German Aerospace Center (DLR), Rutherfordstrasse 2, 12489 Berlin, Germany \and
\label{inst:19} Université de Paris, Institut de physique du globe de Paris, CNRS, F-75005 Paris, France \and
\label{inst:20} Centre for Mathematical Sciences, Lund University, Box 118, 221 00 Lund, Sweden \and
\label{inst:21} Aix Marseille Univ, CNRS, CNES, LAM, 38 rue Frédéric Joliot-Curie, 13388 Marseille, France \and
\label{inst:22} Astrobiology Research Unit, Université de Liège, Allée du 6 Août 19C, B-4000 Liège, Belgium \and
\label{inst:23} Space sciences, Technologies and Astrophysics Research (STAR) Institute, Université de Liège, Allée du 6 Août 19C, 4000 Liège, Belgium \and
\label{inst:24} Leiden Observatory, University of Leiden, PO Box 9513, 2300 RA Leiden, The Netherlands \and
\label{inst:25} Department of Space, Earth and Environment, Chalmers University of Technology, Onsala Space Observatory, 43992 Onsala, Sweden \and
\label{inst:26} Dipartimento di Fisica, Universita degli Studi di Torino, via Pietro Giuria 1, I-10125, Torino, Italy \and
\label{inst:27} University of Vienna, Department of Astrophysics, Türkenschanzstrasse 17, 1180 Vienna, Austria \and
\label{inst:28} Department of Physics, University of Warwick, Gibbet Hill Road, Coventry CV4 7AL, United Kingdom \and
\label{inst:29} Science and Operations Department - Science Division (SCI-SC), Directorate of Science, European Space Agency (ESA), European Space Research and Technology Centre (ESTEC),
Keplerlaan 1, 2201-AZ Noordwijk, The Netherlands \and
\label{inst:30} Konkoly Observatory, Research Centre for Astronomy and Earth Sciences, 1121 Budapest, Konkoly Thege Miklós út 15-17, Hungary \and
\label{inst:32} ELTE E\"otv\"os Lor\'and University, Institute of Physics, P\'azm\'any P\'eter s\'et\'any 1/A, 1117 Budapest, Hungary \and
\label{inst:33} IMCCE, UMR8028 CNRS, Observatoire de Paris, PSL Univ., Sorbonne Univ., 77 av. Denfert-Rochereau, 75014 Paris, France \and
\label{inst:34} Institut d'astrophysique de Paris, UMR7095 CNRS, Université Pierre \& Marie Curie, 98bis blvd. Arago, 75014 Paris, France \and
\label{inst:35} Astrophysics Group, Keele University, Staffordshire, ST5 5BG, United Kingdom \and
\label{inst:36} Department of Astrophysics, University of Vienna, Tuerkenschanzstrasse 17, 1180 Vienna, Austria \and
\label{inst:37} INAF, Osservatorio Astrofisico di Catania, Via S. Sofia 78, 95123 Catania, Italy \and
\label{inst:38} Dipartimento di Fisica e Astronomia "Galileo Galilei", Universita degli Studi di Padova, Vicolo dell'Osservatorio 3, 35122 Padova, Italy \and
\label{inst:39} ETH Zurich, Department of Physics, Wolfgang-Pauli-Strasse 2, CH-8093 Zurich, Switzerland \and
\label{inst:40} Cavendish Laboratory, JJ Thomson Avenue, Cambridge CB3 0HE, UK \and
\label{inst:41} ESTEC, European Space Agency, 2201AZ, Noordwijk, NL \and
\label{inst:42} Center for Astronomy and Astrophysics, Technical University Berlin, Hardenberstrasse 36, 10623 Berlin, Germany \and
\label{inst:43} Institut für Geologische Wissenschaften, Freie UniversitÃ¤t Berlin, 12249 Berlin, Germany \and
\label{inst:44} ELTE Eötvös Loránd University, Gothard Astrophysical Observatory, 9700 Szombathely, Szent Imre h. u. 112, Hungary \and
\label{inst:45} MTA-ELTE Exoplanet Research Group, 9700 Szombathely, Szent Imre h. u. 112, Hungary \and
\label{inst:46} Space Science Data Center (SSDC), ASI, via del Politecnico, s.n.c., 2200, I-00133, Roma, Italy \and
\label{inst:47} INAF, Osservatorio Astronomico di Roma, via Frascati 33, 00078 Monte Porzio Catone (RM), Italy \and
\label{inst:48} Institute of Astronomy, University of Cambridge, Madingley Road, Cambridge, CB3 0HA, United Kingdom \and
\label{inst:49} Kavli Institute for Cosmology, University of Cambridge, Madingley Road, Cambridge CB3 0HA, UK \and
\label{inst:50} Astrophysics Group, University of Exeter, Exeter EX4 2QL, UK \and
\label{inst:50b} STFC Ernest Rutherford Fellow \and
\label{inst:51} INAF-Osservatorio Astrofisico di Torino, Via Osservatorio 20, I-10025, Pino Torinese (TO), Italy  \and
\label{inst:52} Fundación Galileo Galilei-INAF, Rambla J. A. F. Perez, 7, E-38712, S.C. Tenerife, Spain  \and
\label{inst:AP} Center for  Star and  Planet  Formation,  GLOBE  Institute,  University  of  Copenhagen,  OsterVoldgade 5-7, 1350 Copenhagen, Denmark  \and
\label{inst:53} NASA Ames Research Center, Moffett Field, CA 94035, USA  \and
             }

\authorrunning{V.~Bourrier et al.}
\titlerunning{The HD\,3167 system reassessed  }

\offprints{V.B. (\email{vincent.bourrier@unige.ch})}

   \date{Received XXX; accepted XXX}


  \abstract
   {Much remains to be understood about the nature of exoplanets smaller than Neptune, most of which have been discovered in compact multi-planet systems. With its inner ultra-short period planet b aligned with the star and two larger outer planets d-c on polar orbits, the multi-planet system \,HD 3167 features a peculiar architecture and offers the possibility to investigate both dynamical and atmospheric evolution processes. To this purpose we combined multiple datasets of transit photometry and radial velocimetry (RV) to revise the properties of the system and inform models of its planets. This effort was spearheaded by CHEOPS observations of HD\,3167b, which appear inconsistent with a purely rocky composition despite its extreme irradiation. Overall the precision on the planetary orbital periods are improved by an order of magnitude, and the uncertainties on the densities of the transiting planets b and c are decreased by a factor of 3. Internal structure and atmospheric simulations draw a contrasting picture between HD\,3167d, likely a rocky super-Earth that lost its atmosphere through photo-evaporation, and HD\,3167c, a mini-Neptune that kept a substantial primordial gaseous envelope. We detect a fourth, more massive planet on a larger orbit, likely coplanar with HD\,3167d-c. Dynamical simulations indeed show that the outer planetary system d-c-e was tilted, as a whole, early in the system history, when HD\,3167b was still dominated by the star influence and maintained its aligned orbit. RV data and direct imaging rule out that the companion that could be responsible for the present-day architecture is still bound to the HD\,3167 system. Similar global studies of multi-planet systems will tell how many share the peculiar properties of the HD3167 system, which remains a target of choice for follow-up observations and simulations.}


   \maketitle
%

\section{Introduction}

Precise knowledge of a planet mass and radius is essential to infer its internal structure and the presence of an atmosphere. This is especially relevant for small exoplanets (below $\sim$3\,R$_{\oplus}$), which could encompass a wide range of compositions from mini-Neptunes with volatile H/He envelopes, to ocean planets with water mantle and steam atmospheres, to ultra-hot rocky planets with molten lava-rich surfaces and heavyweight envelopes (e.g., \citealt{Winn2018,Otegi2020}). In that respect the HD\,3167 system is of particular interest, as it hosts three known planets (\citealt{Vanderburg2016}; \citealt{Christiansen2017}; \citealt{Gandolfi2017}): HD\,3167b (P = 0.96\,d, R$_\mathrm{p}$ = 1.70$\stackrel{+0.18}{_{-0.15}}$\,R$_{\oplus}$, M$_\mathrm{p}$ = 5.02$\pm$0.38 M$_{\oplus}$), HD\,3167d (P = 8.51\,d, M$_\mathrm{p}$\,sin\,$i$ = 6.90 $\pm$0.71 M$_{\oplus}$), and HD\,3167c (P = 29.84 days, R$_\mathrm{p}$ = 3.01$\stackrel{+0.42}{_{-0.28}}$ \,R$_{\oplus}$, M$_\mathrm{p}$ = 9.80$\stackrel{+1.30}{_{-1.24}}$\,M$_{\oplus}$). Planets b and c are transiting their nearby (47\,pc) and bright (V = 9) K0V star, allowing for detailed measurements of their radius and atmospheric properties. The intermediate planet d is not transiting, but is nonetheless expected to have a low mutual inclination with planet c based on dynamical calculations (\citealt{Dalal2019}). The orbital architecture of the HD\,3167 system is particularly intriguing, because the orbital plane of its innermost planet b is close to the stellar equatorial plane and perpendicular to the orbital planes of the outer planets d and c, on polar orbits around the star (\citealt{Dalal2019}, \citealt{Bourrier2021_3167}).

HD\,3167 c is a mini-Neptune, that is an exoplanet smaller than Neptune still harboring a substantial volatile envelope of hydrogen and helium, or possibly a large fraction of water (eg, \citealt{Mousis2020}). Transit observations in the near-IR with the Wide Field Camera 3 (WFC3) onboard the Hubble Space Telescope (HST), combined with broadband transit measurements with Kepler/K2 and Spitzer/IRAC, revealed molecular absorption in the atmosphere of HD\,3167 c (\citealt{Guilluy2021,mikal-evans_hd3167c}).

HD\,3167 b belongs to the population of ultra-short period planets (USPs, R$<$2\,R$_{\oplus}$; P$<$1\,d). These exotic worlds receive so much energy from their parent star, via irradiation, tides, or even electromagnetic induction (\citealt{Kislyakova2017,Kislyakova2018}) that they cannot retain a volatile atmosphere (eg, \citealt{Lopez2017,Winn2018}). The planetary lithosphere is further expected to weaken and melt, leading to the formation of magma oceans and potential volcanic activity, particularly on the irradiated day side (eg \citealt{Schaefer2009,Barnes2010,Gelman2011,Leger2011,Elkins2012}). When planetary equilibrium temperatures exceed $\sim$1000 K, outgassing can release massive amounts of dust and metals effluents and sustain a secondary atmosphere around the planet (\citealt{Rappaport2012,Rappaport2014,Ito2015}). \citet{Guenther2020} used high-resolution spectroscopy with UVES to search for the absorption lines of a metal-rich envelope during the transit of HD\,3167 b. While they were only able to set upper limits on tracers of this envelope (such as sodium and oxygen, \citealt{Miguel2011}), this does not preclude the possibility that their signatures vary over time (like 55\,Cnc e, \citealt{RiddenHarper2016}) or that the planet is surrounded by an envelope whose refractory content has a broadband optical signature. Measuring with high precision the radius of HD\,3167b, and of USPs in general, in combination with their mass is thus essential to investigate the presence of such exotic atmospheres and gain further insights into their mysterious nature. \\

The CHEOPS satellite (\citealt{Benz2021}) was used to characterize, with a high precision, the transits of the ultra-short period planet HD\,3167b. We combined these observations with published transit photometry of HD\,3167b and c, and with both published and new radial velocity (RV) data of the system (Sect.~\ref{sec:data_red}), to perform a global analysis of all data available on the system (Sect.~\ref{sec:data_ana_glob}), and obtain a complete and refined view of its planets bulk properties (Sect.~\ref{sec:results_sys}). These properties are used to constrain the internal structures of the two transiting planets in the system in Sect.~\ref{sec:int_struct}, and to simulate their past dynamical (Sect.~\ref{sec:dyn_evol}) and atmospheric (Sect.~\ref{sec:atm_evol}) evolution.

\section{Observations and data reduction}
\label{sec:data_red}

\subsection{CHEOPS photometry}

Transits of HD\,3167b were observed with CHEOPS within the frame of the Guaranteed Time Observation (GTO), as part of a sub-program dedicated to measuring with high precision the radius of USPs and better constrain their internal structure. Twelve visits were obtained between 2 August and 14 November 2020. We scheduled three CHEOPS orbits per visit, so that one orbit would cover the transit and two orbits would cover the pre-transit and post-transit phases, allowing us to measure the baseline stellar flux and detrend the light curve. We set an exposure time of 36.5\,s with which we expected to reach a precision of about 7\% on the transit depth in one visit. We eventually obtained an average precision of 15\% per visit due to a higher noise level than anticipated, which we partly attributed to the presence of new hot pixels in the photometric aperture. We note that no transit of planet c was caught by the CHEOPS observations. 

Observations were processed with the CHEOPS DRP (Data Reduction Pipeline, version 13.1.0; \citealt{hoyer_cheops-drp}), which performs aperture photometry and provides four light curves extracted with four different aperture radii. The so-called default aperture, with a radius of 25~pixels, had a consistent lowest rms noise level throughout all visits and we selected this data set for our analysis.

\subsection{K2 photometry}

During the 80-day long campaign 8 of its K2 mission (3 January to 23 March 2016), the Kepler space telescope acquired 30-min cadence photometry of the star HD\,3167, from which \cite{vanderburg_hd3167} discovered the transiting planets b and c.

In addition to the simple aperture photometry (SAP) and the Pre-search Data Conditioning SAP (PDCSAP), the MAST archive\footnote{\url{https://archive.stsci.edu/k2}} provides several high-level science data products (HLSP) based on different photometric extraction techniques. We compared them all and identified the ones leading to light curves with the lowest noise levels: \texttt{K2SC} \citep{aigrain_k2sc} and \texttt{K2SFF} \citep{vanderburg_k2sff}. We selected the light curve provided by \texttt{K2SFF} as it preserves the photometric variability in the data. We fitted the data with a joint model of this variability and of the transit light curves, to properly propagate uncertainties throughout all parameters. \texttt{K2SFF} proposes a best solution between photometry for different aperture shapes and sizes. We compared them one by one, confirmed that the proposed best aperture has the highest signal-to-noise ratio, and subsequently used it for our analysis. \texttt{K2SFF} does not provide uncertainties on the photometric points. We thus investigated the error values generated by the other extraction methods and selected the most conservative (largest) ones for our data set, which were computed by \texttt{K2SC\_SAP}.

We used the Gaia DR2 catalogue \citep{gaia_dr2_2016, gaia_dr2_2018} to check for contamination by nearby stars in the aperture. We only considered objects with G-band magnitude differences smaller than nine with HD\,3167, and found none within the aperture.

\subsection{Spitzer/IRAC photometry} \label{ssec:spitzer_data}

We used Spitzer data from the General Observing (GO) program 13052 (PI: M.~Werner) that contain four observations of the HD\,3167 system at 4.5 $\mu$m (channel 2). These observations covered three transits of planet b (AORs 61072896, 61072640, 68163072) on 22 and 25 October 2016 and 16 October 2018, and one transit of planet c (AOR 61070592) between 31 October and 1 November 2016. The transit of planet c was analyzed in \cite{mikal-evans_hd3167c}.

The data were downloaded from the Spitzer Heritage Archive\footnote{\url{https://sha.ipac.caltech.edu}}. We extracted and pre-processed the photometry following a method described in \cite{demory_55Cnce}, which relies on the modeling of intra-pixel sensitivity of the IRAC instrument \citep{ingalls_irac} using the bilinearly-interpolated subpixel sensitivity (BLISS) mapping technique \citep{stevenson_bliss}. Our method also includes a correction as a linear function of the full-width at half-maximum (FWHM) of the pixel response function (PRF). An additional $\log^2$ ramp as a function of time was added for the pre-processing of the transit of planet c. The uncertainties associated with these corrections were propagated to the error bars of the resulting data points. The four resulting de-trended time series were sampled at a cadence of 27\,sec and we measured a negligible red noise.

\subsection{HST photometry} \label{ssec:hst_data}

We used five transit observations of HD\,3167c collected with HST/WFC3 using the G141 grism configuration (wavelength range 1.1-1.7\,$\mu$m). These observations are part of the GO program 15333 (PI: I. Crossfield) and they were acquired on 22 May 2018, 20 July 2018, 14 June 2019, 12 August 2019, and 5 July 2020. Each of the five visits covers seven HST orbits.

We used the broadband photometric light curves presented in \cite{mikal-evans_hd3167c} that we obatined as a three-column file with the time of observation in JD\_UTC, the normalized flux, and the normalized uncertainty. The light curves were extracted from the sum of all spectra across the full wavelength range. The resulting data set is made of 69.6-second exposures at a cadence of 111 seconds. We also downloaded the raw data from the MAST archive\footnote{\url{https://archive.stsci.edu/hst/wfc3}} to have access to the housekeeping parameters.

We converted the time from JD\_UTC to BJD\_TDB (barycentric Julian date in barycentric dynamical time) using the Python package \texttt{astropy} \citep{astropy1, astropy2} and assuming that the HST spacecraft is located at the center of the Earth. This approximation leads to a timing error of $\pm23.08\,\text{ms}$, which we considered negligible. The correction from JD\_UTC to BJD\_TDB is significant as it leads to correction of up to nearly 10\,minutes.

The light curves feature strong systematics that are typical of HST/WFC3 observations, with one ramp repeatable as a function of HST orbital phase, and one global ramp as a function of time. The flux level also jumps every two points due to the switching from forward to backward scanning of the HST detector between two consecutive exposures. This results in having two mean flux levels that have to be fitted independently with two offsets.

In addition, we carefully checked for the possibility of HD\,3167b transiting during these observations and found out that such a double event occurs in visits 3 and 4, and was not reported in previous analysis of these data sets. We therefore included planet b in the analysis of our HST time series.

\subsection{Radial velocity data}

The RV data analyzed in this work are coming from several instruments. We used data from HARPS programs 097.C-0948 and 098.C-0860, as published in \citealt{gandolfi_hd3167}), data from APF/Levy and Keck/HIRES, as published in \cite{Christiansen2017}, and data from the HARPS-N GTO programme (\citealt{Cosentino2012}), as well as from programs A33TAC\_15 (PI: D. Gandolfi), CAT16B\_61 (PI: H. J. Deeg), A34DDT2 and A36DDT2 (PI: G. H\'ebrard). The HARPS-N data were partly published in \cite{Christiansen2017}, \cite{gandolfi_hd3167} and \citet{Dalal2019}, but we publish here 42 new RV points from the HARPS-N GTO programme.

The HARPS-N RV data were extracted from the instrument raw frames using the latest version of the ESPRESSO data reduction software (DRS, version 2.3.5). Following the work described in \cite{dumusque_harpsn_2021}, the ESPRESSO pipeline has been optimized to work with HARPS-N data. Compared to the HARPS-N DRS version 3.8, the new version of the reduction pipeline, along with the performed optimizations, provide smaller night-to-night variations, estimated to be 0.5 m/s rms compared to 0.8 m/s, and a better stability of the RVs on the long-term, due to a careful selection of Thorium lines used for calibrating the instrument. The new pipeline further extracts RVs from cross-correlation functions computed with improved binary masks (here the G9 mask closer in type to HD\,3167), built with weights more representative of the RV information content of each spectral line (\citealt{Bourrier2021_3167}). We rejected the observation `2018-01-08T20-30-55.308' because it was not possible to correct for the color effect induced by Earth atmospheric diffusion. We removed observation `2017-11-14T21-32-32.784' as well, because the corresponding RV was clearly an outlier of the RV time series, and all stars observed during the same night showed outliers as well. Finally, we removed all observations that were taken at an airmass larger than 1.7, to prevent color dependencies in the RV time series (the ADC corrects for atmospheric extinction up to an airmass of 2.0). This selection gives us a total of 213 HARPN-N RV measurements to analyze.

Merging the data from all four instruments, we obtain a time series of 434 RV data points. To prevent biases induced by the Rossiter-McLaughlin signals from the planets b and c \citep{Dalal2019, Bourrier2021_3167}, we discarded the 111 data points observed during their transits. The filtered RV data set that we included in our analysis represents a total of 323 data points (39 HARPS points, 102 APF/Levy points, 55 Keck/HIRES points, and 127 HARPS-N points) covering more than 5.3 years ($\sim$ 1940 days).

In addition to the RV signals, the HARPS and HARPS-N data sets included several stellar activity indicators. The HARPS data included the bisector inverse slope span (`BIS SPAN') of the cross-correlation function (CCF), the full-width at half maximum (FWHM) of the CCF, and the $\log R_{HK}'$. The activity indicators provided with the HARPS-N data were the `BIS SPAN' of the CCF, the FWHM of the CCF, the contrast, the S$_\text{MW}$-index \citep{2011arXiv1107.5325L}, the H$\alpha$-index \citep{2011A&A...534A..30G}, the $\mathrm{Na\,I}$ lines \citep{2007MNRAS.378.1007D}, and the $\mathrm{Ca\,II}$ lines.


\section{Global analysis of the system}
\label{sec:data_ana_glob}

\subsection{Stellar properties}
\label{sec:star}

We derived the stellar atmospheric parameters ($T_{\mathrm{eff}}$, $\log g$, microturbulence, [Fe/H]) using ARES+MOOG, following the same methodology described in \citet[][]{Santos-13,Sousa-14,Sousa-21}. We used the latest version of ARES \footnote{The last version, ARES v2, can be downloaded at https://github.com/sousasag/ARES} \citep{Sousa-07, Sousa-15} to measure the equivalent widths (EW) of iron lines on the combined HARPS-N spectrum of HD3167. We used a minimization process to find ionization and excitation equilibrium and converge to the best set of spectroscopic parameters. This process makes use of a grid of Kurucz model atmospheres \citep{Kurucz1993} and the radiative transfer code MOOG \citep{Sneden-73}. The same method was also applied to a combined spectrum from HARPS observations, providing a completely compatible set of parameters. The stellar abundances [Mg/H] = 0.07 $\pm$ 0.03 dex and [Si/H] = 0.00 $\pm$ 0.04 dex were derived using the classical curve-of-growth analysis method assuming local thermodynamic equilibrium \citep[e.g.,][]{Adibekyan-12, Adibekyan-15}. The same codes and models were used for the abundance determinations. 

We determine the radius of HD\,3167 using the infrared flux method (IRFM; \citealt{Blackwell1977}) in a Markov-Chain Monte Carlo (MCMC) approach \citep{Schanche2020}. We constructed spectral energy distributions (SEDs) from stellar atmospheric models using the stellar parameters that were derived via the spectral analysis detailed above as priors. These fluxes are compared to observed broadband photometry to derive the apparent bolometric flux, and hence the stellar angular diameter and effective temperature of HD\,3167. To achieve this we retrieve data taken from the most recent data releases for the following bandpasses; {\it Gaia} G, G$_{\rm BP}$, and G$_{\rm RP}$, 2MASS J, H, and K, and {\it WISE} W1 and W2 \citep{Skrutskie2006,Wright2010,GaiaCollaboration2021} and use stellar atmospheric models from the \textsc{atlas} Catalogues \citep{Castelli2003}. The stellar angular diameter is converted to the stellar radius using the offset corrected {\it Gaia} EDR3 parallax \citep{Lindegren2021} from which we obtain $R_{\star}=0.871\pm0.006\, R_{\odot}$.

Stellar mass $M_{\star}$ and age $t_{\star}$ were derived from isochrones starting from $T_{\mathrm{eff}}$, [Fe/H], and $R_{\star}$. To make our final estimates more robust we adopted two different stellar evolutionary models, namely PARSEC\footnote{\textit{PA}dova and T\textit{R}ieste \textit{S}tellar \textit{E}volutionary \textit{C}ode: \url{http://stev.oapd.inaf.it/cgi-bin/cmd}} v1.2S \citep{marigo17} and CLES \citep[Code Liègeois d'Évolution Stellaire,][]{scuflaire08}. In detail, we inferred a first pair of mass and age values by interpolating the input values within pre-computed grids of PARSEC isochrones and tracks through the isochrone placement technique presented in \citet{bonfanti15,Bonfanti2016}. To further improve the convergence we also inputted $v\sin{i}=2.41\pm0.37$ km/s \citep{Bourrier2021_3167} into the code to benefit of the synergy between isochrones and gyrochronology as described in \citet{Bonfanti2016}. The second pair of mass and age, instead, was inferred by injecting the stellar input values into the CLES code, which retrieves the best-fit output values following the Levenberg-Marquardt minimization scheme \citep[see][for the details]{salmon21}.
As thoroughly described in \citet{bonfanti21}, we finally merged the two respective pairs of outcomes after carefully checking their mutual consistency through a $\chi^2$-based criterion and we obtained $M_{\star}=0.852_{-0.015}^{+0.026}\,M_{\odot}$ and $t_{\star}=10.2_{-2.4}^{+1.8}$ Gyr. Relevant stellar parameters are summarized in Tab.~\ref{tab:stellarParam}.

\begin{table}
\caption{HD\,3167: Stellar parameters.}
\label{tab:stellarParam}      
\centering                    
\begin{tabular}{llll}        
\hline\hline
\multicolumn{2}{l}{Parameter} & Value & Method \\
\hline
   $T_{\mathrm{eff}}$ & [K] & $5300\pm73$ & spectroscopy \\
   $\log{g}$ & [cgs]      & $4.47\pm0.12$ & spectroscopy \\\relax
   [Fe/H] & [dex] & $0.037\pm0.048$ & spectroscopy \\\relax
   [Mg/H] & [dex] & $0.07\pm0.03$ & spectroscopy \\\relax
   [Si/H] & [dex] & $0.00\pm0.04$ & spectroscopy \\
   $d$ & [pc] & $47.39\pm0.04$ & Gaia parallax\tablefootmark{(a)}\\
   $\theta$ & [mas] & $0.172\pm0.001$ & IRFM \\
   $R_{\star}$ & [$R_{\odot}$] & $0.871\pm0.006$ & IRFM \\
   $M_{\star}$ & [$M_{\odot}$] & $0.852_{-0.015}^{+0.026}$ & isochrones \\
   $t_{\star}$ & [Gyr]  & $10.2_{-2.4}^{+1.8}$ & isochrones \\
   $L_{\star}$ & [$L_{\odot}$] & $0.537\pm0.031$ & from $R_{\star}$ and $T_{\mathrm{eff}}$\\
   $\rho_{\star}$ & [$\rho_{\odot}$] & $1.289\pm0.041$ & from $R_{\star}$ and $M_{\star}$ \\
\hline
\end{tabular}
\tablefoot{\tablefoottext{a}{Correction from \citet{Lindegren2021} applied}}
\end{table}

\subsection{Joint photometry - velocimetry analysis}

We performed a joint fit combining all the photometry and velocimetry datasets described in Sect.~\ref{sec:data_red}. In the following subsections, we detail how we modeled the planetary signals (transits and RV) consistently in every time series (Sect.~\ref{sssec:planet_models}), and how we corrected the systematics related to each instrument in the CHEOPS, K2, Spitzer, HST and RV data sets (Sect.~\ref{sssec:cheops_analysis} to \ref{sssec:rv_analysis}). Our approach consisted in first performing an analysis of each data set separately to better identify their specificity, and then jointly fitting all the data together (Sect.~\ref{sssec:joint_fit}).

\subsubsection{Planetary signals} \label{sssec:planet_models}

The transits of planets b and c were modeled using the python package \texttt{batman} \citep{kreidberg_batman}. We selected the quadratic law to describe the effect of the stellar limb darkening, and defined a set of two free coefficients for each of the four instrument passbands. We reduced the number of free parameters by one by using the third Kepler's law and by fitting for the stellar density $\rho_\star$ instead of the normalized semi-major axes $a/R_\star$. With this approach, the planetary properties are fitted while accounting for the stellar unicity. We took advantage of exploiting photometry from four instruments to perform broadband transmission spectroscopy, letting the planet-to-star radii ratios of both transiting planets vary between each passband.


We modeled the RV planetary signals with Keplerian functions, and performed the joint fit of all planets while fitting for the following parameters: the time of inferior conjunction $T_0$, the orbital period $P$, the combinations of the eccentricity and the argument of periastron $e\cos\omega$ and $e\sin\omega$, and the RV semi-amplitude $K$. The systemic velocity $v_\gamma$ was fitted independently for each instrument (see Sect.~\ref{sssec:rv_analysis}). Additional free parameters were used for the two transiting planets: the planet-to-star radii ratio $k$, the orbital inclination $i$, and a common stellar density $\rho_\star$. This represents a total of five free parameters per planet, with five more for the transit light curves. We made use of a normal prior on the stellar density $\rho_\star\sim1.289\pm0.048\,\rho_\odot$ that we derived from the stellar properties (see Sect.~\ref{sec:star}).

\subsubsection{CHEOPS photometry} \label{sssec:cheops_analysis}

We discarded 69 out of 3496 (1.97\%) CHEOPS data points flagged by the DRP (DRP `EVENT' flag $>0$). Among the remaining 3427 data points, we identified 48 (1.40\%) outliers by performing a $3\sigma$-clipping visit by visit, and also flagged 121 (3.53\%) points with background levels higher than $4.1\cdot10^5$ electrons beyond which correlations with the flux start to appear \citep[e.g., Fig.~2 of][]{Deline2022_WASP-189b}. 

\begin{figure}
    \centering
    \includegraphics[width=\hsize]{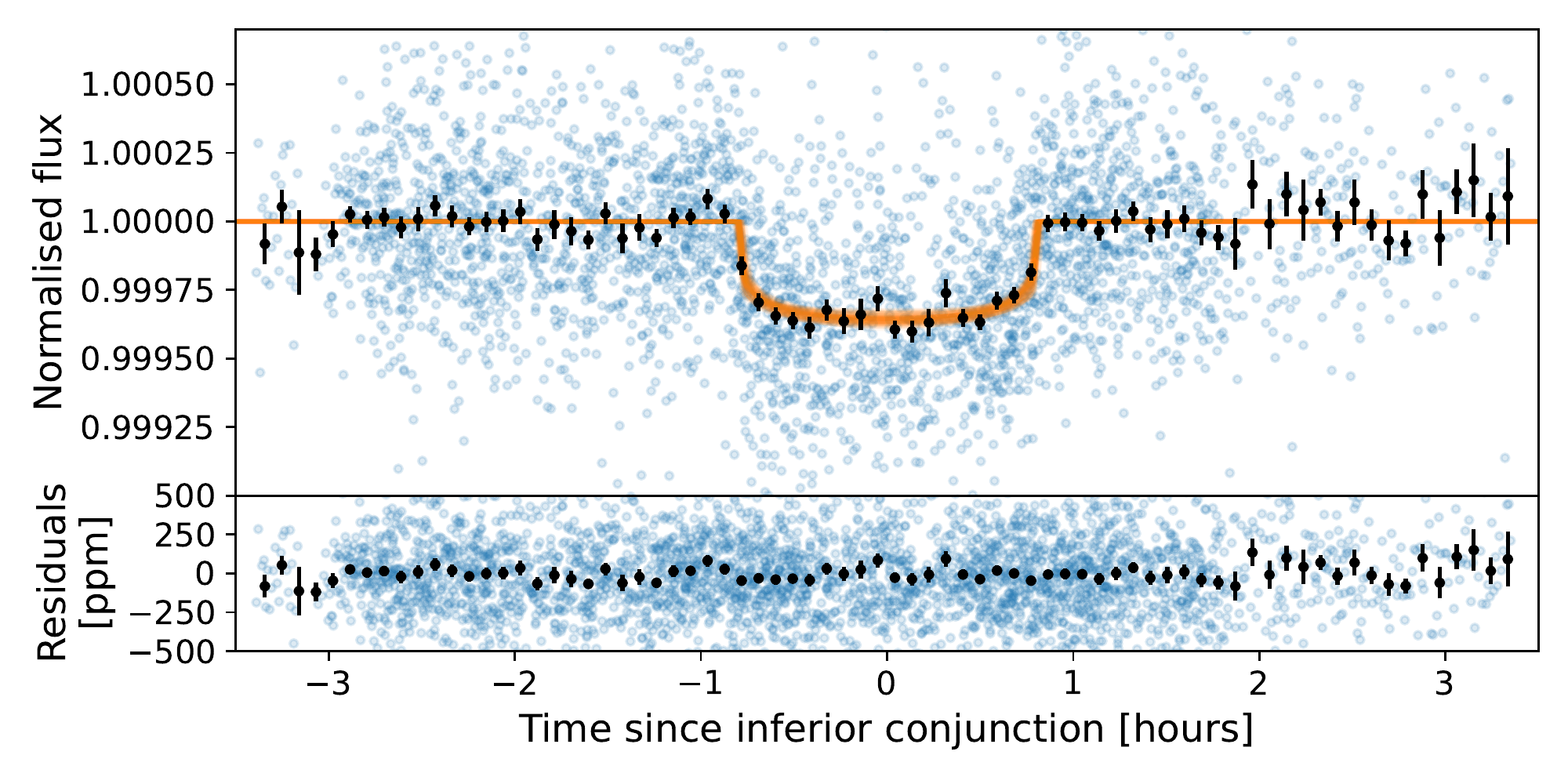}
    \caption{CHEOPS phase-folded transit of HD\,3167\,b. Top panel shows the de-trended transit data points (blue) and the binned data (black) obtained from the joint fit. A sample of 100 transit light curves drawn from the posterior distribution is represented in orange. The lower panel shows the best-fit residuals.}
    \label{fig:cheops_transit}
\end{figure}

After validating that planet c never transits during the several CHEOPS observations, we included a transit model for planet b only using the \texttt{batman} python package \citep{kreidberg_batman}. Sect.~\ref{sssec:planet_models} describes in details the joint modeling of planetary signals. We de-trended each visit with a Gaussian process (GP) as a function of the spacecraft roll angle using a Mat\'ern-3/2 kernel from the \texttt{celerite2} package \citep{foreman-mackey_celerite, foreman-mackey_celerite2}. The hyper-parameter values were the same for all visits, but each visit was fitted independently. We also included a slope in our model for some visits (no.~2,\,3,\,6,\,7,\,8,\,10,\,12) showing a significant linear trend. We quadratically added a jitter term to the error bars of each visit in order to account for underestimation of uncertainties. This leads to a total number of 33 free parameters for the correction of CHEOPS-related systematics, with one flux mean level per visit, one jitter term per visit, a slope for seven visits, and two hyper-parameters for the GP model. The de-trended CHEOPS transit of planet b obtained after the joint-fit are shown in Fig.~\ref{fig:cheops_transit} and the individual light curves are in Fig.~\ref{fig:cheops_raw_data}.

\subsubsection{K2 photometry} \label{sssec:k2_analysis}

We performed a preliminary minimization fit to remove the transits of planets b and c (modeled with the \texttt{batman} package) and the long-term photometric variability (modeled with \texttt{celerite2} GP and a Mat\'ern-3/2 kernel) in order to identify outlying data points from the residuals. We used a $6.5\sigma$-clipping criterion on the residuals to discard 23 outliers out of the 3448 data points (0.67\%). The choice of $6.5\sigma$ was motivated as a good trade-off between efficient clipping and avoiding cutting off non-outliers in the noisiest parts of the time series.

\begin{figure}
    \centering
    \includegraphics[width=\hsize]{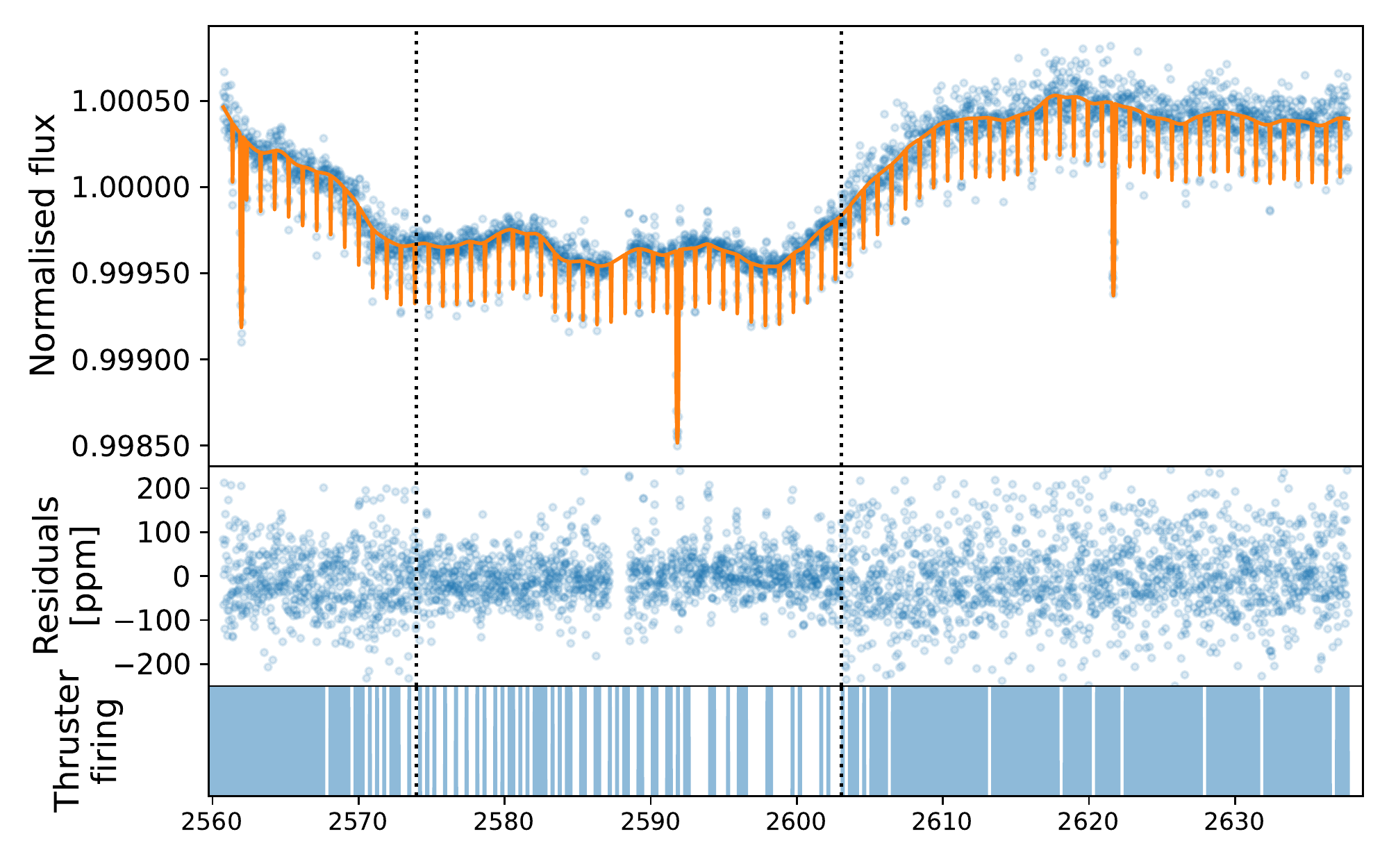}
    \caption{Noise level time ranges in the K2 time series. Top panel: Normalized K2 flux (blue points) with the best-fit model (transits + GP) obtained from minimization. Mid-panel: Residuals data after removing best-fit model. Bottom panel: Flag indicating when the K2 spacecraft is firing its thrusters to correct pointing drifts. In all three panels, the vertical dotted black lines show the noise jump timings providing the best likelihood.}
    \label{fig:k2_noise}
\end{figure}

From the residuals, we spotted by eye some significant changes in the spread of data points indicating differences in the noise level over the 80-day long observations. We identified three time ranges with the middle one having the lowest apparent noise level (see Fig.~\ref{fig:k2_noise}). We investigated the cause of this phenomenon and found that it correlates well with the frequency at which K2's thrusters are firing to correct the pointing drift of the spacecraft. We identified precisely the times at which the noise level changes by selecting the times providing the best likelihood among several minimization fits of the light curve. Each fit was performed using the model described before (GP and transit models for planets b and c) and allocating an individual jitter term per time range. We computed the best-fit likelihood for several pairs of times and selected the time pairs with the maximimum best-fit likelihood. The final jump timings were fixed at BJD$_\mathrm{TDB}$ times of 2\,457\,406.95 and 2\,457\,436.07, respectively. By considering these three time ranges separately and using individual jitter terms, we aimed at limiting the bias induced by noisy regions due the underestimation of error bars, and at maximising the precision obtained from the low-noise middle region.

For the final fit of K2 time series, we used the same model as for the preliminary fit (\texttt{batman} transit models and Mat\'ern-3/2 GP kernel). The transit models are oversampled with respect to the data cadence by a factor 30 (oversampled cadence of about 1\,min), and binned down to the sampling rate of the light curve. This technique allows one to account for distortion effects due to long integration times \citep[e.g., ][]{kipping_binning} with strong effect during ingress and egress especially. The GP model fits for the correlated noise in the data that corresponds to both instrumental systematics and stellar activity. \cite{gandolfi_hd3167} pointed out the presence of the latter in the K2 data with a significant peak in the periodogram matching the stellar rotation period at $\sim$ 24 days. We placed normal priors on both GP hyper-parameters to help convergence of the fit based on the values obtained when analysing the K2 time series alone. The prior values are $\mathcal{N}\!\left(370, 70\right)\,\mathrm{ppm}$ for the GP amlitude $\sigma_\mathrm{GP}$, and $\mathcal{N}\!\left(10, 0.5\right)\,\mathrm{days}$ for the GP correlation time scale $\rho_\mathrm{GP}$, where $\mathcal{N}\!\left(\mu, \sigma\right)$ represents a normal prior of mean $\mu$ and variance $\sigma^2$.

\begin{figure}
    \centering
    \includegraphics[width=\hsize]{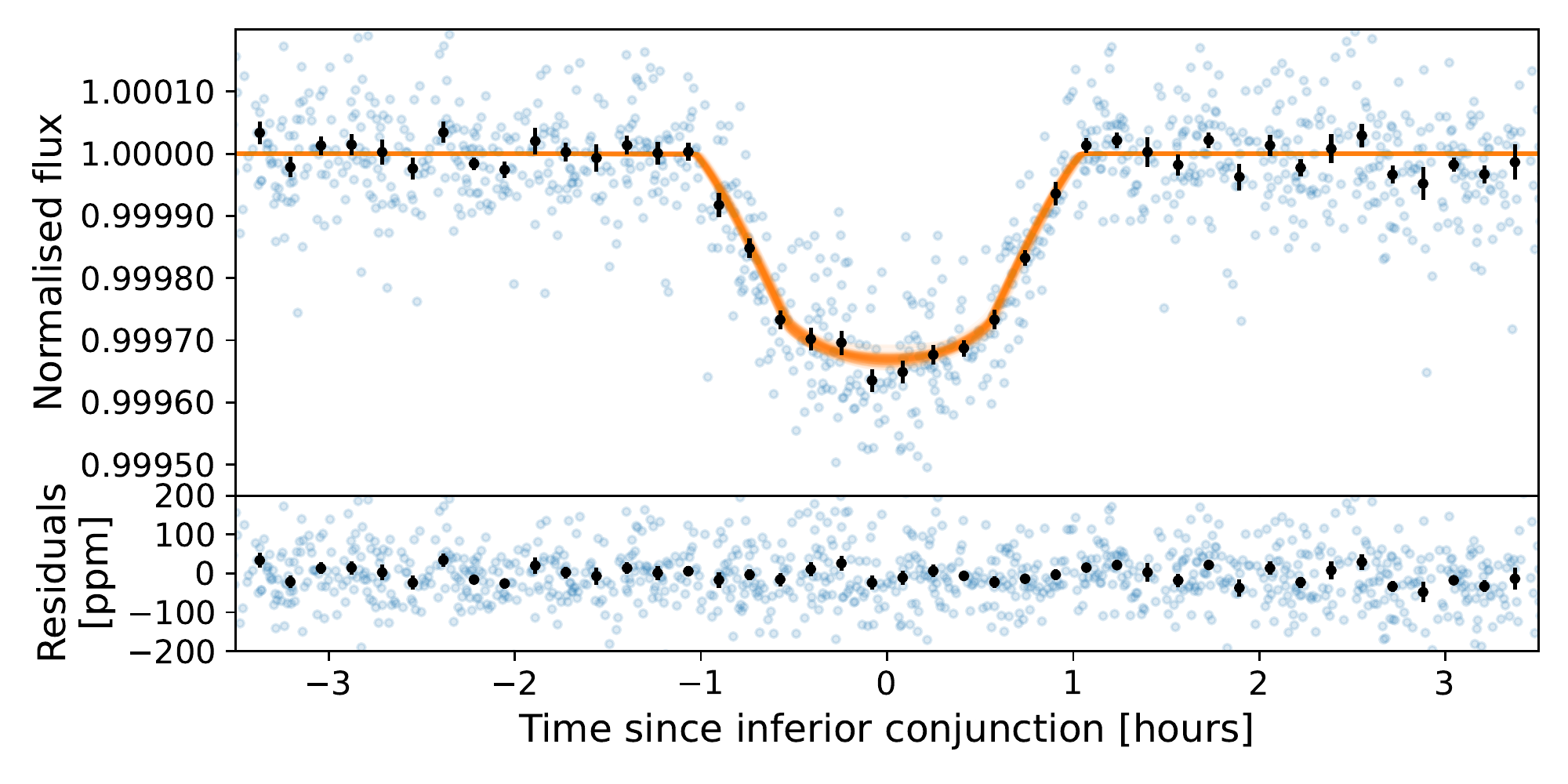}
    \includegraphics[width=\hsize]{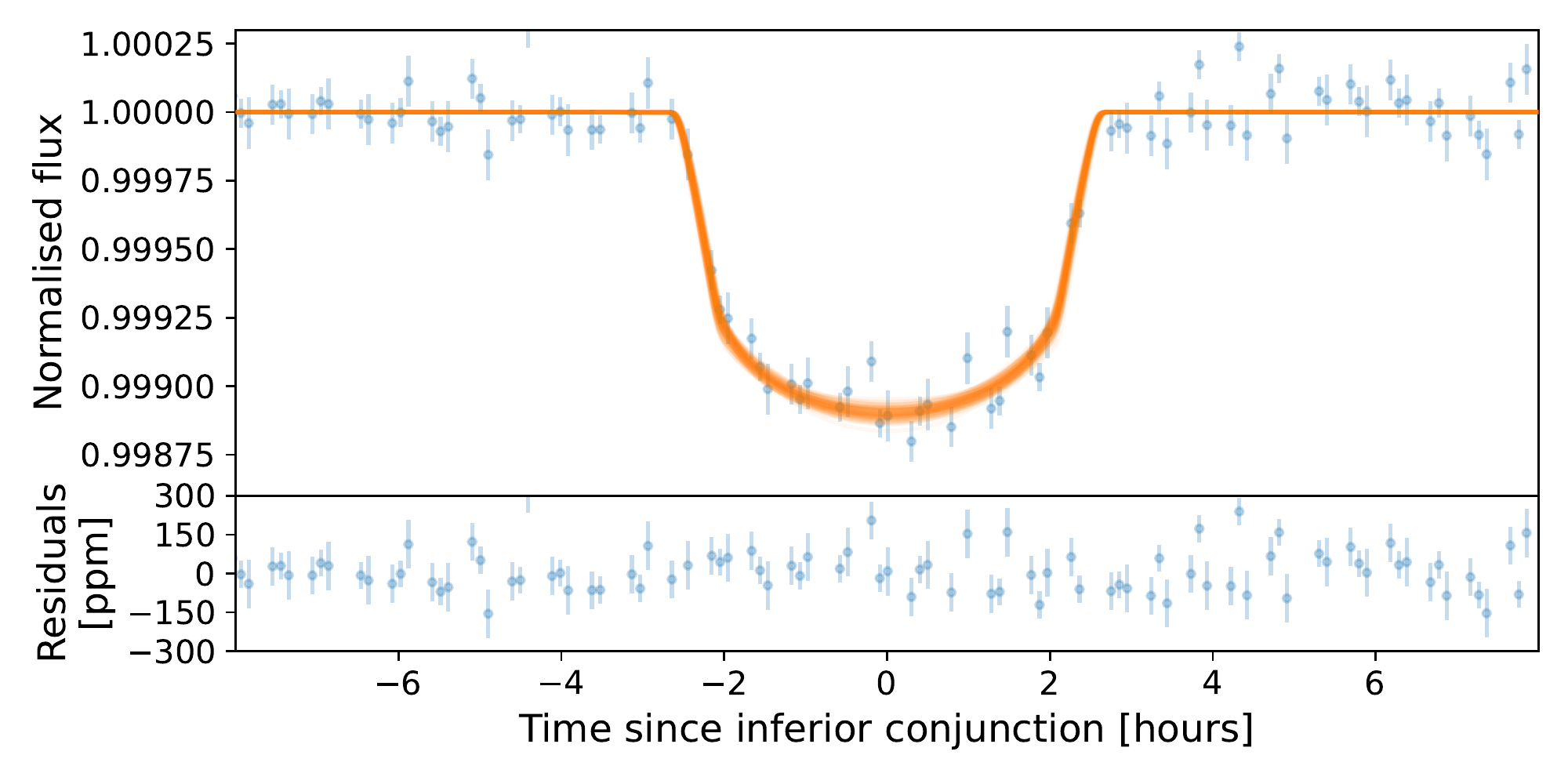}
    \caption{K2 phase-folded transits of planets b (top) and c (bottom). Blue points represent the data after detrending for any other signal. The orange curve are transit models with parameter sets randomly drawn from the posterior distribution. Binned data are represented in the top panel (transit of planet b).}
    \label{fig:k2_transits}
\end{figure}

Fig.~\ref{fig:k2_transits} shows the best joint-fit de-trended K2 transits of planets b and c. The V-shaped transit of HD\,3167\,b is due to the long cadence of the observations that averages the sharp ingress and egress with the flat regions outside and inside the transit \citep[e.g., ][]{kipping_binning}.

\subsubsection{SPITZER photometry} \label{sssec:spitzer_analysis}

\begin{figure}
    \centering
    \includegraphics[width=\hsize]{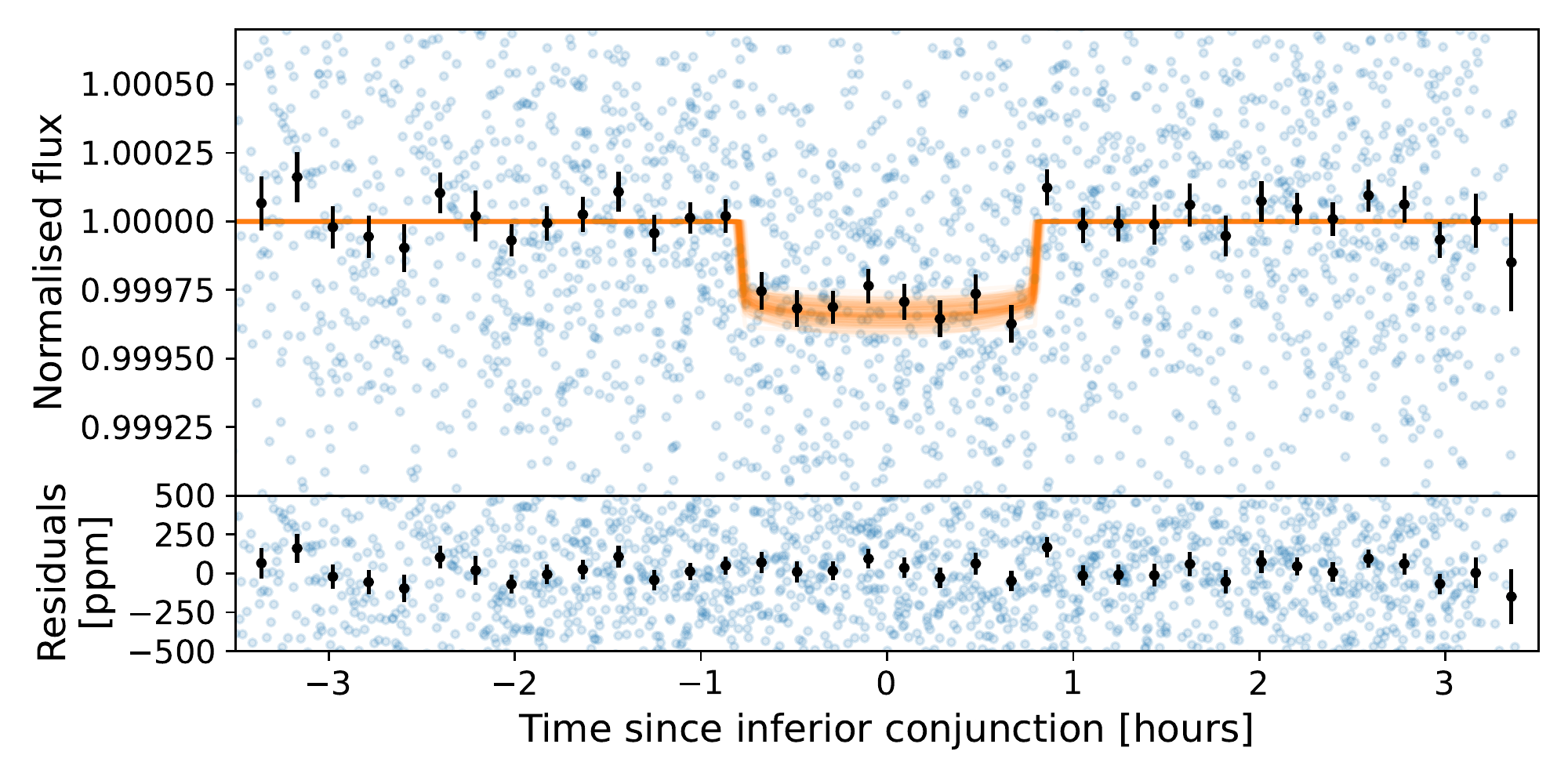}
    \includegraphics[width=\hsize]{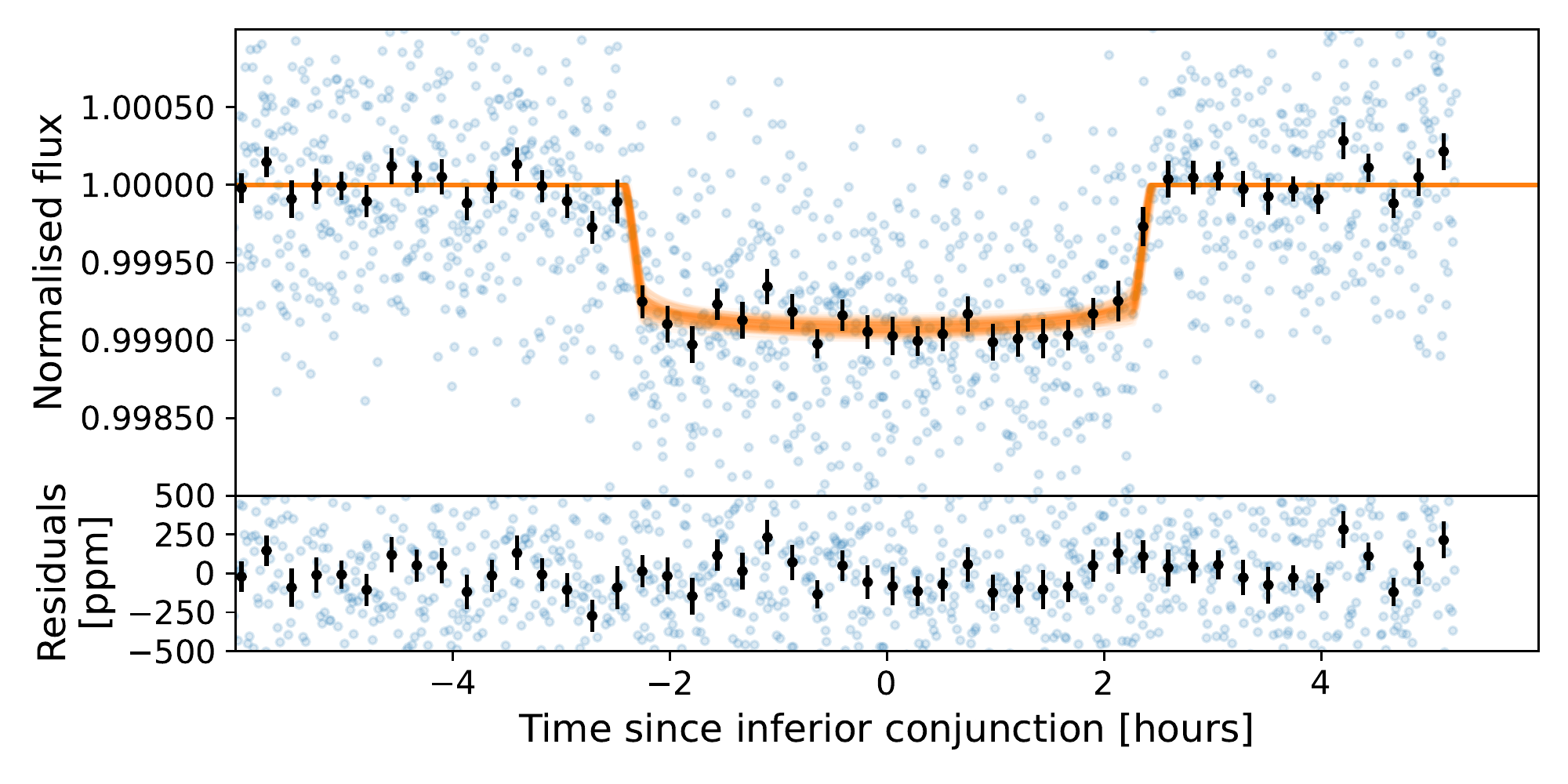}
    \caption{Spitzer phase-folded transits of planets b (top) and c (bottom). Blue and black points are the de-trended and binned data, respectively. The orange shaded area is made of samples drawn from the posterior distribution.}
    \label{fig:spitzer_transits}
\end{figure}

The Spitzer photometry was pre-processed prior to the joint analysis to correct for instrumental systematics (see Sect.~\ref{ssec:spitzer_data}). During the joint fit, the Spitzer observations were fitted with a transit model of planets b and c, and two additional free parameters per observation to account for the flux offset and the underestimation of uncertainties (white noise jitter term). The resulting transit light curves obtained with Spitzer are shown in Fig.~\ref{fig:spitzer_transits}.

\subsubsection{HST photometry} \label{sssec:hst_analysis}

We started by manually flagging one obvious outlier in the fourth orbit of the fifth transit observation. We also discarded 126 out of 722 (17.45\%) points that correspond to the first orbit of every visit, and to the first point of every orbit, following standard approach \citep[e.g., ][]{mikal-evans_hd3167c}.

Given the periodicity of the systematic ramp every HST orbit, we decided to adopt an approach similar to the one typically used with CHEOPS, that is a GP detrending as a function of the spacecraft orbital phase. To properly determine the phase of each data point, we computed a precise orbital period for HST from the housekeeping parameters. We used the spacecraft latitude stored in the jitter files (\textit{*jit.fits} files) and, for each visit, we fitted the latitude variations with a sine wave. We combined the outcome of the fits and computed a precise orbital period of $P_\text{HST} = 95.230^{+0.017}_{-0.009}\,\text{min}$.

The GP correction as a function of the orbital phase was performed using a Mat\'ern-3/2 kernel with a single set of hyper-parameters for all visits, but each visit was fitted individually. We let free the period of HST, while using a strong Gaussian prior $\mathcal{N}\!\left(95.23, 0.02\right)\,\text{min}$ based on the previously derived value. The orbital phase of each data point was therefore computed at each iteration as a function of the HST period value and the JD\_UTC time. For each visit, we also included two flux mean values (forward and backward scans) and a jitter term to account for underestimation of error bars. We found that the addition of a linear trend as a function of time was necessary for all visits, whereas the addition of a quadratic trend was required for visits 2 and 3 only. This leads to a total of 25 parameters for the correction of HST systematics in the data.

\begin{figure}
    \centering
    \includegraphics[width=\hsize]{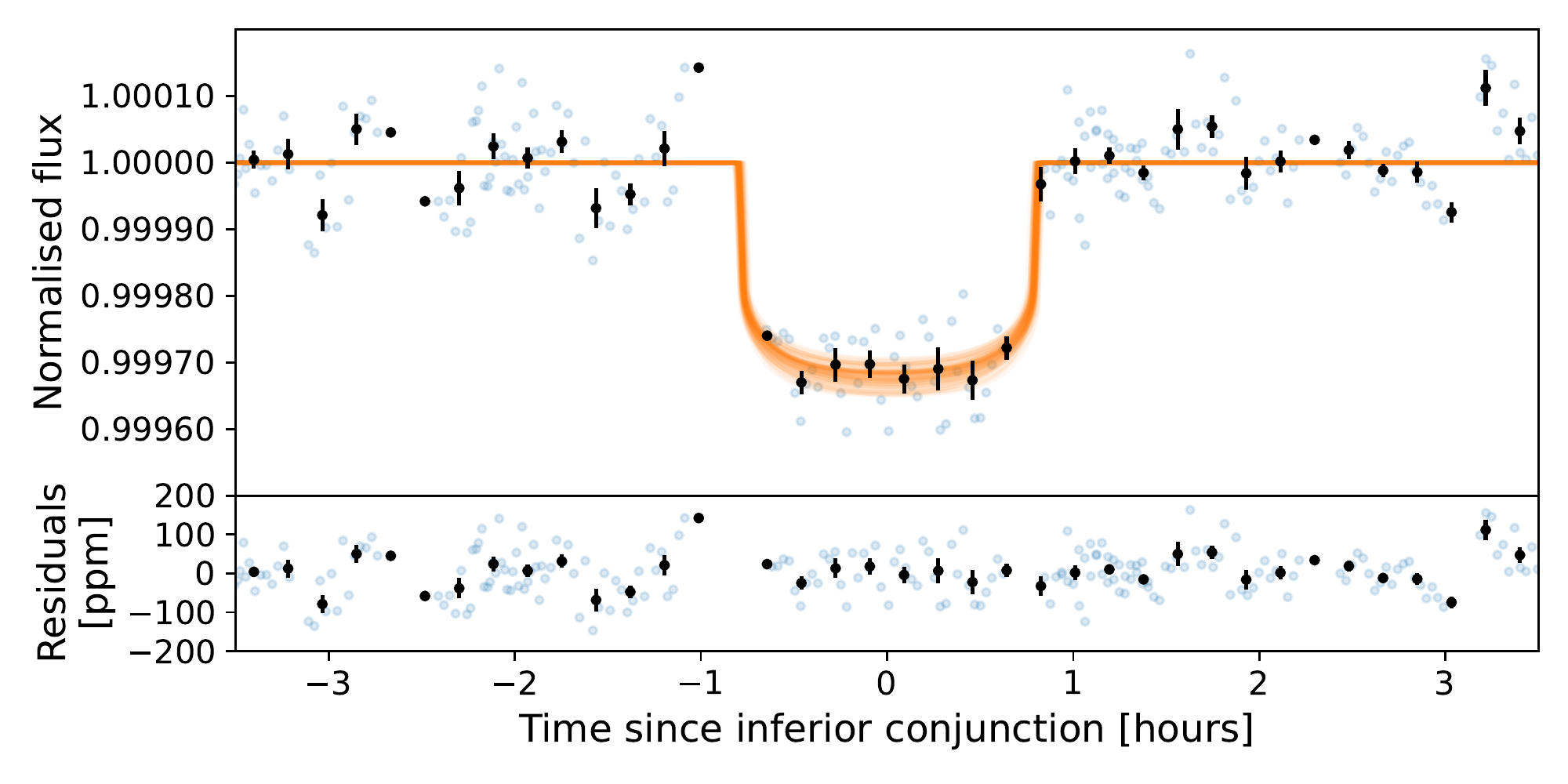}
    \includegraphics[width=\hsize]{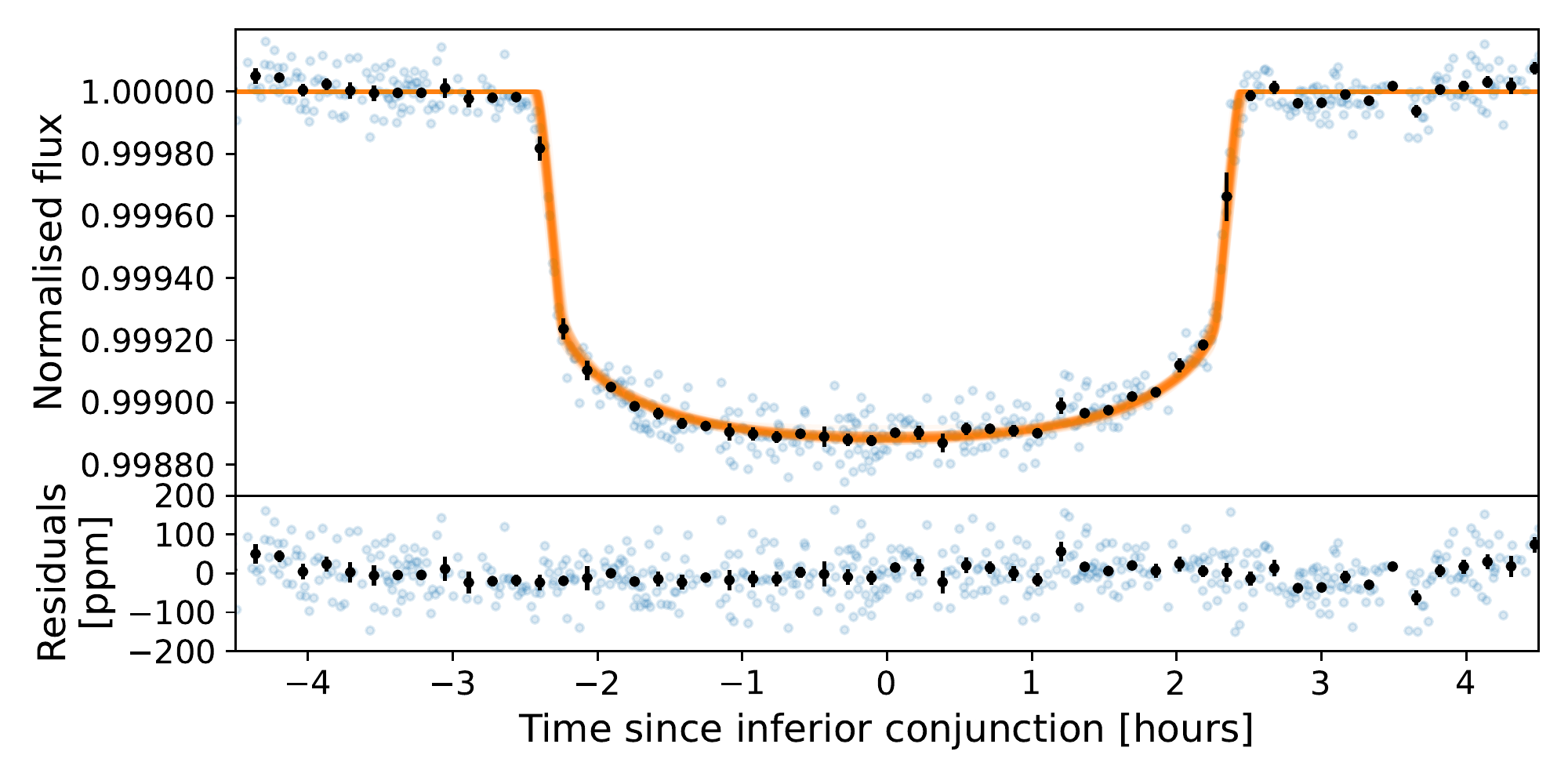}
    \caption{HST phase-folded transits of planets b (top) and c (bottom). De-trended data corrected for the strong periodic instrumental systematics are shown in blue. Binned data points are represented in black. The orange curves are samples from the posterior distribution of the joint analysis.}
    \label{fig:hst_transits}
\end{figure}

We mentioned in Sect.~\ref{ssec:hst_data} that our careful analysis of the HST data lead to the discovery of serendipitous transits of planet b during visits 3 and 4. Therefore, we added a \texttt{batman} transit model accounting for both planets b and c and the best-fit models derived from the joint analysis are represented in Fig.~\ref{fig:hst_transits}. The data of each individual visit are shown in Fig.~\ref{fig:hst_raw_data}.

\subsubsection{Radial velocity} \label{sssec:rv_analysis}

\begin{figure}
    \centering
    \includegraphics[width=\hsize,trim={0.2cm 0.3cm 0.3cm 0.3cm},clip]{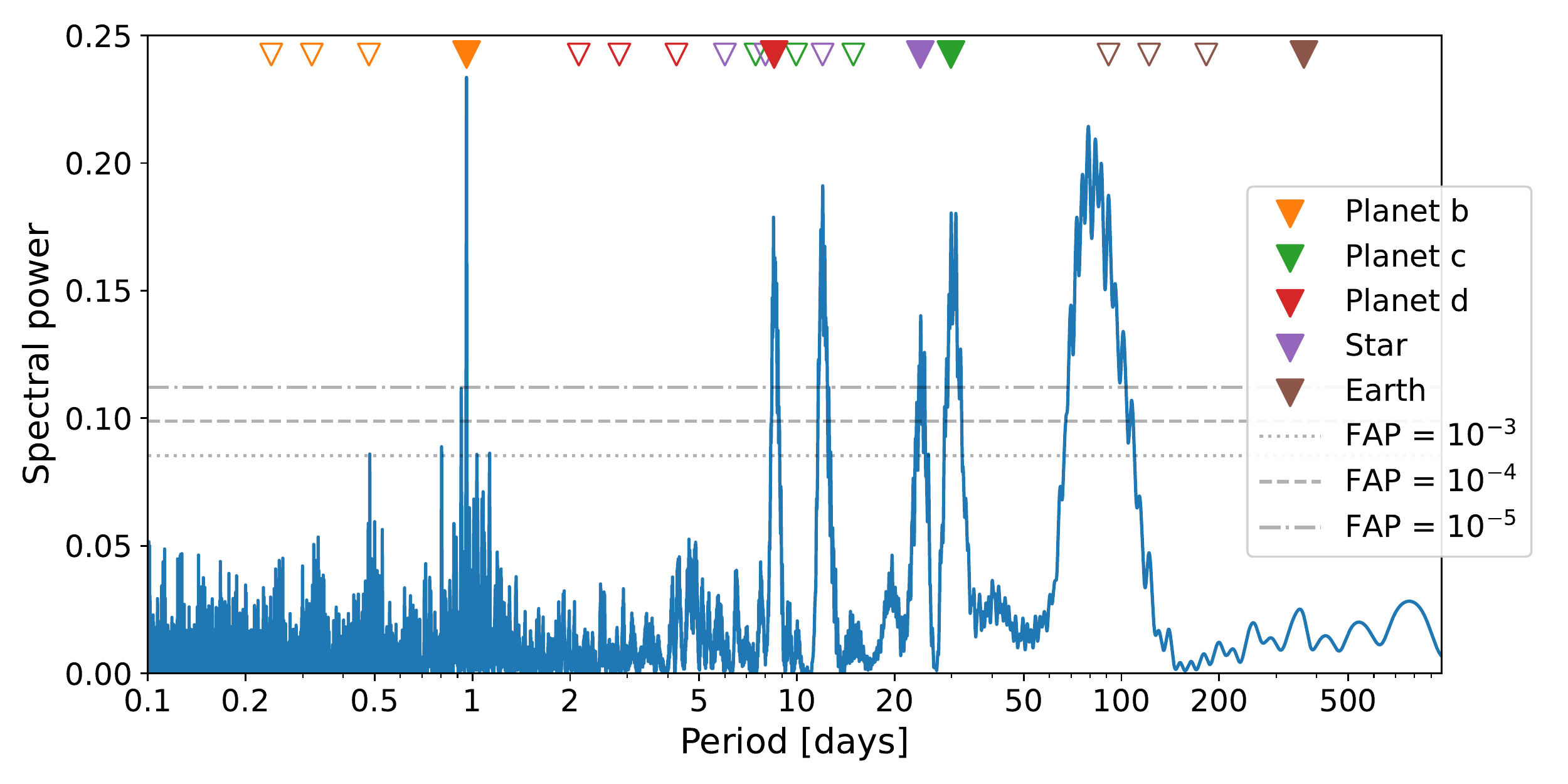}
    \caption{Generalized Lomb-Scargle periodogram of the RV data. The colored triangles at the top represent several periods of interest: orbital periods of the planets b, c and d, expected rotation period of the star ($\sim$ 24 days), and one year. The full triangles show the main periods while the empty ones are the first three harmonics of each period. The horizontal grey lines highlight the False Alarm Probabilities (FAP) of $10^{-3}$, $10^{-4}$ and $10^{-5}$.}
    \label{fig:rv_periodogram}
\end{figure}

We computed the Generalized Lomb-Scargle \citep[GLS --][]{ferraz-mello_1981_gls, zechmeister_2009_gls} periodogram of the selected RV data set after removing an offset (median value of the time series) for each instrument, and we clearly detected the signals from the three planets (see Fig.~\ref{fig:rv_periodogram}). We also have three other significant peaks. Two of them seem to be induced by the stellar rotation with one peak at the rotation period ($\sim$24\,days, also present in the K2 data; \citealt{gandolfi_hd3167}) and another at half of this value (see Sect.~\ref{ssec:results_star} for a detailed discussion). The last very significant peak spans a range of possible periods from 70 to 120 days, and it does not match any of the known objects in the system.

We analyzed the GLS periodograms of the different stellar indices available for the HARPS and HARPS-N data. We found significant peaks at $\sim$ 24 days for the $\mathrm{H}\alpha$ and $\mathrm{Na\,I}$ lines in the HARPS-N data, and nothing around 100 days in either data sets. We also looked at possible correlations between the indices and the RV signals using the Pearson, Kendall and Spearman criteria. We detected significant correlations of the HARPS data with the FWHM ($p\mathrm{-values} < 6\,10^{-3}$) and of the HARPS-N data with the $\mathrm{H}\alpha$ line ($p\mathrm{-values} < 2\,10^{-6}$).

We first designed our RV model using three Keplerian functions for planets b, c and d, and a systemic velocity value for each of the four instruments (APF/Levy, Keck/HIRES, HARPS and HARPS-N). We also included a white-noise term (jitter) for each instrument to account for uncertainty underestimation. Based on the correlation analysis, we jointly fitted for linear functions of the FWHM and the $\mathrm{H}\alpha$ line to correct the HARPS and HARPS-N data, respectively. This modeling choice was motivated by the will to minimize the number of free parameters, even though these correlations may not be as strictly linear as we assume \citep[e.g., ][]{2011A&A...528A...4B, 2014A&A...566A..35S, 2019MNRAS.487.1082C}. We ran a minimization fit and sampled the parameter space with a MCMC approach. We obtained planetary parameters fully consistent with the values from both \cite{Christiansen2017} and \cite{gandolfi_hd3167}. We computed the periodogram of the residuals and retrieved the very significant peak at $\sim$ 100 days, with a false alarm probability (FAP) smaller than $10^{-10}$. 

\begin{figure}
    \centering
    \includegraphics[width=\hsize,trim={0.2cm 0.3cm 0.3cm 0.3cm},clip]{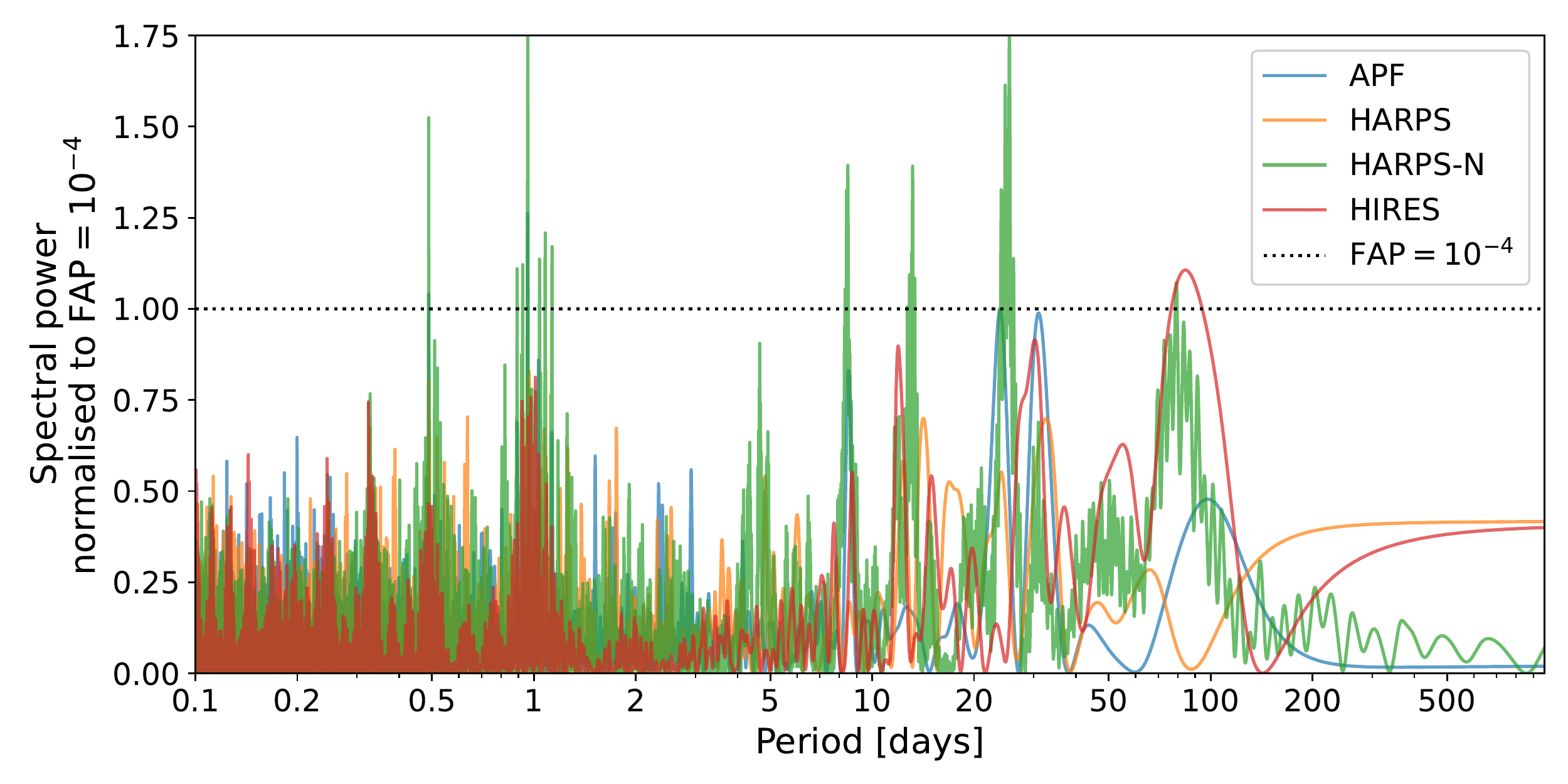}
    \caption{Generalized Lomb-Scargle periodogram of the RV data for each instrument.
    For comparison purposes, each periodogram is normalized by the power corresponding to a False Alarm Probability (FAP) of $10^{-4}$, which is highlighted by the horizontal dotted line.}
    \label{fig:rv_periodogram_inst}
\end{figure}

\begin{figure}
    \centering
    \includegraphics[width=\hsize,trim={0.2cm 0.3cm 0.3cm 0.3cm},clip]{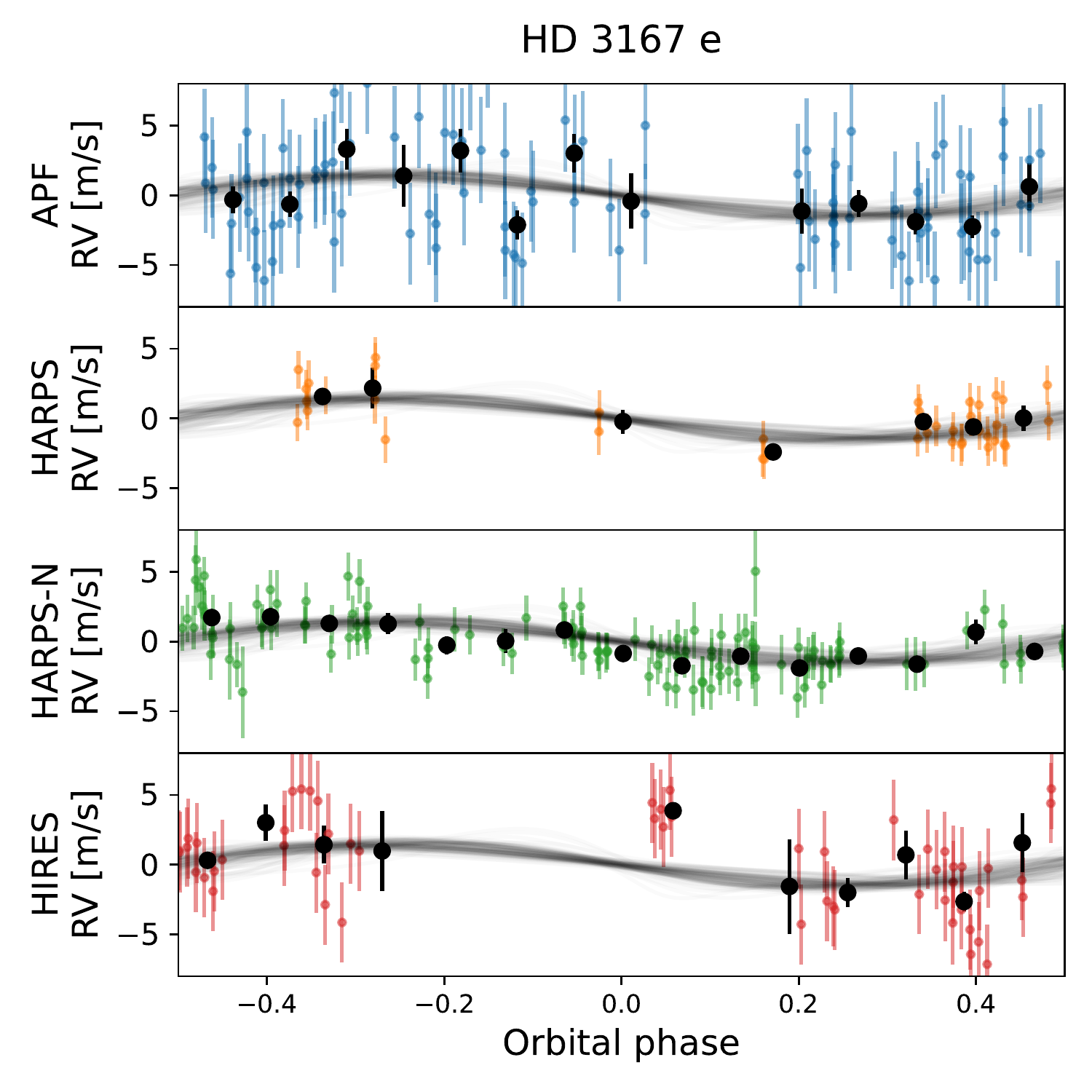}
    \caption{Radial-velocity signals of the planet e measured by each instrument and phase-folded on the best-fit orbital period ($P=102.09$ days). Binned data are shown in black.}
    \label{fig:rv_hd3167e}
\end{figure}

We investigated the possible source of this signal by first looking at indicators of stellar activity at those periods but found no significant signatures. We also compared the periodograms of each spectrograph to search for potential discrepancy that could mean an instrumental origin for the long-period signal (see Fig.~\ref{fig:rv_periodogram_inst}). We found that both HARPS-N and Keck/HIRES data have power in this regime ($\text{FAP}<10^{-4}$). We note that there is also a hint of signal in the APF/Levy time series even though it is less significant. The HARPS data covers only 128 days in total with a poor sampling over this baseline and thus the individual periodogram does not feature any peak around 100~days. The presence of power in several data sets strongly suggested that the signal was not induced by instrumental systematics and might actually have a planetary origin. To further consolidate this hypothesis, we compared the phases of the long-period signal measured by each spectrograph by looking at the RV time series phase-folded on the detected period. Figure~\ref{fig:rv_hd3167e} shows the best result of the joint fit performed in this work where one can visually validate the consistency of the signal phase through all instruments. In the view of these different outcomes, we rejected the stellar or instrumental origins to explain the residual signal and we interpreted it as the RV signature of a fourth planet, whose presence was suggested by \citet{Dalal2019}.

We ran another fit to the data including an additional Keplerian model with a uniform prior on the period spanning a large range from 60 to 200 days. The resulting semi-amplitude of the Doppler signature produced by the new planet was significant by more than 9$\sigma$ and the residual periodogram was not featuring significant peaks anymore. The comparison of the 3-planet and 4-planet models using the Bayesian and Akaike Information Criteria (BIC and AIC) clearly favored the inclusion of the new planet ($\Delta\mathrm{BIC}<-57$ and $\Delta\mathrm{AIC}<-68$). Therefore, the final RV model used in the joint fit with the photometry included a Keplerian model for each of the three known planets b, c and d, and another for the new long-period planet e. We fitted the systemic velocity and a jitter term for each instrument, and added linear correction of the HARPS and the HARPS-N data as functions of the FWHM and the $\mathrm{H}\alpha$ line, respectively. We used ten parameters in total to fit the instrument-related effects (offset, noise and decorrelation).

\begin{figure}
    \centering
    \includegraphics[width=\hsize,trim={0cm 2.3cm 1.8cm 2.3cm},clip]{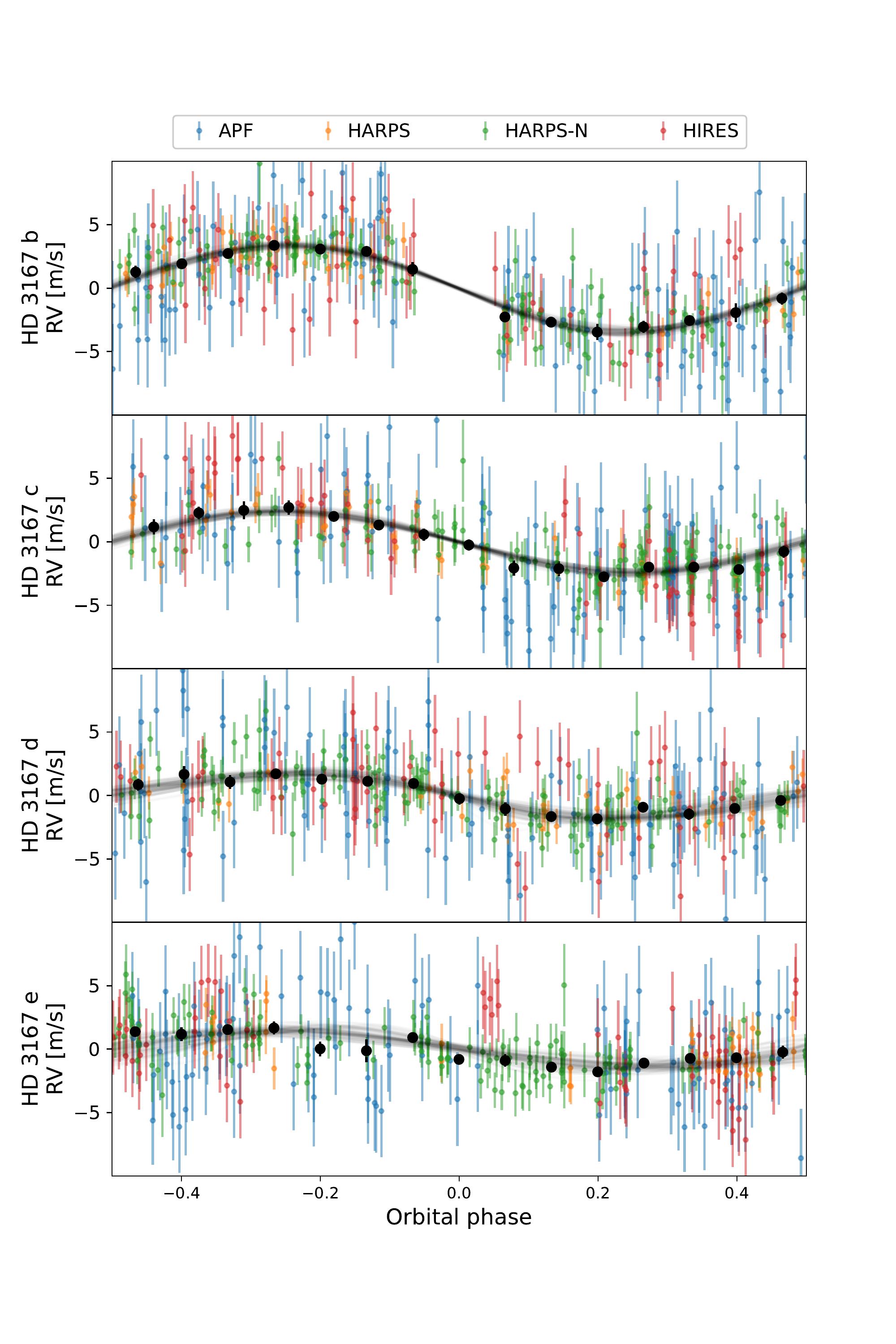}
    \caption{Phase-folded radial-velocity data of the four planets of the HD\,3167 system. The color code of the data points highlights the instrument with which they were observed (APF, HARPS, HARPS-N, HIRES). \rev{The black points show the binned data.} The multiple black curves are samples randomly drawn from the posterior distribution.}
    \label{fig:rv_results}
\end{figure}

The extracted planetary signals fitted from the joint analysis are represented in Fig.~\ref{fig:rv_results}. The gap visible in the RV data phase-folded on the orbital period of HD\,3167\,b is due to the removal of in-transit points to avoid being affected by the Rossiter-McLaughlin effect.

\subsubsection{Joint MCMC fit} \label{sssec:joint_fit}

We used the MCMC algorithm \texttt{emcee} \citep{foreman_emcee} to explore the parameter space simultaneously for all the parameters defining the systematic correction of each instrument and the planetary properties from the transit light curves and the radial-velocity signals. We had a total of 120 parameters and, at each MCMC iteration, we computed the global log-probability by summing the log-probability obtained with each data set (CHEOPS, K2, Spitzer, HST, RV).

The MCMC run was initiated by estimating the best-fit parameters and their uncertainties on each data set independently. We used the posterior distribution to generate the first-guess parameter sets. The 1280 chains of our MCMC joint fit started with a burn-in phase of 145\,000 iterations, and then sampled the parameter space with 260\,000 steps. We kept one iteration every 2000 to reduce the effect of the chain autocorrelation. We checked the convergence of the chains by visually inspecting the trace plots and validated it based on the Gelman-Rubin criterion \citep{gelman-rubin}.


\section{Revision of the system properties}
\label{sec:results_sys}

\subsection{Star} \label{ssec:results_star}

The Rossiter-McLaughlin analysis performed by \citet{Bourrier2021_3167} revealed variations in the contrast of the stellar lines occulted by HD\,3167b and c along their respective transit chords. The authors could explain these variations through a latitudinal dependence of the stellar line contrast, which allowed them to constrain the inclination of the star with respect to the line of sight (i$_{\star}$ = 111.6$\stackrel{+3.1}{_{-3.3}}^{\circ}$ or 68.4$\stackrel{+3.3}{_{-3.1}}^{\circ}$, the two configurations being degenerate) and thus the true equatorial velocity (v$_\mathrm{eq}$ = 2.65$\stackrel{+0.47}{_{-0.42}}$\,km\,s$^{-1}$). Assuming that this scenario is correct, we combined our stellar radius (Sect.~\ref{sec:star}) with the stellar inclination and projected rotational velocity from \citet{Bourrier2021_3167} to update the true equatorial period (P$_\mathrm{eq}$ = 16.63$\stackrel{+3.0}{_{-2.6}}$ d).

\begin{figure}
    \centering
    \includegraphics[width=\hsize]{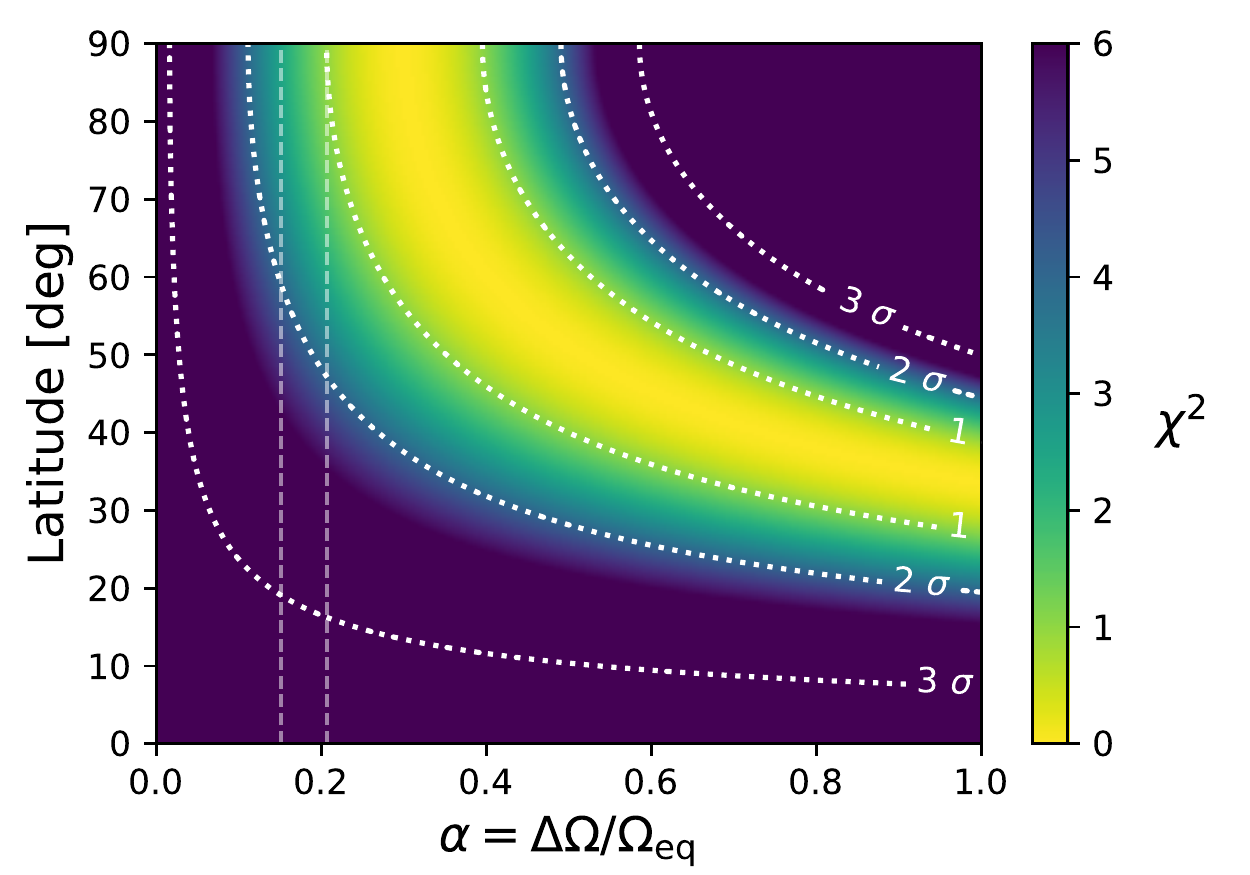}
    \caption{Probability map of the spot latitude attributed to the K2 photometry and RV modulations as a function of the relative differential rotation rate. The colorscale represents the $\chi^2$ probability and the dotted lines highlight the 1-, 2- and 3-$\sigma$ confidence intervals. The vertical dashed lines shows the 1$\sigma$ confidence interval of the expected $\alpha$ value derived from Eq. 2 of \cite{balona_diff_rot}.}
    \label{fig:diff_rot}
\end{figure}

In addition to the planetary signals, we observe consistent peaks in the periodograms of the K2 photometry and RV data at about $\sim$24\,days, which likely trace spots on the rotating stellar surface. The difference between this period, and that derived independently from Rossiter-McLaughlin analysis for P$_\mathrm{eq}$, could indicate that the star is rotating differentially. During their lifetime, spots would spend on average more time in a region located at higher latitudes, rotating more slowly than the equator. Under this hypothesis we can estimate the spot location by assuming a solar-like law for the stellar differential rotation:
\begin{equation}
    P_\text{eq} / P\!\left(\theta\right) = 1-\alpha \sin^2\left(\theta\right),
\end{equation}
with $\alpha$ the relative differential rotation rate between equator and pole, and $\theta$ the stellar latitude. We computed the value of $P\!\left(\theta\right)$ from a Gaussian fit to the periodogram peak in the K2 and RV data sets, yielding $P_\text{K2}=23.4\pm2.2$ days and $P_\text{RV}=24.1\pm1.2$ days. The close agreement between these periodic signals from different datasets give us confidence that they arise from stellar modulation, rather than instrumental variability. We then built a $\chi^2$ map of the stellar latitude as a function of $\alpha$, comparing the theoretical $P_\text{eq} / P\!\left(\theta\right)$ ratio with the measured values (Fig.~\ref{fig:diff_rot}). The relative differential rotation rate for HD\,3167 can be estimated independently from Eq.~2 of \cite{balona_diff_rot}, which is based on photometric modulations in a wide sample of \textit{Kepler} stars. Using $T_\text{eff}=5300\pm73\,\text{K}$ and $\Omega_{eq}=2\pi/P_\text{eq}$, we derive $\alpha=0.179\pm0.028$. Within 2$\sigma$, this value is consistent with the measurements of $P_\text{eq}$ and $P\!\left(\theta\right)$ for stellar latitudes $\gtrsim50\deg$. Spots on HD\,3167 would thus be located closer to its poles than on the Sun, where spots appear at latitudes of $\sim$35$
^{\circ}$ at the beginning of a new cycle and converge toward the equator over 11 years.

\subsection{Planets}
\label{sec:pl_results}

We derived the parameter values and uncertainties from the posterior distribution of the MCMC run. We computed the median and the 68.3\% confidence interval for each of the fitted parameters (see Table~\ref{tab:result_fitted}). We combined the MCMC chains to calculate the values and uncertainties on a series of useful parameters that were not fitted directly (Table~\ref{tab:result_derived}). Some of these derived parameters were obtained using the stellar mass or radius, and we propagated the uncertainties on these parameters by drawing random values from normal distributions based on their estimated values: $M_\star=0.852\pm0.026\,M_\odot$ and $R_\star=0.871\pm0.006\,R_\star$ (Sect.~\ref{sec:star}).

\begin{table*}
\caption{Fitted stellar and planetary parameters of the HD\,3167 system.}
\label{tab:result_fitted}
\centering
\begin{tabular}{lcccr}
\toprule
\toprule
Fitted parameters & Symbols & Values & Priors & Units \\
\midrule
\midrule
Planet b &&&& \\
\quad Time of inferior conjunction & $T_{0,\,b}$ & ${2\,457\,394.37421}_{-{0.00035}}^{+{0.00036}}$ & - & $\text{BJD}_\text{TDB}$ \\
\quad Orbital period & $P_b$ & ${0.95965428}_{-{0.00000029}}^{+{0.00000030}}$ & - & days \\
\quad Planet-to-star radii ratios &&&& \\
\qquad CHEOPS passband & $k_b^\text{CHEOPS}$ & ${0.01765}\pm{0.00036}$ & - & - \\
\qquad K2 passband & $k_b^\text{K2}$ & ${0.01657}_{-{0.00021}}^{+{0.00022}}$ & - & - \\
\qquad HST/WFC3/G141 passband & $k_b^\text{HST}$ & ${0.01687}_{-{0.00036}}^{+{0.00035}}$ & - & - \\
\qquad Spitzer/IRAC/Ch2 passband & $k_b^\text{Spitzer}$ & ${0.01789}_{-{0.00081}}^{+{0.00076}}$ & - & - \\
\qquad All passbands & $k_b$ & ${0.01712}_{-{0.00059}}^{+{0.00086}}$ & - & - \\
\quad Orbital inclination & $i_b$ & ${87.59}_{-{1.75}}^{+{1.63}}$ & $\mathcal{U}\!\left(0, 90\right)$ & deg \\
\multirow{2}{*}{\quad Eccentricity / argument of periastron} & $e_b\,\cos\omega_b$ & ${-0.015}_{-{0.023}}^{+{0.022}}$ & - & - \\
& $e_b\,\sin\omega_b$ &  ${0.030}_{-{0.034}}^{+{0.022}}$ & - & - \\
\quad RV semi-amplitude & $K_b$ & ${3.425}_{-{0.185}}^{+{0.176}}$ & - & m/s \\
\midrule
Planet c &&&& \\
\quad Time of inferior conjunction & $T_{0,\,c}$ & ${2\,457\,394.97778}_{-{0.00057}}^{+{0.00056}}$ & - & $\text{BJD}_\text{TDB}$ \\
\quad Orbital period & $P_c$ & ${29.8464948}_{-{0.0000154}}^{+{0.0000157}}$ & - & days \\
\quad Planet-to-star radii ratios &&&& \\
\qquad K2 passband & $k_c^\text{K2}$ & ${0.03073}_{-{0.00046}}^{+{0.00047}}$ & - & - \\
\qquad HST/WFC3/G141 passband & $k_c^\text{HST}$ & ${0.03176}\pm{0.00028}$ & - & - \\
\qquad Spitzer/IRAC/Ch2 passband & $k_c^\text{Spitzer}$ & ${0.02967}_{-{0.00058}}^{+{0.00055}}$ & - & - \\
\qquad All passbands & $k_c$ & ${0.03075}_{-{0.00112}}^{+{0.00103}}$ & - & - \\
\quad Orbital inclination & $i_c$ & ${89.421}_{-{0.071}}^{+{0.130}}$ & $\mathcal{U}\!\left(0, 90\right)$ & deg \\
\multirow{2}{*}{\quad Eccentricity / argument of periastron} & $e_c\,\cos\omega_c$ & ${-0.007}_{-{0.032}}^{+{0.031}}$ & - & - \\
& $e_c\,\sin\omega_c$ & ${-0.014}_{-{0.051}}^{+{0.055}}$ & - & - \\
\quad RV semi-amplitude & $K_c$ & ${2.461}_{-{0.174}}^{+{0.180}}$ & - & m/s \\
\midrule
Planet d &&&& \\
\quad Time of inferior conjunction & $T_{0, d}$ & ${2\,457\,585.20}\pm{0.22}$ & - & $\text{BJD}_\text{TDB}$ \\
\quad Orbital period & $P_d$ & ${8.4783}\pm{0.0025}$ & - & days \\
\multirow{2}{*}{\quad Eccentricity / argument of periastron} & $e_d\,\cos\omega_d$ & ${-0.023}_{-{0.088}}^{+{0.084}}$ & - & - \\
& $e_d\,\sin\omega_d$ & ${0.120}_{-{0.111}}^{+{0.102}}$ & - & - \\
\quad RV semi-amplitude & $K_d$ & ${1.793}_{-{0.167}}^{+{0.165}}$ & - & m/s \\
\midrule
Planet e &&&& \\
\quad Time of inferior conjunction & $T_{0, e}$ & ${2\,457\,643.6}_{-{5.3}}^{+{4.0}}$ & - & $\text{BJD}_\text{TDB}$ \\
\quad Orbital period\tablefootmark{$\dagger$} & $P_e$\tablefootmark{$\dagger$} & ${102.09}_{-{0.50}}^{+{0.52}}$\tablefootmark{$\dagger$} & - & days \\
\multirow{2}{*}{\quad Eccentricity / argument of periastron} & $e_e\,\cos\omega_e$ & ${0.012}_{-{0.108}}^{+{0.113}}$ & - & - \\
& $e_e\,\sin\omega_e$ & ${-0.089}_{-{0.185}}^{+{0.219}}$ & - & - \\
\quad RV semi-amplitude & $K_e$ & ${1.536}_{-{0.180}}^{+{0.186}}$ & - & m/s \\
\midrule
Star &&&& \\
\quad Stellar density & $\rho_\star$ & ${1.284}_{-{0.047}}^{+{0.046}}$ & $\mathcal{N}\!\left(1.289, 0.048\right)$ & $\rho_\odot$ \\
\quad Limb-darkening coefficients: &&&& \\
\multirow{2}{*}{\qquad CHEOPS passband} & $u_1^\text{CHEOPS}$ & ${0.276}_{-{0.187}}^{+{0.255}}$ & - & - \\
 & $u_2^\text{CHEOPS}$ & ${0.44}_{-{0.37}}^{+{0.28}}$ & - & - \\
\multirow{2}{*}{\qquad K2 passband} & $u_1^\text{K2}$ & ${0.513}_{-{0.208}}^{+{0.193}}$ & - & - \\
 & $u_2^\text{K2}$ & ${0.11}_{-{0.29}}^{+{0.31}}$ & - & - \\
\multirow{2}{*}{\qquad HST/WFC3/G141 passband} & $u_1^\text{HST}$ & ${0.203}_{-{0.084}}^{+{0.080}}$ & - & - \\
 & $u_2^\text{HST}$ & ${0.306}_{-{0.118}}^{+{0.120}}$ & - & - \\
\multirow{2}{*}{\qquad Spitzer/IRAC/Ch2 passband} & $u_1^\text{Spitzer}$ & ${0.143}_{-{0.102}}^{+{0.159}}$ & - & - \\
 & $u_2^\text{Spitzer}$ & ${0.049}_{-{0.130}}^{+{0.163}}$ & - & - \\
\bottomrule
\bottomrule
\end{tabular}
\tablefoot{
Uniform priors between $a$ and $b$ are represented by $\mathcal{U}\!\left(a, b\right)$. normal priors with mean $\mu$ and variance $\sigma^2$ are represented by $\mathcal{N}\!\left(\mu, \sigma\right)$.
The limb-darkening coefficients correspond to the quadratic model \citep{1977A&A....61..809M}: $I\!\left(\mu\right)/I_0 = 1 - u_1\left(1-\mu\right)-u_2\left(1-\mu\right)^2$, where $\mu=\sqrt{1-x^2}$ and $x$ is the normalized radial coordinate on the stellar disk ($x=0$ at the center, $x=1$ at the limb).
\tablefoottext{$\dagger$}{Note that the marginalized posterior distribution of $P_e$ has several modes spanning a large range from 79\,days to 125\,days, and that the error bars of~$\sim\!0.5\,\text{days}$ are dominated by the main mode around 102\,days.}
}
\end{table*}

\begin{table*}
\caption{Derived planetary parameters of the HD\,3167 system.}
\label{tab:result_derived}
\centering
\begin{tabular}{lccr}
\toprule
Derived parameters & Symbols & Values & Units \\
\midrule
\midrule
Planet b &&& \\
\quad Optimal time of inferior conjunction & $T_{0,\,b}^\text{opt}$ & ${2\,458\,269.57891}_{-{0.00024}}^{+{0.00026}}$ & $\text{BJD}_\text{TDB}$ \\
\quad Impact parameter & $b_b$ & ${0.181}_{-{0.123}}^{+{0.141}}$ & $R_\star$ \\
\quad Transit duration & $T_{14,\,b}$ & ${1.6092}_{-{0.0144}}^{+{0.0172}}$ & hours \\
\quad Eccentricity\tablefootmark{$\dagger$} & $e_b$\tablefootmark{$\dagger$} & $<0.10$\tablefootmark{$\dagger$} & - \\
\multirow{2}{*}{\quad Semi-major axis} & $a_b/R_\star$ & ${4.450}_{-{0.055}}^{+{0.053}}$ & - \\
& $a_b$ & ${0.01802}\pm{0.00025}$ & AU \\
\quad Mass & $M_b$ & ${4.73}_{-{0.29}}^{+{0.28}}$ & $M_\oplus$ \\
\quad Radii &&& \\
\qquad CHEOPS passband & $R_b^\text{CHEOPS}$ & ${1.677}\pm{0.036}$ & $R_\oplus$ \\
\qquad K2 passband & $R_b^\text{K2}$ & ${1.575}_{-{0.023}}^{+{0.024}}$ & $R_\oplus$ \\
\qquad HST/WFC3/G141 passband & $R_b^\text{HST}$ & ${1.602}_{-{0.036}}^{+{0.035}}$ & $R_\oplus$ \\
\qquad Spitzer/IRAC/Ch2 passband & $R_b^\text{Spitzer}$ & ${1.700}_{-{0.078}}^{+{0.074}}$ & $R_\oplus$ \\
\qquad All passbands & $R_b$ & ${1.627}_{-{0.058}}^{+{0.083}}$ & $R_\oplus$ \\
\quad Bulk densities &&& \\
\qquad CHEOPS passband & $\rho_b^\text{CHEOPS}$ & ${5.50}_{-{0.49}}^{+{0.52}}$ & g/cm$^3$ \\
\qquad K2 passband & $\rho_b^\text{K2}$ & ${6.64}_{-{0.51}}^{+{0.52}}$ & g/cm$^3$ \\
\qquad HST/WFC3/G141 & $\rho_b^\text{HST}$ & ${6.30}_{-{0.55}}^{+{0.61}}$ & g/cm$^3$ \\
\qquad Spitzer/IRAC/Ch2 passband & $\rho_b^\text{Spitzer}$ & ${5.28}_{-{0.71}}^{+{0.87}}$ & g/cm$^3$ \\
\quad Equilibrium temperature\tablefootmark{$\ddagger$} & $T_{\text{eq},\,b}$\tablefootmark{$\ddagger$} & ${1777}\pm{27}$\tablefootmark{$\ddagger$} & K \\
\midrule
Planet c &&& \\
\quad Optimal time of inferior conjunction & $T_{0,\,c}^\text{opt}$ & ${2\,458\,439.605096}_{-{0.000147}}^{+{0.000149}}$ & $\text{BJD}_\text{TDB}$ \\
\quad Impact parameter & $b_c$ & ${0.451}_{-{0.120}}^{+{0.078}}$ & $R_\star$ \\
\quad Transit duration & $T_{14,\,c}$ & ${4.869}_{-{0.025}}^{+{0.026}}$ & hours \\
\quad Eccentricity\tablefootmark{$\dagger$} & $e_c$\tablefootmark{$\dagger$} & $<0.15$\tablefootmark{$\dagger$} & - \\
\multirow{2}{*}{\quad Semi-major axis} & $a_c/R_\star$ & ${44.01}_{-{0.54}}^{+{0.52}}$ & - \\
& $a_c$ & ${0.1783}\pm{0.0025}$ & AU \\
\quad Mass & $M_c$ & ${10.67}_{-{0.81}}^{+{0.85}}$ & $M_\oplus$ \\
\quad Radii &&& \\
\qquad K2 passband & $R_c^\text{K2}$ & ${2.919}_{-{0.048}}^{+{0.049}}$ & $R_\oplus$ \\
\qquad HST/WFC3/G141 passband & $R_c^\text{HST}$ & ${3.017}_{-{0.033}}^{+{0.034}}$ & $R_\oplus$ \\
\qquad Spitzer/IRAC/Ch2 passband & $R_c^\text{Spitzer}$ & ${2.819}_{-{0.058}}^{+{0.056}}$ & $R_\oplus$ \\
\qquad All passbands & $R_c$ & ${2.923}_{-{0.109}}^{+{0.098}}$ & $R_\oplus$ \\
\quad Bulk densities &&& \\
\qquad K2 passband & $\rho_c^\text{K2}$ & ${2.35}_{-{0.21}}^{+{0.23}}$ & g/cm$^3$ \\
\qquad HST/WFC3/G141 & $\rho_c^\text{HST}$ & ${2.133}_{-{0.177}}^{+{0.187}}$ & g/cm$^3$ \\
\qquad Spitzer/IRAC/Ch2 passband & $\rho_c^\text{Spitzer}$ & ${2.61}_{-{0.25}}^{+{0.28}}$ & g/cm$^3$ \\
\quad Equilibrium temperature\tablefootmark{$\ddagger$} & $T_{\text{eq},\,c}$\tablefootmark{$\ddagger$} & ${565.0}_{-{8.5}}^{+{8.6}}$\tablefootmark{$\ddagger$} & K \\
\midrule
Planet d &&& \\
\quad Optimal time of inferior conjunction & $T_{0,\,d}^\text{opt}$ & ${2\,457\,797.16}\pm{0.21}$ & $\text{BJD}_\text{TDB}$ \\
\quad Eccentricity\tablefootmark{$\dagger$} & $e_d$\tablefootmark{$\dagger$} & $< 0.44$\tablefootmark{$\dagger$} & - \\
\multirow{2}{*}{\quad Semi-major axis} & $a_d/R_\star$ & ${19.02}\pm{0.23}$ & - \\
& $a_d$ & ${0.07703}_{-{0.00108}}^{+{0.00106}}$ & AU \\
\quad Minimum mass & $M_d\,\sin i_d$ & ${5.03}\pm{0.50}$ & $M_\oplus$ \\
\quad Equilibrium temperature\tablefootmark{$\ddagger$} & $T_{\text{eq},\,d}$\tablefootmark{$\ddagger$} & ${859.5}_{-{12.9}}^{+{13.0}}$\tablefootmark{$\ddagger$} & K \\
\midrule
Planet e &&& \\
\quad Eccentricity\tablefootmark{$\dagger$} & $e_e$\tablefootmark{$\dagger$} & $<0.60$\tablefootmark{$\dagger$} & - \\
\multirow{2}{*}{\quad Semi-major axis} & $a_e/R_\star$ & ${99.93}_{-{1.59}}^{+{1.65}}$ & - \\
& $a_e$ & ${0.4048}_{-{0.0074}}^{+{0.0077}}$ & AU \\
\quad Minimum mass & $M_e\,\sin i_e$ & ${9.74}_{-{1.15}}^{+{1.20}}$ & $M_\oplus$ \\
\quad Equilibrium temperature\tablefootmark{$\ddagger$} & $T_{\text{eq},\,e}$\tablefootmark{$\ddagger$} & ${374.8}_{-{7.3}}^{+{7.1}}$\tablefootmark{$\ddagger$} & K \\
\midrule
\bottomrule
\end{tabular}
\tablefoot{
\tablefoottext{$\dagger$}{Upper limits on the orbital eccentricities are computed with a confidence probability of 99.73\%.}
\tablefoottext{$\ddagger$}{Equilibrium temperatures are derived from the equation $T_\text{eq}=T_\text{eff}/\sqrt{2a/R_\star}$, which assumes black-body emissions for both the planet and the star, a Bond albedo $A_B=0$, and a perfect heat redistribution in the planetary atmosphere (uniform temperature).}
}
\end{table*}

\begin{figure}
    \centering
    \includegraphics[width=\hsize, trim={0.2cm 0.4cm, 0.2cm 0.2cm}, clip]{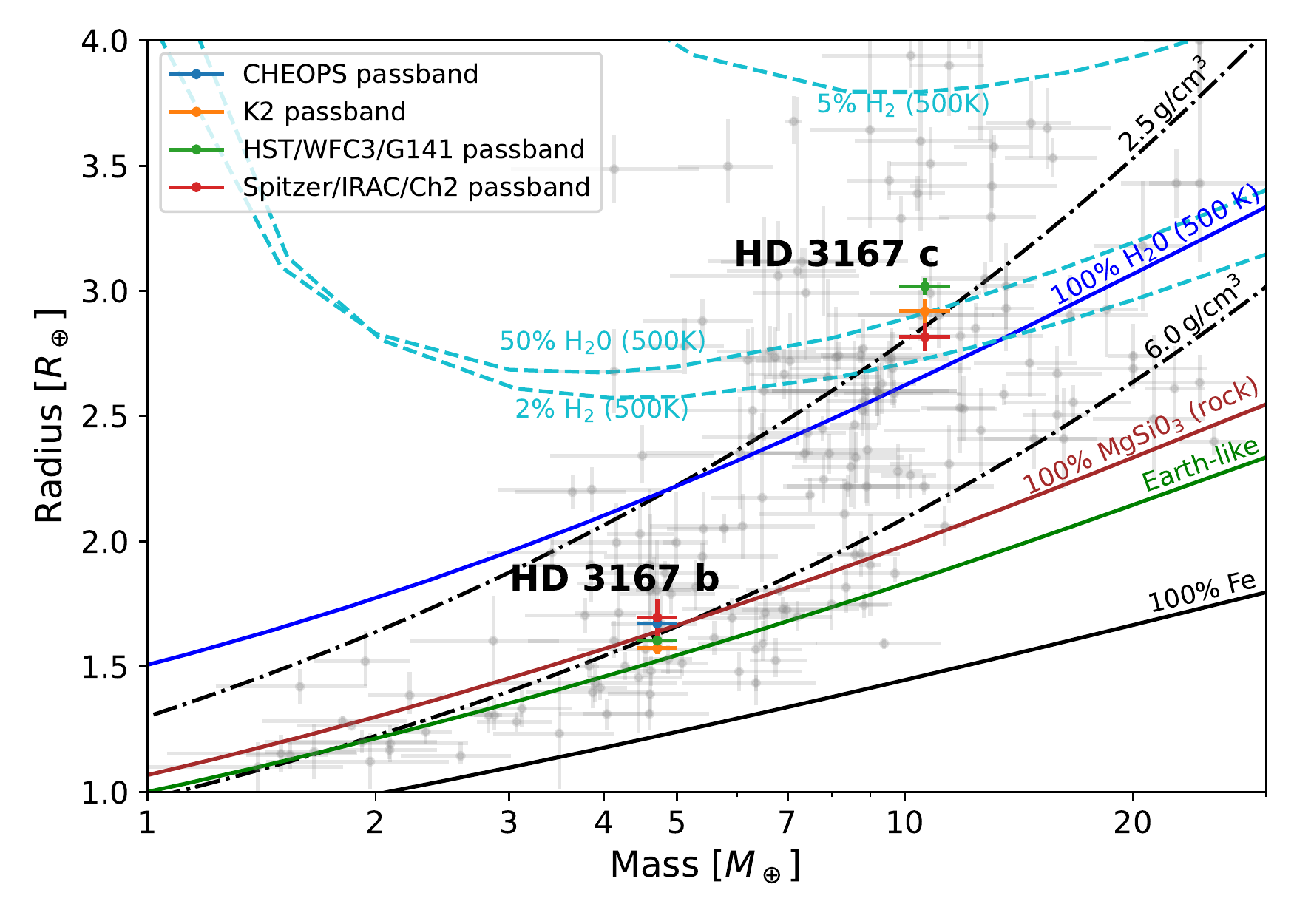}
    \caption{Mass-radius diagram. The grey point represent all the exoplanets from the Extrasolar Planets Encyclopaedia (\protect\url{http://www.exoplanet.eu/}) as of 1\,March\,2022, with masses and radii known with a precision better than 30\%. HD\,3167\,b and c are represented in the different instrument passbands analyzed in this work (CHEOPS, K2, HST/WFC3/G141 and Spitzer/IRAC/Ch2).
    The black dash-dotted lines indicate two iso-density profiles matching the planets b and c.
    The colored solid and dashed lines are indicative chemical compositions as computed by \cite{zeng_2019}.}
    \label{fig:mass-radius}
\end{figure}

All the fitted and derived parameter values are consistent with the ones reported by \cite{Christiansen2017} and \cite{gandolfi_hd3167}. The inclusion of the CHEOPS, HST and Spitzer data sets allows us to improve significantly the precision on the orbital periods of planets b, c, and d by factors of $\sim40$, $>50$ and $\sim17$, respectively. We obtain a better precision on HD\,3167\,b and c planet-to-star radii ratios in the K2 passband analyzed in \cite{Christiansen2017} and \cite{gandolfi_hd3167}, and we improve by more than a factor two the absolute planetary size thanks to the smaller uncertainty on the stellar radius. We also reduce the errors on the absolute and minimum masses of b, c and d thanks to the improved RVs reduction and additional datapoints. These improvements on the planets mass and radius lead to an overall reduction of the uncertainty on the bulk densities of HD\,3167\,b and c by more than a factor three (see Fig.~\ref{fig:mass-radius}).

\begin{figure}
    \centering
    \includegraphics[width=\hsize, trim={0.1cm 0.2cm, 0.25cm 0.1cm}, clip]{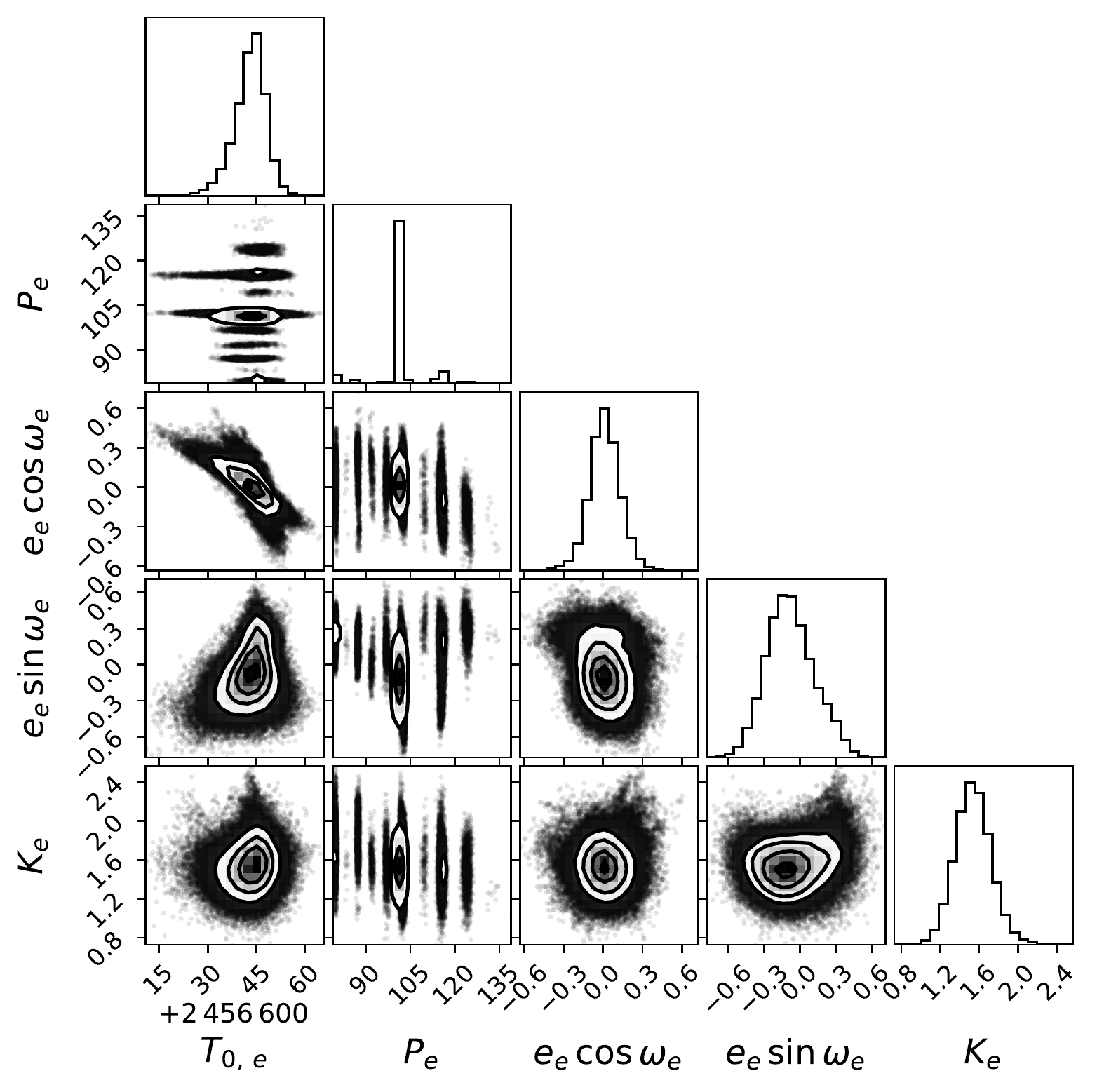}
    \caption{Corner plot of the posterior distribution of the planetary parameters of HD\,3167\,e. The orbital period shows a very peculiar distribution that is not significantly correlated to any other parameter.}
    \label{fig:corner_plot_hd3167e}
\end{figure}

We detect the fourth planet HD\,3167\,e with a semi-amplitude significance $>8\sigma$ and a minimum mass of ${9.73}_{-{1.15}}^{+{1.17}} M_\oplus$. This planet was hinted in the RV data previously available to \cite{Dalal2019}, as a $0.03M_J$ outer companion with an orbital period of 78\,days that could explain the peculiar orbital architecture of the system. The order of magnitude of both the mass and the period we derive matches well their original estimates. We note that the orbital period of this new planet has a very peculiar marginalized posterior distribution. Indeed, the uncertainty of about 0.5\,days listed in Table~\ref{tab:result_fitted} is dominated by the main mode of the distribution. However, the MCMC solution does also explore other possible orbital periods that are less likely but nevertheless span a large range from 79\,days to 125\,days (see Fig.~\ref{fig:corner_plot_hd3167e}). 
We explain this distribution and its invariance with respect to the time of inferior conjunction $T_{0,\,e}$ by the fact that the data are strongly unevenly sampled. Most of the data points (94\%) were taken during the first year of observation (over about 200 days) and the remaining 6\% are HARPS-N data spread over four years. Therefore, $T_{0,\,e}$ is strongly constrained by the bulk distribution of the first year that lasts less than two periods of planet e. The multi-modal distribution of $P_e$ reflects the uneven sampling of the RV time series by highlighting the periods that best match the scattered data. The other orbital parameters of HD\,3167\,e are well defined and not correlated with $P_e$.

The orbits of the four planets are fully consistent with circular configurations with upper limits (99.73\% confidence) on their eccentricities of $e_b<0.11$, $e_c<0.15$, $e_d<0.45$ and $e_e<0.61$ for planets b, c, d and e, respectively.

\begin{figure}
    \centering
    \includegraphics[width=\hsize, trim={2.5cm 0.2cm, 2.2cm 0.5cm}, clip]{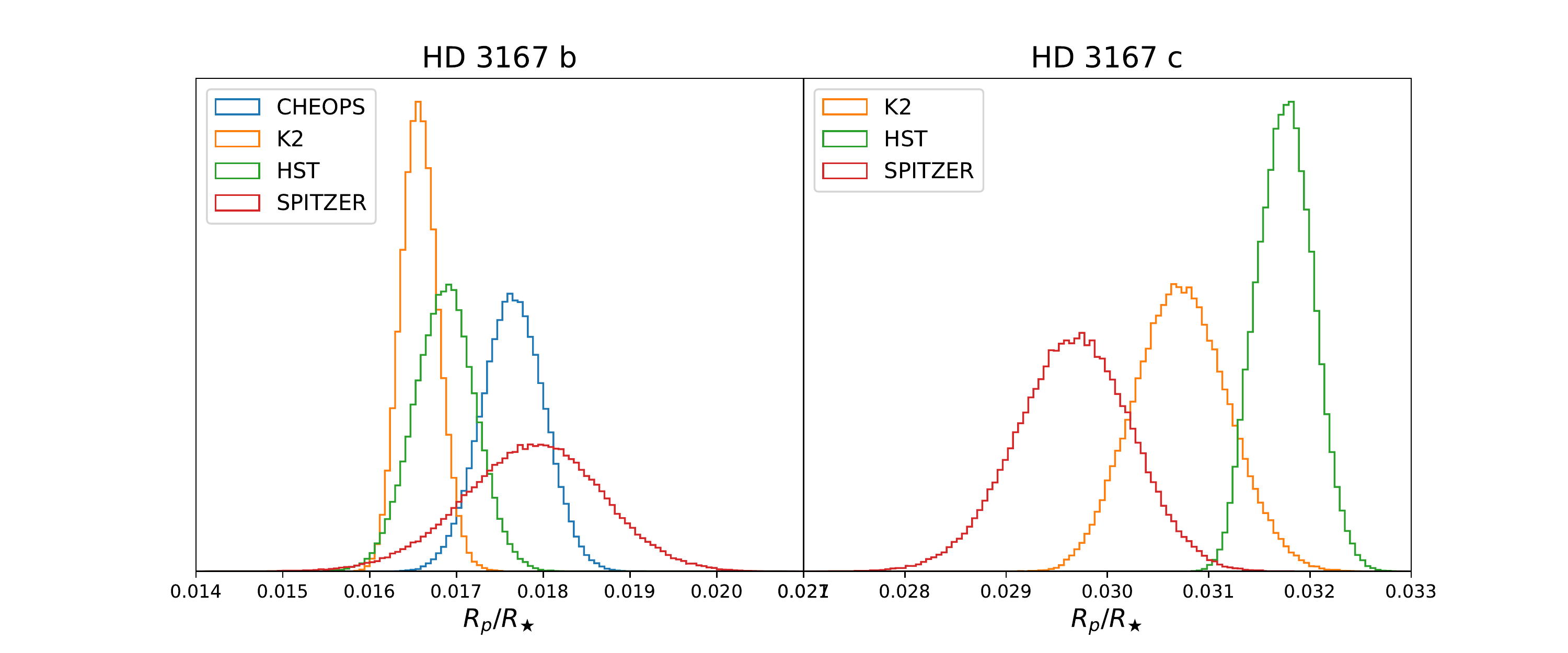}
    \caption{Posterior distributions of the planet-to-star radii ratios of planets b and c measured in the four instrument passbands.}
    \label{fig:planet_radii}
\end{figure}

We provide average transit depths and planetary radii for planets b and c, obtained from the merged distributions over the four available instrumental passbands (CHEOPS, K2, HST/WFC3/G141, Spitzer/IRAC/Ch2). To further characterize the system, we allowed the radii of the two transiting planets to vary independently in those passbands. We measure consistent radii for the inner planet b, which is expected from an USP planet unable to retain any volatile atmosphere. However, we note a significant difference ($>3.5\sigma$) between the radius obtained for planet c in the HST/WFC3/G141 ($\lambda \sim 1.4 \mu m$) and the Spitzer/IRAC/Ch2 ($\lambda \sim 4.5 \mu m$) passbands (see Fig.~\ref{fig:planet_radii}). This difference could arise from broadband variations in the optical depth of the planet atmosphere linked to its chemical composition and physical structure (\citealt{Guilluy2021,mikal-evans_hd3167c}). 

\subsection{Stellar companions}
\label{sec:compa}

\begin{figure}
    \centering
    \includegraphics[width=\hsize, trim={0cm 0cm, 0cm 0cm}, clip]{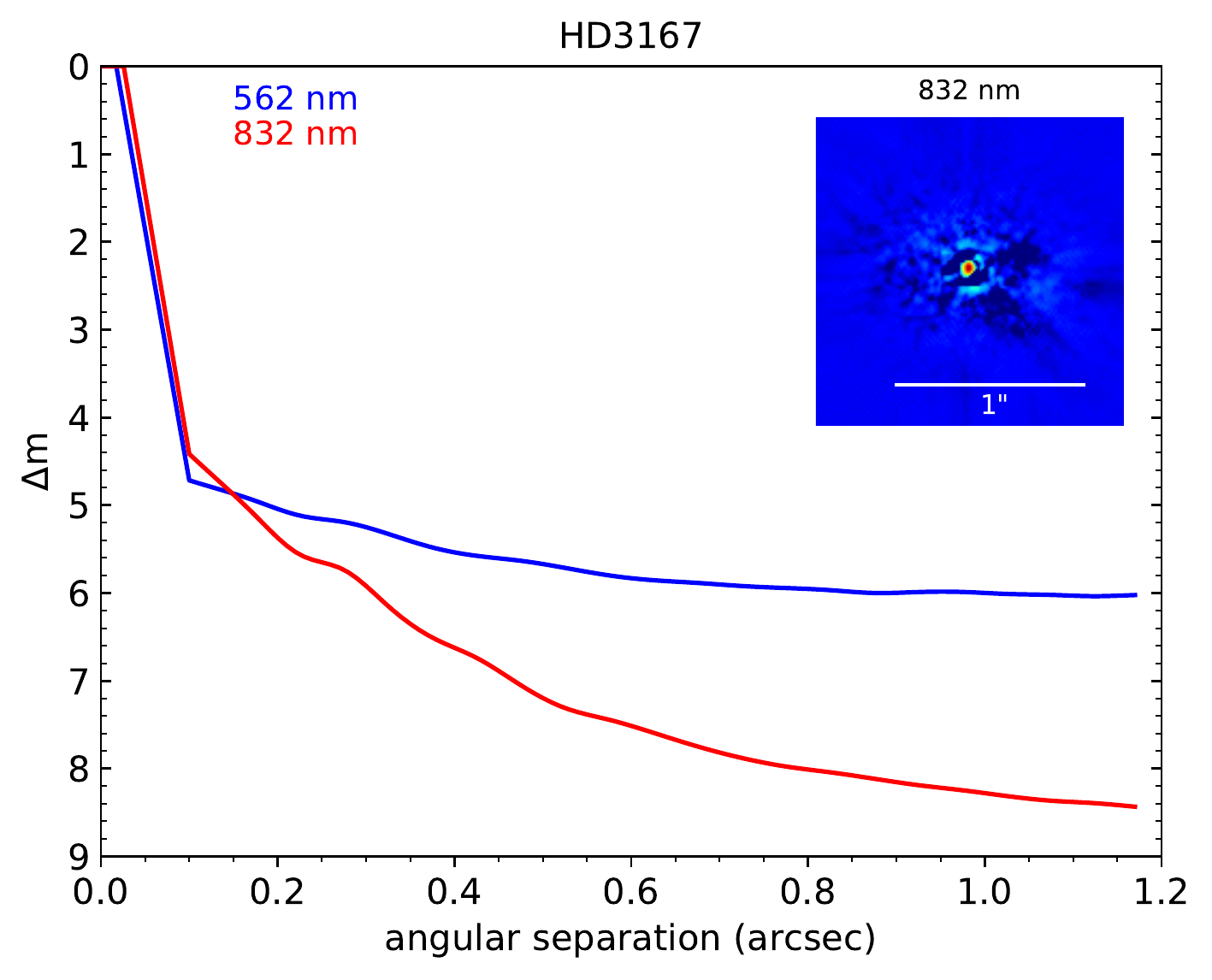}
    \caption{5-$\sigma$ contrast curves of HD\,3167 from Gemini North/‘Alopeke at 562\,nm (blue) and 832\,nm (red). The insert shows our reconstructed 832 nm speckle image.}
    \label{fig:imaging}
\end{figure}

In order to check for stellar companions lying within the environment of HD\,3167, high angular resolution optical speckle interferometric imaging was performed. HD\,3167 was observed on 2021 June 28 UT using the ‘Alopeke speckle instrument on Gemini North (\citealt{Scott2021}). ‘Alopeke provides simultaneous speckle imaging in two narrow bands (562\,nm and 832\,nm) with output data products including a reconstructed image and robust contrast limits on companion detections (\citealt{Howell2011,Howell2016}). The night had clear skies and good seeing ($<$1.0 arcsec) during the observations. As shown in Figure~\ref{fig:imaging}, we detect no stellar companions which are brighter than two delta-magnitudes within 0.1” and no companions brighter than five to 8.5 magnitudes within the angular separation limits of 0.1” to 1.2”. Using a distance of d = 47\,pc for HD\,3167, these angular and luminosity limits on stellar companions correspond to main sequence stellar types of K6V (at 0.94 au) and M2.5V to M4.5V between 4.7 to 56.4 au. 

\section{Planetary internal structures}
\label{sec:int_struct}

Using the derived stellar and planetary properties (in particular the mass and average transit depths reported in Table~\ref{tab:result_derived}), we computed the internal structure of both transiting planets using a Bayesian analysis, and following the method described in \citet{Leleu2021}. We recall here the two main elements in this method: the assumed priors and the forward model. 

Our forward model computes the radius of planets as a function of some hidden parameters: mass of the solid Fe/S core, fraction of Fe in the core,  mass of the silicate mantle and its composition (Si, Mg and Fe molar ratios), mass of the water layer, mass of the gas envelope (composed in this model of pure H/He), equilibrium temperature of the planet, and age (which is supposed to be the same as the age of the star). We assume in our model that the Si/Mg/Fe ratio in the bulk planet is the same as in the star (\citet{Dorn2015}, \cite{Thiabaud2015}). Note that recently, \citet{Adibekyan2021} have shown that these ratio are indeed correlated but may not follow a 1-to-1 correlation. Including this in the model is the subject of future work. Regarding the priors, the core, mantle and water mass fraction (relative to the non-gas part) follow a uniform prior (subject to the constraint that they add up to one), whereas the mass fraction of the H/He layer follows a prior which is uniform in log. We finally note that the gaseous (H/He) part of the planet does not influence, in our model, the `non-gas' part of the planet (core, mantle and water layer). This means that the innermost layers of the planet are not modified by the potential compression and thermal isolation effect from the gas envelope. 

\begin{figure*}
    \subfloat{{\includegraphics[width = 8.5cm]{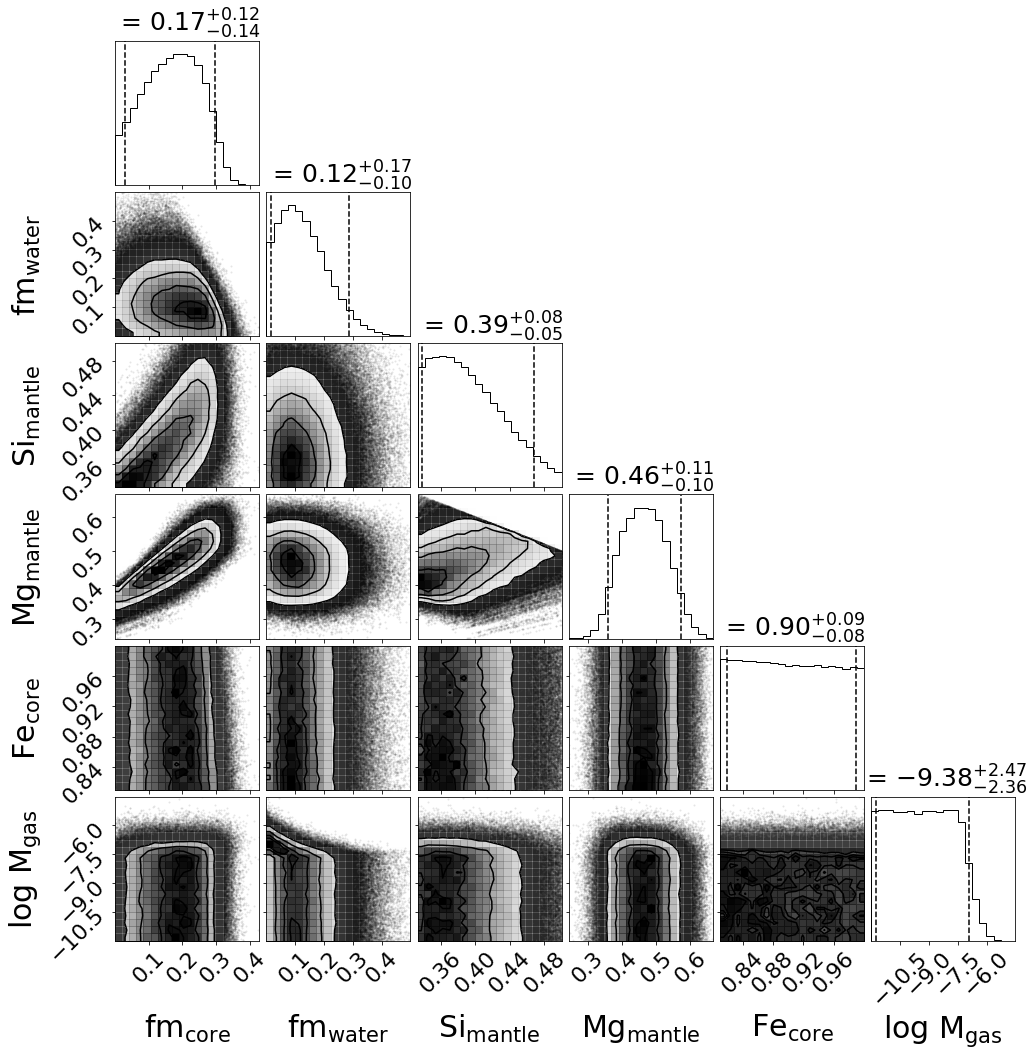}}}
    \qquad
    \subfloat{{\includegraphics[width = 8.5cm]{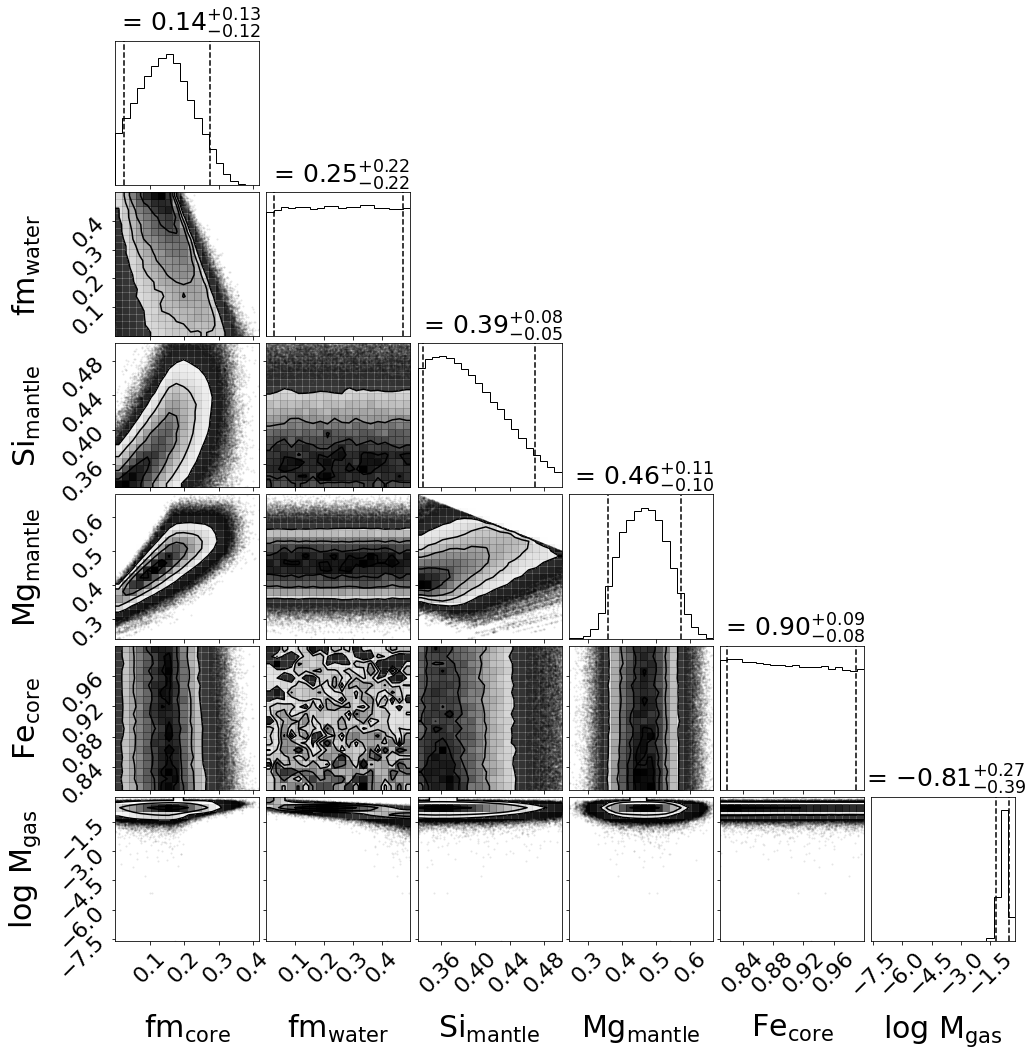} }}
  \caption{Corner plot showing the results on the interior composition models of HD3167b (left) and HD3167c, (right). The vertical dashed lines and the 'error bars' given at the top of each column represent the 5~\% and 95~\% percentiles. The internal structure parameters are the mass fraction of the core and the water layer with respect to the solid planet, the molar fraction of Si and Mg in the mantle, the molar fraction of Fe in the inner core and the logarithm of the gas mass in Earth masses.}
  \label{fig:corner_IS}
\end{figure*}

Fig.~\ref{fig:corner_IS} shows the resulting internal structure of both planets presented as corner plots and summarized in Table~\ref{tab:intstructure}. Planet c hosts a substantial gaseous envelope, weighing a little less than 0.2 $M_\oplus$, whereas its fraction of water is unconstrained. We emphasize that this result depends on the assumed priors. In particular, the resulting planetary model would be more gas rich and less water-rich if the H/He layer followed a uniform prior. One of our main findings is that planet b mass and radius seem to be inconsistent with a pure iron-core and silicate-mantle structure whose composition would reflect the Fe/Si/Mg ratio in the star. Indeed, the density of the planet is smaller than what would be expected for such a model. Since our model assumes the inner layers of the planet to be unaffected by the influence of the gas envelope, for planet b we underestimate the temperature of the ‘non-gas’ part of the planet. If we increase the temperature of the core and mantle layers in our model, we do observe an increase in radius of up to 2\% for pure iron-core and silicate-mantle structures matching the Fe/Si/Mg ratio of the star. However, this effect alone is not enough to explain the observed radius of the planet. We hence conclude that a light element must be present in the planet.

Our fit converges toward a negligible mass fraction of gas, which is expected considering that the intense irradiation of this USP would lead a H/He atmosphere to be lost extremely fast. However, the mass fraction of water for our model of HD\,3167 b is quite well constrained and non zero. With an equilibrium temperature in excess of 1600\,K (Table~\ref{tab:result_derived}) any water layer would be made of steam, which has been shown to be much more resilient to atmospheric loss (e.g., \citealt{Lopez2012}). It should be kept in mind that our model assumes a fully differentiated planet. It is possible that water is mixed with a magma ocean covering HD\,3167b, in which case its actual mass fraction of water would be reduced compared to the one we derive (see \cite{Dorn2021}). More detailed internal structure model accounting for this mixing, and for the existence of a dust- and metal-rich envelope, are required to better constrain the true nature of this planet.

\begin{center}
\begin{table}
    \centering
    \caption{Interior structure properties of planets b and c. The errors correspond to the 5\% and 95\% percentiles.}
    \label{tab:intstructure}
    \begin{tabularx}{0.67\columnwidth}{ l  l  X }
    \toprule
    \textbf{Property (unit)} & \multicolumn{2}{c}{\textbf{Values}} \\
    \hline
       & \textbf{HD3167b}  &\textbf{HD3167c}   \\
    M\textsubscript{\textit{core}}/M\textsubscript{\textit{total}}    & \Mcoreb  & \Mcorec \\
    M\textsubscript{\textit{water}}/M\textsubscript{\textit{total}}  & \Mwaterb & \Mwaterc   \\
    log(M\textsubscript{\textit{gas}})   & \Mgasb   & \Mgasc   \\
    Fe\textsubscript{\textit{core}} & \Fecoreb  & \Fecorec   \\
    Si\textsubscript{\textit{mantle}} & \Simantleb  & \Simantlec   \\
    Mg\textsubscript{\textit{mantle}} & \Mgmantleb  & \Mgmantlec   \\
    \bottomrule
    \end{tabularx}
\end{table}
\end{center}

\section{Dynamical evolution}
\label{sec:dyn_evol}

With planet b aligned with its host star (\citealt{Bourrier2021_3167}) and the distant planet c on a polar transit (\citealt{Dalal2019}), the HD\,3167 system is particularly rich and interesting for dynamical studies. Its planets have wide orbital separations and are far from mean motion resonances, so that their dynamics is fully secular.
Moreover the age of the system suggests that its orbital configuration is dynamically stable.

The dynamical analysis by \citet{Dalal2019} suggested that planet b could have stayed aligned with the equatorial plane of the star, which has since been confirmed by \citet{Bourrier2021_3167}. They further showed that planets c and d have a low mutual inclination, and that the three planets known at that time could not, by themselves, have caused the polar orbit of planet c. \citet{Dalal2019} thus proposed that the orbits of planets c and d could have been tilted with respect to the star due to a massive outer companion, whose existence we have confirmed in the present study (Sect.~\ref{sec:pl_results}). It is thus natural to investigate whether this planet e could indeed explain the polar orbit of planet c.

To gain some insights onto the dynamics of the system, we consider an analytical framework describing the precession of the orbits \citep{Boue2006}.
Following \cite{Boue2014}, we compute the characteristic frequencies $\nu^{k/j}$ that represents the relative influence of the body $k$ over the direction of the angular momentum of body $j$.
We refer to \citet[][Sec. 5.2.]{Dalal2019} for a more precise description of the model used here.
To summarize, if $\nu^{k/j}\ll\nu^{j/k}$, the angular momentum direction of $j$ is almost constant while the angular momentum of $k$ precesses around.

\begin{table}
    \centering
    \caption{Characteristic precession frequencies for different interactions in the system for the current configuration as well as during the system formation. The typical relative uncertainty is~10\%.
      \label{tab:frequencies}}
    \begin{tabularx}{0.8\columnwidth}{l c c}
         \toprule
                & {\bf Old star} & {\bf Young star}\\
        \midrule
         $P_S$ & 18 d & 3 d\\
         $k_2$ & 0.018 & 0.18\\
         \midrule
        $\nu^{b/S}\ {(\rm rad.yr^{-1})}$  & $3.94 \times 10^{-5}$ & $1.26 \times 10^{-2}$\\
        $\nu^{S/b}\ {(\rm rad.yr^{-1})}$  & $2.28 \times 10^{-6}$ & $1.29 \times 10^{-4}$\\
        \midrule
        $\nu^{b/d}\ {(\rm rad.yr^{-1})}$  & \multicolumn{2}{c}{$1.04 \times 10^{-4}$ } \\
        $\nu^{d/b}\ {(\rm rad.yr^{-1})}$  & \multicolumn{2}{c}{$2.33 \times 10^{-4}$ } \\
        $\nu^{d/c}\ {(\rm rad.yr^{-1})}$  & \multicolumn{2}{c}{$1.45 \times 10^{-4}$ } \\
        $\nu^{c/d}\ {(\rm rad.yr^{-1})}$  & \multicolumn{2}{c}{$4.51 \times 10^{-4}$ } \\
        $\nu^{e/c}\ {(\rm rad.yr^{-1})}$  & \multicolumn{2}{c}{$1.30 \times 10^{-4}$ } \\
        $\nu^{c/e}\ {(\rm rad.yr^{-1})}$  & \multicolumn{2}{c}{$9.26 \times 10^{-5}$ } \\
         \bottomrule
    \end{tabularx}
\end{table}

We study the dynamics of the system at two different epochs. 
First, in its current configuration, with a stellar rotation period of about 17 d. 
Second, right after the system's formation, when the star was rotating fast and its spin had a stronger influence onto planet b.
The different characteristic frequencies\footnote{A similar analysis was performed by \cite{Dalal2019} but we found a typo in the code computing the frequencies. The main conclusions of \cite{Dalal2019} are unchanged but the planet-planet interactions were underestimated which means that planet b is not as strongly coupled to the star as stated in the paper.} in these two configurations are summarized in Table~\ref{tab:frequencies}.
While the precession frequencies ruling the planet interactions are the same in both settings, there is a significant change for the interaction between the star's spin and the planets as we can see on the frequencies \(\nu^{b/S}\) and \(\nu^{S/b}\).
The change is not only due to the shorter rotation period of the star early on, but also because the second fluid Love number $k_2$ can be significantly larger for a fast rotating star \citep{Becker2020}.
We adopt a value of $k_2=0.18$ for the fast rotating star, an order of magnitude larger than the expected value for HD 3167 today.

\subsection{System stability}

\begin{figure}
    \includegraphics[width=0.9\columnwidth]{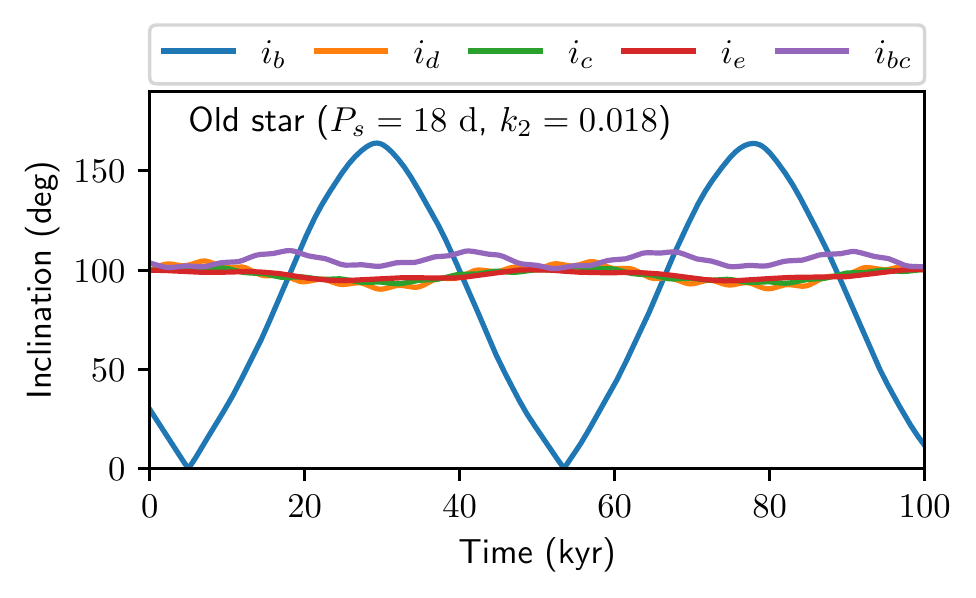}
    \caption{Inclination evolution in the current system configuration, assuming the planets c,d and e are close to coplanar.\label{fig:bprec}}
\end{figure}

In the present system configuration we have \(\nu^{b/d},\nu^{d/b}\gg\nu^{S/b}\gg\nu^{b/S}\), which indicates that planet b's orbit precesses around the angular momentum of the outer planets and that the star plays a negligible role. Moreover, the stellar spin is dynamically unaffected by the planets.
We confirm this hypothesis by running an N-body simulation using the integrator \texttt{WHFast} \citep{Rein2015a} from the library \texttt{Rebound} \citep{Rein2012a}.
We include relativistic corrections as well as the influence of the stellar $J_2$ using the library \texttt{Reboundx} \citep{Tamayo2019}.
The stellar spin is fixed and along the $z$-axis.
For this particular simulation, the initial orbits are assumed circular, the planets c, d and e are assumed coplanar.
We use an initial condition compatible with the 3D configuration determined by \cite{Bourrier2021_3167}, \(i_b=30^\circ\), \(\Omega_b=100^\circ\) and \(i_{d,c,e}=100^\circ,\Omega_{d,c,e}=0^\circ\).
As a result, the mutual inclination between planet b and the rest of the system is $i_{bc}=103^\circ$.
This approximate initialization is sufficient because we only want to illustrate the typical dynamics at play.
In reality, there is likely a non zero inclination between planet d, c, and e since planets d and e are not transiting.
We integrate the system over $100\ {\rm kyr}$ and plot on Figure~\ref{fig:bprec} the planetary inclinations with respect to the star, as well as the mutual inclination between planet b and c.
During this short integration, we observe no evolution of the eccentricities.
We see that the mutual inclination between b and c, as well as the inclination of the outer planets are almost constant while planet b orbital plane precesses around the orbital plane of the outer planets.
We conclude that, while planet b is not strongly coupled with the star today, a primordial misalignment of the outer planets with respect to the star and planet b leads to a stable configuration, compatible with the observations. However, early in the history of the system, we have \(\nu^{S/b}\gg\nu^{b/d},\nu^{d/b},\nu^{b/S}\) which means that planet b could have been coupled with the star instead of the outer planets.
If planets c and d gained their large obliquities early on, planet b would have stayed close to the stellar equator.

\subsection{Spin-orbit misalignment by planet e}

\begin{figure}
    \includegraphics[width=\linewidth]{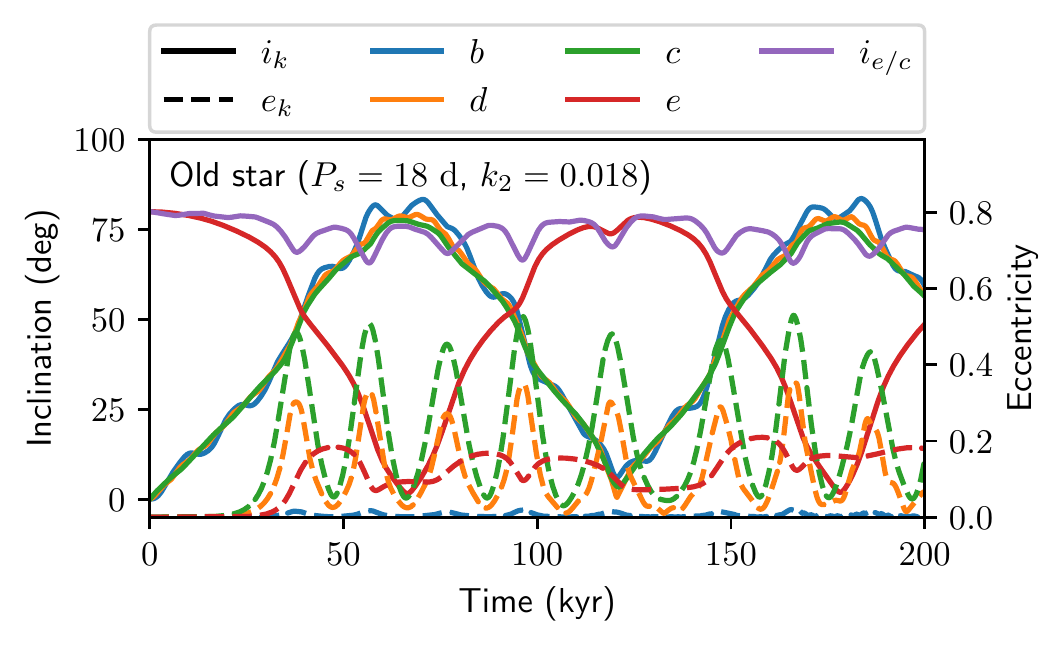}
    \caption{Evolution of the inclinations and eccentricities assuming the current star properties and a planet e initially misaligned with the system.\label{fig:KL}}
\end{figure}

We now investigate whether planet e could be the cause of the polar orbits of planets c and d.
The characteristic frequency \(\nu^{e/c}\) is not negligible, which suggests that planet e could tilt planets c and d if it was originally inclined with respect to the star.
\cite{Boue2014} have determined that an external companion can tilt a planetary system as a whole if the coefficient
\begin{equation}
    \beta_{{\rm KL,dc}} = \frac{m_{\rm comp}}{m_{\rm d}}\left(\frac{a_{\rm c}}{a_{\rm d}}\right)^2\left(\frac{a_{\rm c}}{b_{\rm comp}}\right)^3\ll1,
    \label{eq:KLcondition}
\end{equation}
where \(b_{\rm comp}=a_{\rm comp}\sqrt{1-e_{\rm comp}^2}\).
However, planet e is too close to c to tilt the system without triggering Kozai-Lidov oscillations for planet d and c \citep{Kozai1962,Lidov1962}.
Indeed, we have $\beta_{\rm KL}=0.88\pm0.13$. As a result, a large mutual inclination of planet e with respect to c excites the orbital eccentricities, eventually leading to the system destruction.
We run a numerical simulation starting with the inner planet b, d and c on coplanar, circular orbits within the star equatorial plane and a planet e on an orbit tilted by $80^\circ$.
The simulation lasts 200 kyr and we plot on Figure \ref{fig:KL} the eccentricities and inclinations as a function of time.
As expected planets c, d and e enter Kozai-Lidov oscillations and the eccentricities grow to values close to 0.5,
which is excluded by the observations (at 2-$\sigma$, $e<$0.3 for planets d and e and $e<$0.08 for c) and is enough to trigger the dynamical instability of the system. Moreover, in that scenario planet b would remain in the plane of planets d and c, which confirms that the outer system had to get misaligned early-on when the coupling between planet b and the star was stronger. Planet e thus cannot explain the polar orbit of planet c.

\subsection{Spin-orbit misalignment by an outer companion.}

\begin{figure}
    \centering
    \includegraphics[width=\linewidth]{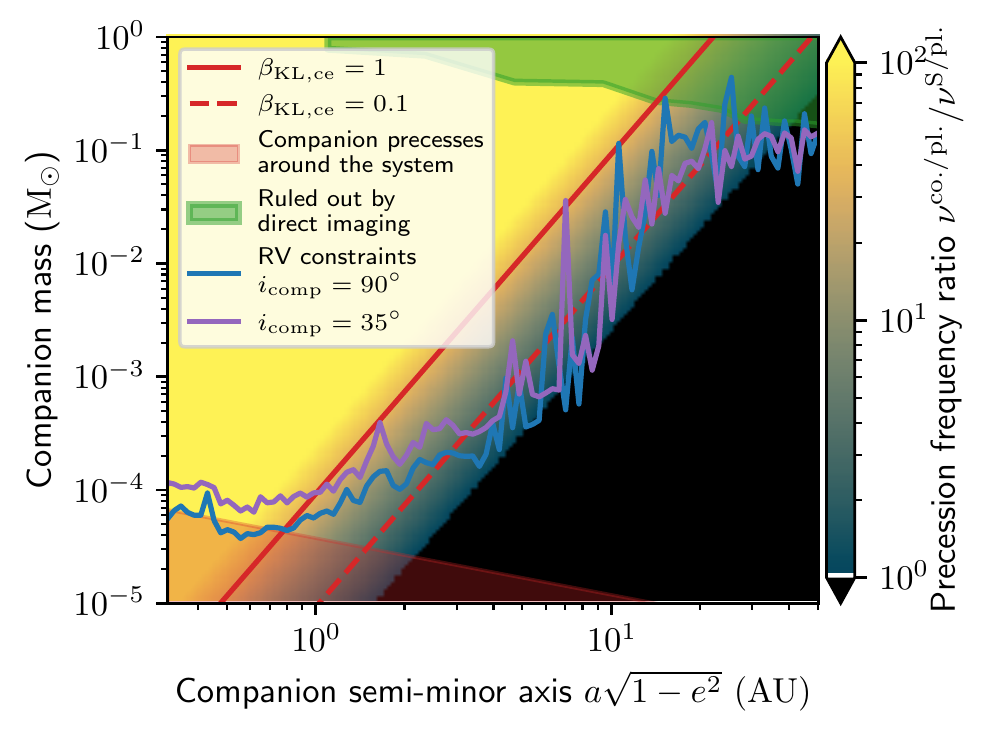}
    \caption{Range of masses and semi-minor axis that allows a companion to tilt the outer planets of the system while preserving its mutual inclinations. The system is not destroyed for companions below the line $\beta_{{\rm KL,ce}}=1$ and tilting is possible for a frequency ratio larger than 1.
    The companion has a significant influence on the system if its angular momentum is larger than the system's angular momentum.
    Otherwise, the companion's orbit precesses around the system angular momentum and does not induce large obliquity for the system.
    The green region is the 5-$\sigma$ limit set by direct imaging constraints. The purple and blue curve are the 5-$\sigma$ limit set by RVs, assuming a circular orbit and inclinations of 90 and 35$^{\circ}$.}
    \label{fig:usefulcomp}
\end{figure}

While a misaligned planet e leads to the instability of the system, we explore whether a more distant companion could explain the present configuration.
Such an hypothetical companion should lie in a range of masses and semi-major axes that verify the Kozai-Lidov condition
\begin{equation}
    \beta_{{\rm KL,ce}} = \frac{m_{\rm comp}}{m_{\rm c}}\left(\frac{a_{\rm c}}{a_{\rm e}}\right)^2\left(\frac{a_{\rm e}}{b_{\rm comp}}\right)^3\ll1,
    \label{eq:KLcondition-ce}
\end{equation}
while being able to tilt the system as a whole.
We plot on Figure \ref{fig:usefulcomp} the ratio of the precession frequency $\nu^{\rm co./pl.}/\nu^{\rm S/pl.}$ that represents the relative influence of a companion onto the outer planets with respect to the influence of the star onto the system as a function of the semi-minor axis and the mass of a potential companion.
We also plot the levels of $\beta_{{\rm KL,ce}}=1$  and $\beta_{{\rm KL,ce}}=0.1$.
A companion can tilt the outer planets if condition \eqref{eq:KLcondition-ce} is met and precession frequencies verify $\nu^{\rm co./pl.} \gg \nu^{\rm S/pl.}$.
Additionally, the companion has a significant influence on the inclination of the system if its angular momentum is larger than the system's angular momentum. 
Otherwise, the system orbital plane remains unchanged and the companion orbital plane  precesses around.
We plot this theoretical constraint in Fig.~\ref{fig:usefulcomp}.

We further included in Fig.~\ref{fig:usefulcomp} the constraints derived from our direct imaging measurements (Sect.~\ref{sec:compa}). We converted luminosity differences with the K0V type star HD\,3167 into spectral types for various separations, and then used Table 5 from \citet{Pecaut2013} to assign mean masses to these spectral types.
We also assumed circular orbits to estimate the semi-minor axes (which is the most conservative assumption).
While the masses adopted for a given dwarf subtype are tentative, it provides an approximate upper limit on the companion mass. 
The direct imaging constraints impose that it orbits within $\sim$30\,au from the star and cannot be more massive than about 0.1 solar mass.
Finally we included in Fig.~\ref{fig:usefulcomp} the stringent constraints from our RV dataset.
The constraints change little with the unknown orbital inclination of the companion unless it is seen nearly pole-on.
The RV constraints rule out most, if not all, the configurations where an outer companion could lead to a significant misalignment of the system.
We conclude that the polar orbits of the outer planets are most likely due to an early misalignment during the system formation that did not rely on a companion still present in the system.
Mechanisms that do not rely on binary companion or secular interactions have been proposed such as magnetic coupling between the young star and the disk \citep{Lai2011,Foucart2011,Romanova2021}.

\begin{figure*}
    \includegraphics[width=\linewidth]{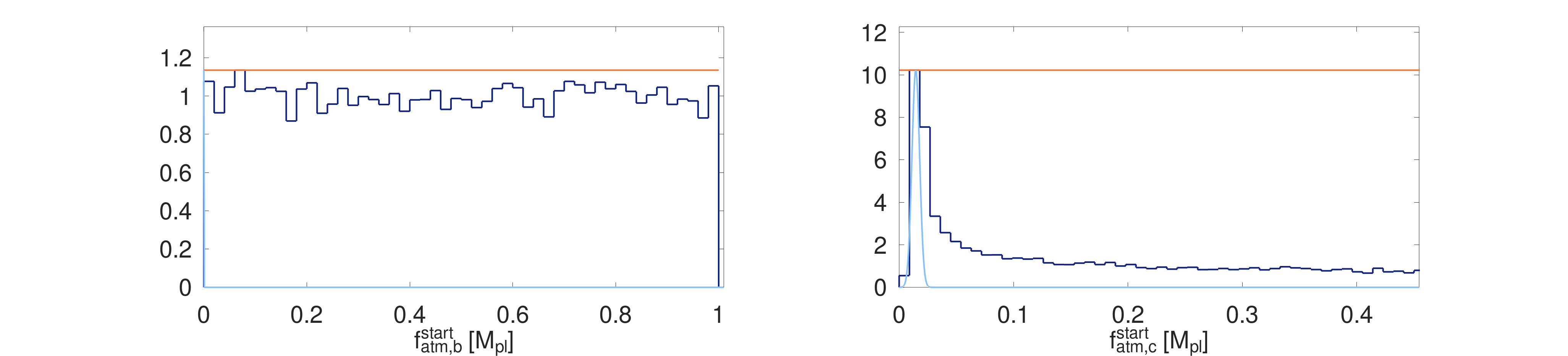}
    \caption{Posterior distributions of the initial atmospheric mass fractions for planets HD\,3167 b and c derived by \texttt{PASTA}. The light-blue line represents the distribution of the estimated present-day atmospheric mass-fraction. The orange horizontal lines indicate the uninformative prior distributions. \label{fig:fatm}}
\end{figure*}

\section{Atmospheric evolution}
\label{sec:atm_evol}

To constrain the atmospheric evolution of the planets in the HD\,3167 system and the stellar rotation history of the host star, we employed the tool \texttt{P}lanetary \texttt{A}tmospheres and \texttt{S}tellar Ro\texttt{T}ation R\texttt{A}tes \citep[\texttt{PASTA};][]{bonfanti21b}. \texttt{PASTA} uses the measured system parameters and the present-day atmospheric mass fractions determined by the internal structure modeling (see Section \ref{sec:int_struct}) to return posterior probability distributions for the initial atmospheric mass-fraction of each planet, further constraining the history of the stellar rotation rate. Because of the need of an estimate of the present-day atmospheric mass fraction, or at least of a radius measurement, to constrain the evolution of the planetary atmosphere, this tool in its full capability can only be employed on the two transiting planets HD\,3167 b and c. Since the present-day stellar rotation rate is not well defined, we employed a uniform prior ranging between 15 and 20 days.

\subsection{Transiting planets HD\,3167 b and c}
Planet HD\,3167 b orbits very close to its host star, resulting in it having being subject to large amounts of X-ray and extreme ultraviolet (XUV) irradiation, particularly in the early phases of its evolution. \texttt{PASTA} predicts that this planet has lost its primary H/He-dominated atmosphere at some point in the past, and thus the code is unable to constrain the initial atmospheric mass-fraction, resulting in a uniform posterior distribution (left panel of Figure \ref{fig:fatm}). 

\begin{figure}
    \centering
    \includegraphics[width=\hsize]{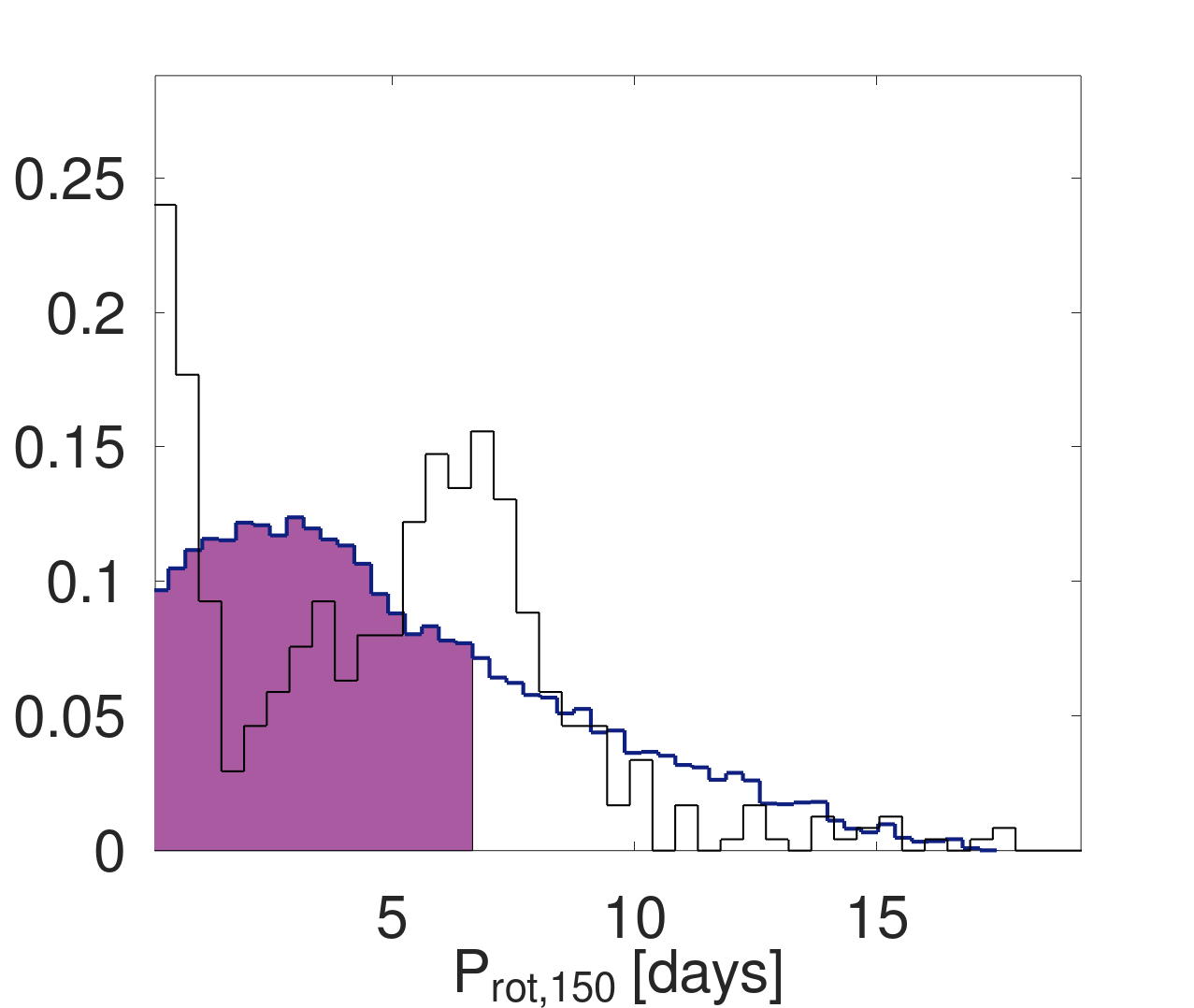}
    \caption{Posterior distribution (blue line) of the stellar rotation rate of HD\,3167 after 150 Myrs derived by \texttt{PASTA}. The purple area represents the highest posterior density (HPD) interval of the distribution. The black line represents the distribution of the stellar rotation rate of young open cluster stars with mass comparable to that of HD\,3167 based on the collection of data provided by \citet{Johnstone2015c}.}
    \label{fig:pRotIni}
\end{figure}

For planet HD\,3167 c, \texttt{PASTA} prefers evolutionary tracks for which atmospheric mass loss did not play a significant role. This is represented by the posterior distribution of the initial atmospheric mass fraction that peaks around the present-day value (right panel of Figure \ref{fig:fatm}). However, the results also indicate that evolutionary tracks characterized by significant mass loss, though less likely, are not completely excluded. In case the host star was a particularly fast rotator when it was young, it would indeed have emitted a significant amount of XUV radiation \citep{Sanz-Forcada2011}. Therefore, we explored this possibility more thoroughly. Unfortunately, the characteristics of HD\,3167 c, which is the only transiting planet in the system still holding its primordial H/He-dominated atmosphere, do not enable \texttt{PASTA} to constrain the rotation history of the star. This is represented by a rather flat posterior distribution of the stellar rotation rate after 150 Myr that we use as proxy to illustrate the evolution of the stellar rotation rate (Figure \ref{fig:pRotIni}). The rotation rate distribution of stars with a mass comparable to that of HD\,3167 and member of open clusters with ages of about 150 Myr is bimodal \citep[e.g.,][]{Johnstone2015c}: one peak represents fast rotators at a rotation rate close to one day, while the other peak represents moderate rotators at roughly six days. The observed distribution is shown in Figure \ref{fig:pRotIni}. We have performed additional runs with \texttt{PASTA}, imposing priors on the stellar rotation rate at 150 Myr corresponding to Gaussian fits to each of the two peaks of the distribution. Assuming the host star was a fast rotator when it was young, the relative occurrence of evolutionary tracks presenting some significant atmospheric loss increases compared to when not constraining the stellar rotation type at all. However, we still obtain a posterior distribution of the initial atmospheric mass fraction which peaks close to the present-day atmospheric mass fraction, indicating that most likely atmospheric loss has not played a major role in the evolution of this planet independently of the evolutionary history of the stellar rotation rate.


\subsection{Non-transiting planets HD\,3167 d and e}
Since HD\,3167 d and e are not transiting, it is not possible to estimate their current atmospheric mass fraction, and thus it is not possible to use \texttt{PASTA} to infer the initial atmospheric mass fraction. To estimate the current atmospheric content of both planets, we start from assuming that the planets accreted a primordial H/He-dominated atmosphere of \citep{mordasini2020}
\begin{equation}
    \frac{M_{\rm env,0}}{M_{\oplus}} = 0.024 \left( \frac{M_{\rm c}}{M_{\oplus}} \right)^{2.23} \left( \frac{\rm a}{1 \rm AU} \right)^{0.72}\,,
    \label{eq:mordasini}
\end{equation}
where $M_{\rm env,0}$ is the envelope-mass, $M_{\rm c}$ the core-mass, and $a$ the planetary orbital separation. We used \texttt{PASTA} to compute the atmospheric evolution of HD\,3167 d and e starting it with the atmospheric mass fraction given by Equation~(\ref{eq:mordasini}). The simulations also require an estimate of the evolution of the rotation rate of the host star. Since \texttt{PASTA} has been unable to constrain the stellar rotation history using planets b and c, we further assumed a value of the rotation rate of the host star at 150 Myr of 5.44 days. This value corresponds to the mean of the distribution of stellar rotation rates of young open cluster stars with mass comparable to HD\,3167 (Figure \ref{fig:pRotIni}).

As HD\,3167 d orbits quite close to its host star and has a low mutual inclination with the transiting HD\,3167c (Sect.~\ref{sec:dyn_evol}), we assumed the measured lower mass limit $M_d sin(i)$ to be a good approximation of the core mass. Through Equation (\ref{eq:mordasini}), we then estimate an initial atmospheric mass fraction of 0.029. \texttt{PASTA}'s evolution simulation predicts for this planet to have lost all of its primordial H/He-dominated envelope via photo-evaporation. The atmospheres of planets b and d therefore seem to have had very similar evolutionary paths.

HD\,3167 e on the contrary orbits significantly further away than HD\,3167 c, for which we already determined that hydrodynamic mass loss was only important if the star would have been a very fast rotator. Therefore, we assumed for the measured lower mass limit $M_e sin(i)$ to be a good approximation of the initial total mass of the planet after the dispersion of the protoplanetary nebula. With these assumptions, Equation (\ref{eq:mordasini}) leads to an initial atmospheric mass fraction of 0.147. Using the estimates of the initial total mass and atmospheric mass fraction, estimates on the core-mass and envelope-mass after the dispersion of the nebula can be provided. Assuming an earth-like density for the core, this results in an estimate of the average density of the planet at the beginning of its atmospheric evolution. This average density then changes due to atmospheric loss as the evolution progresses. As expected, \texttt{PASTA}'s evolution simulation predicts no significant mass loss for this planet, resulting in a Saturn-like density of about 0.67 g/cm$^3$. This value is however heavily dependent on Equation (\ref{eq:mordasini}), as it estimates the initial atmospheric mass fraction and therefore the initial core- and envelope-mass.

\subsection{Comparison with previous studies}
\citet{Kubyshkina2019} applied an earlier version of \texttt{PASTA} and \citet{bonfanti21b} the same version of \texttt{PASTA} used here on the HD\,3167 system considering the system parameters available at the time. \citet{bonfanti21b} focused on the two transiting planets b and c, while \citet{Kubyshkina2019} did also investigate the atmospheric evolution of the non-transiting planet d. For the two close-in planets b and d the two previous studies agree with our conclusion that the planets should have lost all of their primary H/He-dominated envelopes. 

Both \citet{Kubyshkina2019} and \citet{bonfanti21b} ran their fits considering the planetary radii, instead of the current atmospheric mass fractions, which led them to assume slightly larger atmospheric mass fractions compared to what was used here. Furthermore, both studies used significantly longer present-day stellar rotation rates, with \citet{Kubyshkina2019} using a prior peaking at roughly 25 days, while \citet{bonfanti21b} used an even larger peak value of about 50 days, based on gyrochronological considerations. \citet{bonfanti21b} obtained that the star was likely to be a slow rotator, which might be related to the large value of the current stellar rotation period they considered. Along the lines of our results, they concluded that planet c has most likely retained most of its primary H/He-dominated atmosphere and, therefore, has not undergone significant mass loss. In contrast, \citet{Kubyshkina2019} concluded that the young star was a moderate-to-fast rotator, in agreement with our hypothesis.




\section{Conclusions}

\begin{figure}
    \centering
    \includegraphics[width=\hsize]{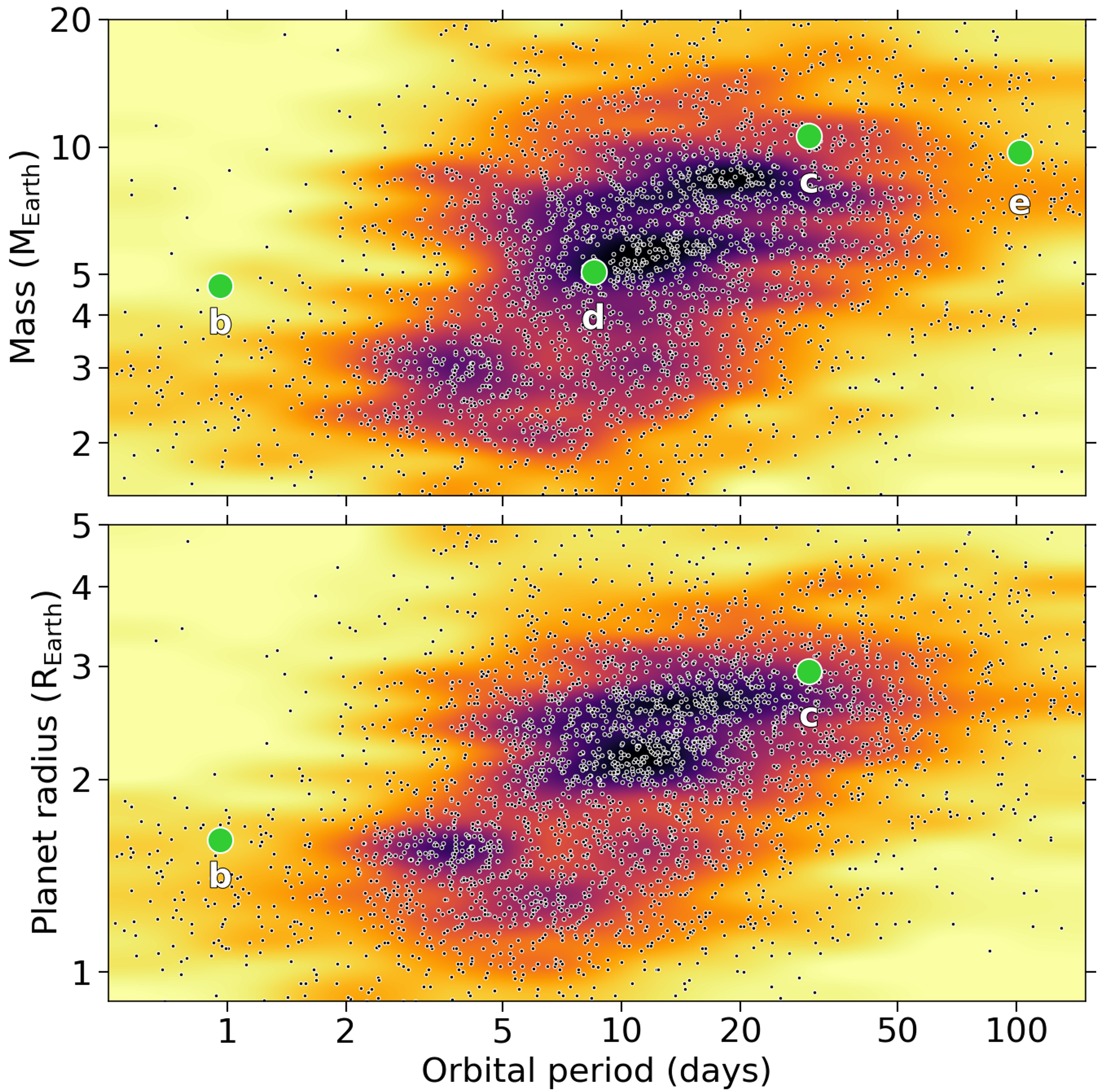}
    \caption{Mass-period (top panel) and mass-radius (bottom) diagrams of the exoplanet population in the Earth-Neptune range. HD\,3167 planets are shown as green disks (HD\,3167d and e are positioned at their measured minimum mass). }
    \label{fig:Population}
\end{figure}

We performed a joint analysis of transit photometry (CHEOPS, K2, HST/WFC3, Spitzer/IRAC) and radial velocimetry (HARPS-N, HARPS, APF/Levy, Keck/HIRES) to refine the bulk and orbital properties of the planets orbiting the bright and nearby star HD\,3167. New CHEOPS photometry and RV measurements were added to the published datasets, which were re-analyzed using improved techniques. 

We first revised the stellar age, mass and radius. The discrepancy found between the activity signal measured in the K2/RV data and the equatorial period derived from RM analysis of HD\,3167b and c can be tentatively attributed to stellar differential rotation (at a rate of $\sim$18\%) and spots located at higher latitudes than $\approxsup$50$^{\circ}$.

We confirmed the RV drift measured by \citealt{Dalal2019} as HD\,3167e, a fourth non-transiting planet with minimum mass of $\sim$10\,M$_{\oplus}$ and a 102\,day period. This discovery sheds a new light on the peculiar orbital architecture of the HD3167 system. The present-day system is dynamically stable, with the orbits of its four planets consistent with circular configurations, but HD3167b orbits close to the stellar equatorial plane while HD3167d and c are on nearly coplanar, polar orbits. Using an analytical approach to investigate the secular dynamical history of the system, we showed that the tilting of the outer planets must have happened early in the system history or planet b, initially coupled with the star, would have become coupled with planets d-c and followed their misalignment. Yet planet e cannot explain the polar orbits of planets d-c without destroying the system through Kozai-Lidov oscillations, which further implies that these three planets have low mutual inclinations. We finally explored the possibility that an unknown, more distant companion tilted the outer planetary system. Our analytical estimates, combined with constraints from velocimetry and direct imaging data (Gemini North / ‘Alopeke), however rule out this possibility unless the companion is a massive sub-solar body with a highly inclined orbit, or a star in the birth cluster that later unbound from the system. 

Our revision of the system properties improves the precision on the transiting planets radii by a factor two. When combined with the refined planet masses, this allows us to reduce the uncertainty on their density by a factor three, which is of particular interest to our understanding of their nature. 
Our internal structure retrievals show that the low density of HD\,3167b cannot be explained by a pure iron-core and silicate-mantle, suggesting that the planet contains a substantial fraction of lighter elements. It could be water, mixed with a magma ocean and/or in a steam atmosphere resilient to evaporation, or it could be a more exotic envelope made of dust and metals. In contrast we find that HD\,3167c is a mini-Neptune hosting a substantial volatile envelope, with a gaseous mass fraction of $\sim$0.2\,M$_{\oplus}$ or larger if the planet is water-poor. 
The different passbands used for transit observations allowed us to search for broadband spectral variations in these two planet radii. We measure consistent sizes for HD\,3167b, as expected from the absence of a volatile envelope. In contrast we measure significant spectral variations in the size of HD\,3167c, with a smaller radius in the infrared. 
These results strengthen the interest and amenability of HD\,3167c for follow-up observations at all wavelengths, to better determine its atmospheric structure and catalog its chemical content. Transit follow-up at high spectral resolution in the ultraviolet-optical domains, or phase curve measurements in the infrared, would also be of interest to disentangle the possible scenarios for the interior of HD\,3167b. We emphasize that our data analysis improves the precision on the planetary orbital periods by more than one order of magnitude, which will greatly help future transit follow-up.

Finally, we use atmospheric simulations to bring additional insight into the history and nature of the HD3167 planets. Due to its strong irradiation, HD\,3167b lost any primordial volatile envelope shortly after its formation. In contrast, we find that atmospheric loss did not play a significant role in the evolution of HD\,3167c, regardless of the stellar evolutionary history, so that its present-day atmosphere may still trace its primordial composition. With reasonable assumptions on the current atmospheric mass fraction of HD\,3167d-c, we further find that planet d likely lost all of its atmosphere through photo-evaporation while planet e was unaffected and retains a substantial gaseous envelope.

To summarize, our revised picture of the HD\,3167 system (Fig.~\ref{fig:Population}) consists in :
\begin{itemize}
    \item HD\,3167 : an old K-type star, initially a moderate-to-fast rotator
    \item HD\,3167b : a transiting ultra-short period planet with a heavyweight envelope, initially coupled with the star and thus still orbiting near the stellar equatorial plane
    \item HD\,3167d : a non-transiting super-Earth with no gaseous envelope, which followed a similar atmospheric evolution as planet b but a similar dynamical evolution as planet c,e
    \item HD\,3167c : a massive transiting mini-Neptune, which likely kept its primordial envelope and was tilted its present polar orbit early in the system history
    \item HD\,3167e : a non-transiting planet that likely followed the same atmospheric and dynamical evolution as planet c, which implies that it orbits in a nearby plane and is similar in mass.
\end{itemize}
In-depth characterization of the HD\,3167 system and comparison with other multi-planet systems will shed more light on its origins and evolution.

\begin{acknowledgements}
We thank the referee for their appreciative and constructive review.
We warmly thank Thibault Kuntzer for his early analysis of the HD\,3167 photometry, and Tom Mikal-Evans for sharing the HST/WFC3 broadband photometry reduced in \cite{mikal-evans_hd3167c}. This work has been carried out in the frame of the National Centre for Competence in Research PlanetS supported by the Swiss National Science Foundation (SNSF). DE acknowledges financial support from the Swiss National Science Foundation for project 200021\_200726. This project has received funding from the European Research Council (ERC) under the European Union's Horizon 2020 research and innovation programme (project {\sc Spice Dune}, grant agreement No 947634; project {\sc Four Aces}, grant agreement 724427; project {\sc SCORE}, grant agreement No 851555).  

CHEOPS is an ESA mission in partnership with Switzerland with important contributions to the payload and the ground segment from Austria, Belgium, France, Germany, Hungary, Italy, Portugal, Spain, Sweden, and the United Kingdom. The CHEOPS Consortium would like to gratefully acknowledge the support received by all the agencies, offices, universities, and industries involved. Their flexibility and willingness to explore new approaches were essential to the success of this mission. 
ACC acknowledges support from STFC consolidated grant numbers ST/R000824/1 and ST/V000861/1, and UKSA grant number ST/R003203/1. 
S.G.S. acknowledge support from FCT through FCT contract nr. CEECIND/00826/2018 and POPH/FSE (EC). 

ACC and TGW acknowledge support from STFC consolidated grant numbers ST/R000824/1 and ST/V000861/1, and UKSA grant number ST/R003203/1. 
YA, MJH and JAE acknowledge the support of the Swiss National Fund under grant 200020\_172746. 
We acknowledge support from the Spanish Ministry of Science and Innovation and the European Regional Development Fund through grants ESP2016-80435-C2-1-R, ESP2016-80435-C2-2-R, PGC2018-098153-B-C33, PGC2018-098153-B-C31, ESP2017-87676-C5-1-R, MDM-2017-0737 Unidad de Excelencia Maria de Maeztu-Centro de Astrobiologí­a (INTA-CSIC), as well as the support of the Generalitat de Catalunya/CERCA programme. The MOC activities have been supported by the ESA contract No. 4000124370. 
S.C.C.B. acknowledges support from FCT through FCT contracts nr. IF/01312/2014/CP1215/CT0004. 
XB, SC, DG, MF and JL acknowledge their role as ESA-appointed CHEOPS science team members. 
ABr was supported by the SNSA. 
This project was supported by the CNES. 
The Belgian participation to CHEOPS has been supported by the Belgian Federal Science Policy Office (BELSPO) in the framework of the PRODEX Program, and by the University of Liège through an ARC grant for Concerted Research Actions financed by the Wallonia-Brussels Federation. 
L.D. is an F.R.S.-FNRS Postdoctoral Researcher. 
This work was supported by FCT - Fundação para a Ciência e a Tecnologia through national funds and by FEDER through COMPETE2020 - Programa Operacional Competitividade e Internacionalizacão by these grants: UID/FIS/04434/2019, UIDB/04434/2020, UIDP/04434/2020, PTDC/FIS-AST/32113/2017 \& POCI-01-0145-FEDER- 032113, PTDC/FIS-AST/28953/2017 \& POCI-01-0145-FEDER-028953, PTDC/FIS-AST/28987/2017 \& POCI-01-0145-FEDER-028987, O.D.S.D. is supported in the form of work contract (DL 57/2016/CP1364/CT0004) funded by national funds through FCT. 
B.-O.D. acknowledges support from the Swiss National Science Foundation (PP00P2-190080). 
R.D.H. is funded by the UK Science and Technology Facilities Council (STFC)'s Ernest Rutherford Fellowship (grant number ST/V004735/1).
MF and CMP gratefully acknowledge the support of the Swedish National Space Agency (DNR 65/19, 174/18). 
DG gratefully acknowledges financial support from the CRT foundation under Grant No. 2018.2323 ``Gaseousor rocky? Unveiling the nature of small worlds''. 
M.G. is an F.R.S.-FNRS Senior Research Associate. 
SH gratefully acknowledges CNES funding through the grant 837319. 
KGI is the ESA CHEOPS Project Scientist and is responsible for the ESA CHEOPS Guest Observers Programme. She does not participate in, or contribute to, the definition of the Guaranteed Time Programme of the CHEOPS mission through which observations described in this paper have been taken, nor to any aspect of target selection for the programme. 
This work was granted access to the HPC resources of MesoPSL financed by the Region Ile de France and the project Equip@Meso (reference ANR-10-EQPX-29-01) of the programme Investissements d'Avenir supervized by the Agence Nationale pour la Recherche. 
ML acknowledges support of the Swiss National Science Foundation under grant number PCEFP2\_194576. 
PM acknowledges support from STFC research grant number ST/M001040/1. 
GSc, GPi, IPa, LBo, VNa and RRa acknowledge the funding support from Italian Space Agency (ASI) regulated by “Accordo ASI-INAF n. 2013-016-R.0 del 9 luglio 2013 e integrazione del 9 luglio 2015 CHEOPS Fasi A/B/C”. 
This work was also partially supported by a grant from the Simons Foundation (PI Queloz, grant number 327127). 
IRI acknowledges support from the Spanish Ministry of Science and Innovation and the European Regional Development Fund through grant PGC2018-098153-B- C33, as well as the support of the Generalitat de Catalunya/CERCA programme. 
GyMSz acknowledges the support of the Hungarian National Research, Development and Innovation Office (NKFIH) grant K-125015, a PRODEX Institute Agreement between the ELTE E\"otv\"os Lor\'and University and the European Space Agency (ESA-D/SCI-LE-2021-0025), the Lend\"ulet LP2018-7/2021 grant of the Hungarian Academy of Science and the support of the city of Szombathely. 
V.V.G. is an F.R.S-FNRS Research Associate. 

This work has made use of data from the European Space Agency (ESA) mission
{\it Gaia} (\url{https://www.cosmos.esa.int/gaia}), processed by the {\it Gaia}
Data Processing and Analysis Consortium (DPAC,
\url{https://www.cosmos.esa.int/web/gaia/dpac/consortium}). Funding for the DPAC
has been provided by national institutions, in particular the institutions
participating in the {\it Gaia} Multilateral Agreement. Some of the observations in the paper made use of the High-Resolution Imaging instrument ‘Alopeke obtained under Gemini LLP Proposal Number: GN/S-2021A-LP-105. ‘Alopeke was funded by the NASA Exoplanet Exploration Program and built at the NASA Ames Research Center by Steve B. Howell, Nic Scott, Elliott P. Horch, and Emmett Quigley. Alopeke was mounted on the Gemini North (and/or South) telescope of the international Gemini Observatory, a program of NSF’s OIR Lab, which is managed by the Association of Universities for Research in Astronomy (AURA) under a cooperative agreement with the National Science Foundation. on behalf of the Gemini partnership: the National Science Foundation (United States), National Research Council (Canada), Agencia Nacional de Investigación y Desarrollo (Chile), Ministerio de Ciencia, Tecnología e Innovación (Argentina), Ministério da Ciência, Tecnologia, Inovações e Comunicações (Brazil), and Korea Astronomy and Space Science Institute (Republic of Korea).
\end{acknowledgements}

\bibliographystyle{aa} 
\bibliography{biblio} 

\begin{appendix}

\section{CHEOPS observations}
    \begin{@twocolumnfalse}
    \begin{figure*}[ht]
    \parbox{\textwidth}{
    \centering
    \includegraphics[width=.9\hsize]{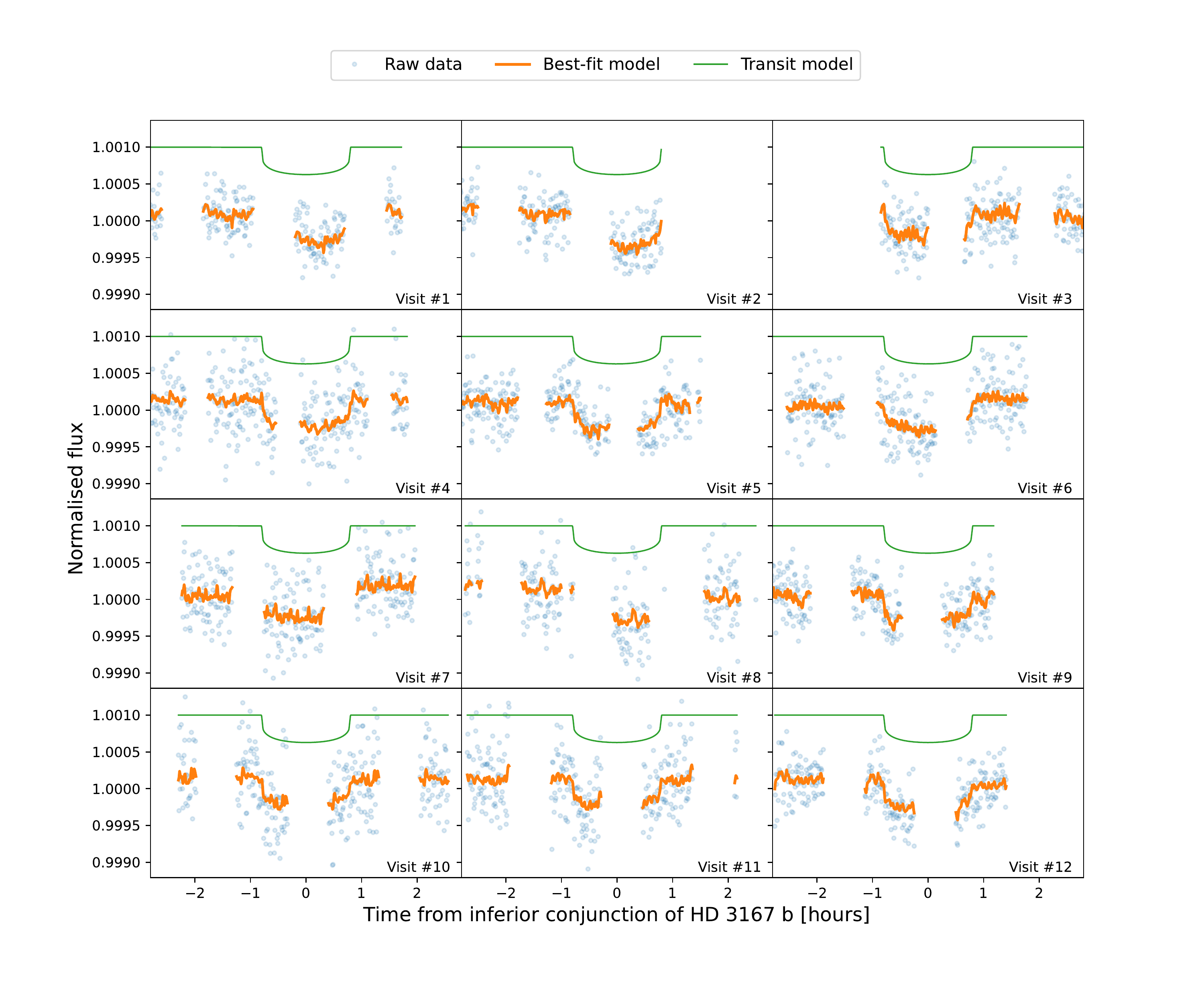}
    \caption{CHEOPS observations of HD\,3167\,b. The raw data extracted by the DRP is shown in blue and the best-fit model is represented by the orange line. The transit model is shown in green with an offset for better visibility.}
    \label{fig:cheops_raw_data}       
    }                 
    \end{figure*}
    \end{@twocolumnfalse}
\clearpage

\section{HST observations}
    \begin{@twocolumnfalse}
    \begin{figure*}[ht]
    \parbox{\textwidth}{
    \centering
    \includegraphics[width=.9\hsize]{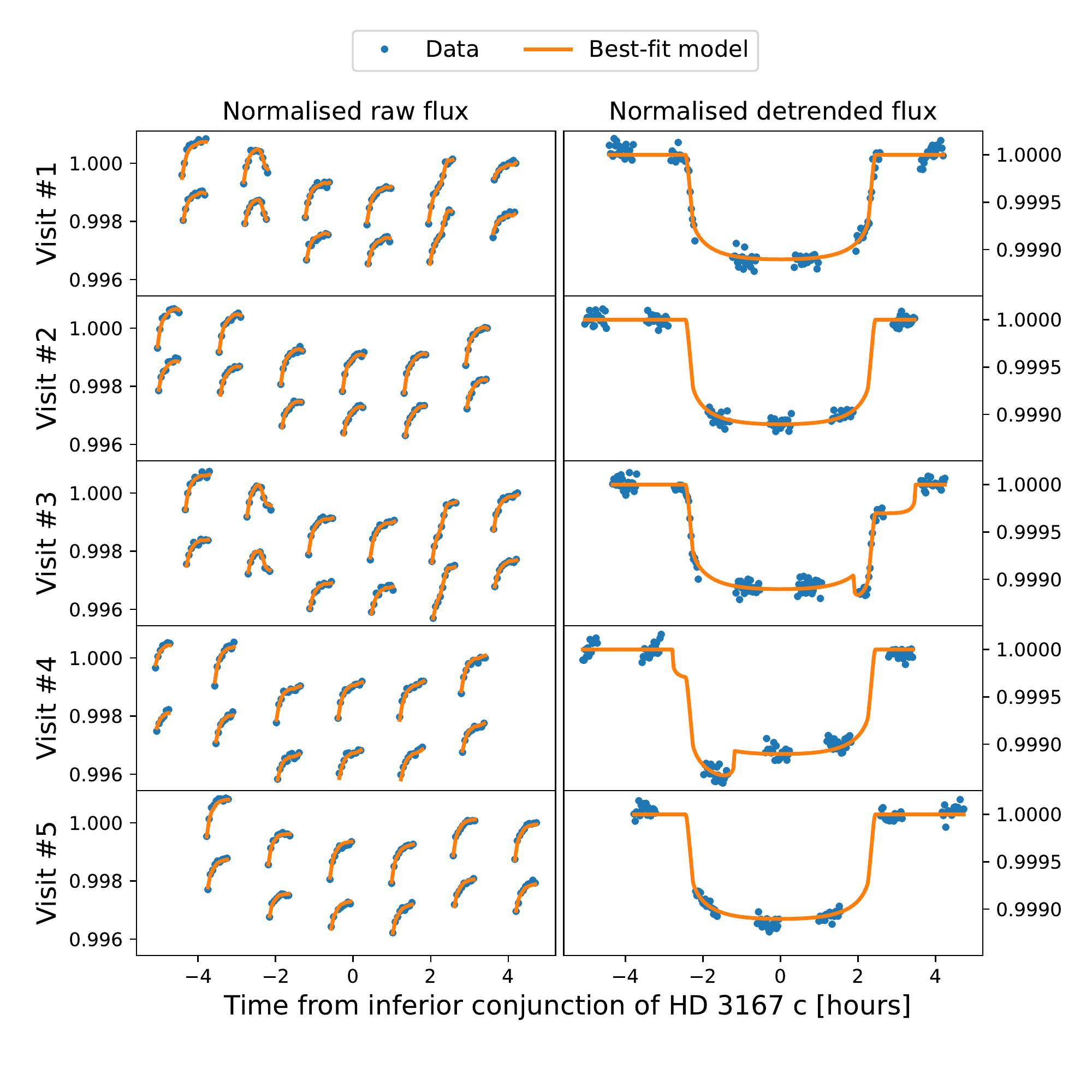}
    \caption{HST observations of HD\,3167\,c. The left column shows the raw data in blue with the best model in orange for the 5 visits. The right column represents the detrended data with the transit model in orange. Visit 3 and 4 are affected by a serendipitous transit of HD\,3167\,b.}
    \label{fig:hst_raw_data}       
    }                 
    \end{figure*}
    \end{@twocolumnfalse}
\clearpage

\end{appendix}

\end{document}